\documentclass[a4paper,11pt]{article}
\usepackage{jcappub}
\usetikzlibrary{decorations.markings}
\usepackage{subcaption}

\title{Quantum corrections to symmetron fifth-force profiles}
\author{Peter Millington}
\author{and Michael Udemba}
\affiliation{Department of Physics and Astronomy, University of Manchester, Manchester M13 9PL, U.K.}

\emailAdd{peter.millington@manchester.ac.uk}
\emailAdd{michael.udemba@manchester.ac.uk}

\abstract{Nonlinear scalar-tensor theories of gravity have been considered as candidates for dark matter and dark energy. Often, they possess screening mechanisms, which allow the fifth force mediated by the additional scalar degree(s) of freedom to evade detection from local experiments. Their classical behaviour is well studied, but their quantum nature is relatively unexplored. We outline a Green's function method for obtaining the leading-order quantum corrections to the classical symmetron field profile, in the vicinity of a spherically symmetric extended source, in the planar limit. For parameters that experiments had previously ruled out, our calculations indicate that the symmetron force may be weaker than the classical field suggests.}

\arxivnumber{2508.16726}

\begin{document}
\tikzset{->-/.style={decoration={
  markings,
  mark=at position .5 with {\arrow{>}}}, postaction={decorate}}}

\maketitle
\section{Introduction}
Scalar-tensor theories of gravity are ubiquitous in modern theoretical physics. Their study is motivated by many of the most popular extensions to the Standard Model --- including extra dimensions \cite{PhysRevD.96.123530}, supersymmetry \cite{PhysRevD.88.085038} and string theory \cite{CICOLI20241} --- and they make natural candidates to explain observations related to dark matter \cite{dm1, dm2}, dark energy \cite{de1, de2} and inflation \cite{inf, dong_symmetron_2014}. The scalar fields they introduce mediate new ``fifth forces'' between matter particles, a phenomenon which cannot be avoided without appealing to scale \cite{scale1, scale2} or conformal invariance \cite{conformal1, conformal2}. Consequently, measurements of gravity on Earth and in the Solar System would have immediately ruled out theories that give rise to fifth forces were it not for nonlinearities in their field equations, which can screen them from local tests. The precise form of these nonlinearities, and thus screening mechanisms, depends on the specific features of the model. For instance, in theories such as the chameleon model \cite{PhysRevLett.93.171104}, the mass \(m\) of the scalar field is background-dependent, becoming high in environments at least as dense as the Solar System and low in environments less dense than interstellar space. Consequently, the range of the fifth force \(m^{-1}\) is inversely related to the background density, and local observations remain consistent with the predictions of general relativity. For further details, and a review of other screening mechanisms, see ref.~\cite{brax_2021}.

The symmetron model \cite{khoury_2010}, which is similar in some ways to the chameleon model, is the scalar field under consideration in this work. In regions of space where the density of matter is higher than some critical value, the vacuum expectation value (VEV) of the field vanishes and the dynamics remain invariant under a \(\mathbb Z_2\) symmetry \(\phi\rightarrow -\phi\). In regions of space where the matter density is lower than the critical density, the symmetry is broken, and the field acquires a nonzero VEV. The coupling strength of the symmetron's fifth force is proportional to its VEV, and so it too is nonzero in low-density regions and negligible in high-density regions. With an appropriate choice of parameters, the symmetron model may take on the characteristics of conventional modified gravity models or even have sufficient energy density to contribute to the gravitational potential of a galaxy as a dark matter component \cite{PhysRevD.99.043539}. Searches for symmetrons use a wide variety of techniques, with tests including atom interferometry \cite{Burrage_2015, PhysRevD.101.083501, PhysRevLett.123.061102} and measurement of the electron magnetic dipole moment \cite{PhysRevD.97.084050}. The most recent constraints \cite{yin_experimental_2025} are derived from experiments using magnetically levitated sensors and improve on previous bounds by over six orders of magnitude. See also ref.~\cite{universe10070297} for the recent constraints covering different areas of the parameter space, including forecasts for future experiments.

Exact solutions have been found in 1D \cite{PhysRevD.97.064015, PhysRevD.103.084013, PhysRevD.99.024045} and are immediately applicable to experiments which effectively vary along one spatial dimension. These include Casimir experiments \cite{cannex}, torsion pendulum experiments \cite{PhysRevD.86.102003, PhysRevLett.110.031301} and bouncing neutron experiments \cite{cronenberg_acoustic_2018, PhysRevLett.107.111301, PhysRevD.88.083004}. The solutions have also found application in the context of the Hubble tension and the cosmic distance ladder. In particular, some authors find that the cosmic distance ladder is a probe comparable in strength to Solar System tests, and that the symmetron generally increases the Hubble tension \cite{PhysRevD.108.024007}. Interestingly, if the symmetron does not couple universally to matter and instead has a Yukawa coupling to the electron, its contribution to the electron mass could instead alleviate the tension \cite{PhysRevD.105.103536}.

Despite the obvious upshot of leaving the laws of gravity unchanged exactly where we expect them to be, the nonlinearities in screened gravity models can present significant technical challenges. For instance, unless one makes certain simplifying assumptions, the classical field equations are often very difficult, if not impossible, to solve in closed form. Technical difficulty may partially explain why much of the literature deals exclusively with the dynamics of classical fields, as may the conventional wisdom regarding quantum corrections, given formally by Erick Weinberg \cite{weinberg_2012} (see also ref.~\cite{burrage_2021}). To briefly summarise his argument, one can meaningfully talk about a classical field on some macroscopic length scale \(L\) if quantum fluctuations averaged over \(L\) are small compared to the variation of the classical field over \(L\). Simultaneously, \(L\) must be small compared to the characteristic length scale of the classical field configuration. This line of reasoning boils down to an inequality, relating Lagrangian parameters to the length scale \(L\), which must be satisfied for the classical field profile not to suffer from large quantum corrections. Applied to ``normal'' field theories, this argument tends to suggest that quantum corrections become relevant at high energies or, equivalently, small length scales. This reasoning breaks down for very light scalar fields, which have the potential to mediate quantum phenomena over astrophysical distances.

Strong hints towards a need to move beyond tree-level have existed in the fifth force literature for some time. A recent work by Burrage et al.~\cite{burrage_2021} uses Weinberg's argument and a generalisation of Derrick's theorem \cite{derrick_comments_1964} to show that the classical approximation to the symmetron field is insufficient to describe fifth forces due to highly compact sources. An earlier work by Upadhye et al.~\cite{PhysRevLett.109.041301} demonstrates that if one wants the assumption of small quantum corrections to hold for the chameleon, then theoretical considerations place very strong constraints on the mass of the scalar field. More robust treatments of the quantum nature of fifth force models have also emerged in recent years. Some derive quantum forces from path integrals in chameleon and chameleon-like models \cite{PhysRevD.99.104049, BRAX2023101294}. Others have used methods developed in open quantum systems to quantise fluctuations on spatially homogeneous background solutions of the symmetron model \cite{KADING2025101788, openquantum}. We aim to build upon these results by quantising fluctuations on spatially varying background solutions.

The aim of this paper is to determine an analytical estimate of the relevance of quantum corrections to the symmetron field. This will allow us to make the first estimation of the theoretical uncertainty on fifth-force calculations that arise from neglecting radiative effects. We begin with a spherical problem but, for tractability, we work in its planar limit. This makes the dynamics effectively one-dimensional, and so our results are also relevant for experiments with planar geometry, such as CANNEX \cite{cannex, universe7070234}.

The paper is arranged as follows. A summary of our quantisation method, as well as a Green's function method for solving the quantum equation of motion, is given in section \ref{app:a}. It will essentially provide us with a list of things to calculate: the classical field configuration, the Green's function and the renormalised tadpole contribution. Section \ref{sec:2} contains a calculation of the exact classical field profile around a non-relativistic spherical source of radius \(R\), in the limit where \(R\) is much greater than the field's Compton wavelength (analogous to Coleman's thin-wall approximation for vacuum decay \cite{callan_1977, garbrecht_2015_a}). Section \ref{sec:4} contains our computation of the Green's function, as well as its coincidence limit. This is done in the so-called planar-wall approximation, so that the sum over angular momentum numbers is approximated by an integral over a continuum \cite{garbrecht_2015_a}. Finally, in section \ref{sec:5}, making use of the Coleman--Weinberg effective potential, we renormalise the tadpole contribution. This step is performed by numerically integrating the frequency-domain Green's function along with pseudo-counterterms \cite{garbrecht_2015_a}. Once obtained, we compute the leading-order quantum correction, as well as the one-loop field and force profiles.

We find, in agreement with expectations from ref.~\cite{garbrecht_2015_a}, that the general feature of quantum corrections is a flattening of the field profile, which leads to a weakening of the force across the parameter space. The strength of this correction grows almost linearly with the strength of the self-coupling. Smaller self-couplings yield one-loop predictions that are mostly in line with tree-level predictions, as expected. However, we note that quantum corrections due to couplings to the Standard Model are expected to become the dominant contribution in this region of parameter space. Furthermore, in contrast to loop corrections on spatially homogeneous backgrounds, the shift due to loop corrections here depends on position. These are fundamental modifications to the behaviour of the theory and cannot simply be fine-tuned away. The one-loop approximation is technically invalid for non-perturbative couplings. However, the fact that the magnitude of the correction increases with the coupling strength --- already to an appreciable degree in the perturbative regime --- may change how we interpret current and future constraints on the parameter space. We discuss the implications of these results in more detail in section \ref{sec:6}.

We employ similar techniques previously applied in the context of vacuum decay \cite{garbrecht_2015_a, garbrecht_2015_b}, as well as for the Fubini-Lipatov instanton \cite{garbrecht_2018} and Higgs-Yukawa theory \cite{ai_2018}. In these contexts, the fluctuation operator has at most one negative eigenmode, a signal of the expected instability. On the other hand, for static solutions to the symmetron field equations, we expect a positive definite spectrum, and we show this to be the case in appendix \ref{sec:eigenmodes}. We opt to compute the Green's function directly instead of representing it as a sum over eigenfunctions. The process is somewhat involved, so we detail a simpler version of it in appendix \ref{sec:piecewise_con} before a more complete analysis in appendix \ref{sec:green_technical}.

Throughout this paper, we work with natural units and a mostly-minus metric signature \((+, -, -, -)\).

\section{One-loop equation of motion}
\label{app:a}
The starting point for the method that we use is the quantum effective action \cite{goldstone}, which will provide a consistent framework within which to determine the leading quantum corrections to the fifth-force profiles. In this section, we review the derivation of the quantum effective action, following the approach of ref.~\cite{GARBRECHT2016105}, and summarise the resulting one-loop equation of motion needed for our subsequent analysis.

The effective action is defined by a Legendre transform,
\begin{equation}
    \Gamma\left[\phi\right] = \max_J\left(W[J] - \int\text{d}^4x\,J(x)\phi(x)\right)\equiv \max_J\Gamma_J[\phi]\:,
\end{equation}
where \(J\) is a local source,
\begin{equation}
    W[J] = - i\hbar \ln Z[J]
\end{equation}
is the generating functional of connected correlation functions,
\begin{equation}
    Z[J] = \int\left[\text{d}\Phi\right]\,\exp\left[\frac{i}{\hbar}\left(S[\Phi] + \int\text{d}^4x\,J(x) \Phi(x)\right)\right]
\end{equation}
is the generating functional of all correlation functions, and \(S\) is the classical action. The particular source \(\mathcal J\) which extremises the effective action, hereafter called the external source, satisfies
\begin{equation}
    \left.\frac{\delta\Gamma_J[\phi]}{\delta J(x)}\right\vert_{J = \mathcal J} = 0\:.
\end{equation}
On performing this extremisation, the effective action takes the more familiar form
\begin{equation}
    \label{eq:effective_action}
    \Gamma\left[\phi\right] = W[\mathcal J] - \int\text{d}^4x\,\mathcal J(x)\phi(x)
\end{equation}
The source-dependent one-point function is
\begin{equation}
    \label{eq:one-point}
    \phi(x) \equiv \langle\hat\phi(x)\rangle = \left.\frac{\delta W[J]}{\delta J(x)}\right\vert_{J = \mathcal J}\:,
\end{equation}
source-dependent in the sense that it is a functional of the external source \(\mathcal J\). Similarly, the above expression also defines the external source \(\mathcal J\) as a functional of \(\phi\), for which we adopt the notation \(\mathcal J(x)[\phi]\). Finally, using the expression above, one can show that
\begin{equation}
    \frac{\delta\Gamma[\phi]}{\delta\phi(x)} = - \mathcal J(x)[\phi]\:,
\end{equation}
which yields the quantum-corrected equation of motion for the one-point function \(\phi\).

To leading order in \(\hbar\), the path integral in eq.~\eqref{eq:effective_action} is dominated by the solution that makes the exponent stationary. We denote this solution by \(\varphi\), and it is defined by
\begin{equation}
    \label{eq:stationary}
    \left.\frac{\delta S[\Phi]}{\delta\Phi(x)}\right\vert_{\Phi = \varphi} = -\mathcal J(x)[\phi]\:.
\end{equation}
Much like the source-dependent one-point function \(\phi\), the stationary solution \(\varphi\) is to be regarded as a functional of \(\mathcal J\). We expand around this solution by writing \(\Phi = \varphi + \sqrt\hbar\,\tilde\Phi\), with the factor \(\sqrt\hbar\) added for bookkeeping purposes. The exponent of the path integral can then be expanded as
\begin{align}
    \label{eq:action_expansion}
    \nonumber S[\Phi] + \int\text{d}^4x\,\mathcal J(x)[\phi]\Phi(x) =&\, S[\varphi] + \sqrt\hbar\int\text{d}^4x\,\left.\frac{\delta S[\Phi]}{\delta\Phi(x)}\right\vert_{\Phi=\varphi}\tilde\Phi(x)\\
    \nonumber &+\frac{\hbar}{2}\iint\text{d}^4x\,\text{d}^4y\,\tilde\Phi(x)\mathcal G^{-1}(x, y)[\phi]\tilde\Phi(y)\\
    \nonumber &+ \int\text{d}^4x\,\mathcal J(x)[\phi] \left(\phi(s) + \sqrt\hbar\,\tilde\Phi(x)\right)\\
    \nonumber =&\,S[\varphi] + \int\text{d}^4x\,\mathcal J(x)[\phi]\varphi(x)\\
    &+ \frac{\hbar}{2}\iint\text{d}^4x\,\text{d}^4y\,\tilde\Phi(x)\mathcal G^{-1}(x, y)[\phi]\tilde\Phi(y)\:,
\end{align}
wherein we defined the inverse two-point function
\begin{equation}
    \label{eq:inverse_two-point}
    \mathcal G^{-1}(x, y)[\phi] = G^{-1}(x, y)[\varphi] \equiv \left.\frac{\delta^2S[\Phi]}{\delta\Phi(x)\delta\Phi(y)}\right\vert_{\Phi = \varphi} = -\left[\square + V''(\varphi)\right]\delta^{(4)}(x - y)\:,
\end{equation}
which we will refer to as the fluctuation operator, where \(V\) is the classical potential and \(\square = \partial_\mu\partial^\mu\) is the d'Alembertian. The rest of our analysis assumes that the discrete spectrum of this operator is positive definite, which we will prove in appendix \ref{sec:eigenmodes}.

By substituting the expansion in eq.~\eqref{eq:action_expansion} back into the effective action and computing the functional integral over \(\tilde\Phi(x)\), we obtain
\begin{equation}
    \Gamma[\phi] = S[\varphi] + \hbar\,\Gamma_1[\varphi] + \int\text{d}^4x\,\mathcal J(x)[\phi]\left(\varphi(x) - \phi(x)\right)\:,
\end{equation}
where the one-loop contribution to the effective action is given by
\begin{equation}
    \Gamma_1[\varphi] = \frac{i}{2}\tr\log G^{-1}(x, y)[\varphi]\:.
\end{equation}
We now eliminate \(\phi\) in favour of \(\varphi\) by writing \(\varphi(x) = \phi(x) - \hbar\,\delta\varphi(x)\) and expanding the left-hand side as
\begin{equation}
    \Gamma[\phi] = \Gamma[\varphi] + \int\text{d}^4x\,\left.\frac{\delta \Gamma[\phi]}{\delta\varphi(x)}\right\vert_{\phi = \varphi}\left(\phi(x) - \varphi(x)\right)\:,
\end{equation}
up to one loop, while the right-hand side is
\begin{equation}
    \Gamma[\phi] = S[\varphi] + \hbar\,\Gamma_1[\varphi] + \int\text{d}^4x\,\frac{\delta \Gamma[\phi]}{\delta\phi(x)}\left(\phi(x) - \varphi(x)\right)\:.
\end{equation}
Immediately, it follows that
\begin{equation}
    \Gamma[\varphi] = S[\varphi] + \hbar\,\Gamma_1[\varphi]\:.
\end{equation}
Note that, by following the approach of ref.~\cite{GARBRECHT2016105}, the one-point function here is the same as that which appears in the saddle-point approximation to the path integral. As a result, the source is constrained such that \(\varphi\) solves the quantum equation of motion,
\begin{equation}
    \left.\frac{\delta\Gamma[\phi]}{\delta\phi(x)}\right\vert_{\phi = \varphi} = 0\:.
\end{equation}

We interpret the stationary solution \(\varphi\) as the one-loop quantum field. It is given by the sum of the classical field plus fluctuations, \(\varphi = \varphi_\text{cl} + \delta\phi\). In terms of the classical action, the quantum equation of motion reads
\begin{equation}
    \frac{\delta S[\varphi]}{\delta\varphi(x)} + \hbar\left.\frac{\delta\Gamma_1[\phi]}{\delta\phi(x)}\right\vert_{\phi = \varphi} = 0\:.
\end{equation}
The stationarity condition in eq.~\eqref{eq:stationary} implies that the above may be rendered in the form
\begin{equation}
    \mathcal J(x)[\phi] = \hbar\left.\frac{\delta\Gamma_1[\phi]}{\delta\phi(x)}\right\vert_{\phi = \varphi}\:.
\end{equation}
It is this relation which constrains the external source. One can solve this condition to find that the consistent choice of an external source for our desired solution \(\varphi\) is
\begin{equation}
    \mathcal J(x)[\phi] = 3i\lambda \hbar G(x)[\varphi_\text{cl}]\varphi_\text{cl}(x) \equiv \hbar\Pi(x)[\varphi_\text{cl}]\varphi_\text{cl}(x)\:,
\end{equation}
where \(\Pi = 3i\lambda G\) is the tadpole contribution.

At the one-loop level, the quantum equation of motion therefore reads
\begin{equation}
    \label{eq:quantum_eom}
    \square\varphi + V_\text{qu}'(\varphi) = 0\:,
\end{equation}
where the derivative effective potential for the one-loop quantum field is defined by
\begin{equation}
    V_\text{qu}'(\varphi) = V'(\varphi) - \hbar \Pi(x)[\varphi_\text{cl}]\varphi_\text{cl}(x)\:.
\end{equation}
When expressed in terms of the classical field to first order in \(\delta\phi\), the equation of motion for \(\varphi\) reduces to an equation for \(\delta\phi\) alone,
\begin{equation}
    G^{-1}(x)[\varphi_\text{cl}]\delta\phi = \Pi(x)[\varphi_\text{cl}]\varphi_\text{cl}(x)\:,
\end{equation}
which may readily be solved as
\begin{equation}
    \label{eq:q_correction}
    \delta\phi(x) = \int\text{d}^4y\,G(x, y)[\varphi_\text{cl}]\Pi(y)[\varphi_\text{cl}] \varphi_\text{cl}(y)\:.
\end{equation}
Computing the one-loop correction to the field profile of the symmetron and the resulting one-loop correction to the fifth force is the aim of this work. The integral above gives us a clear order of operations to carry out. First, we calculate the classical field profile \(\varphi_\text{cl}\). Then, we compute the Green's function of the fluctuation operator. Finally, we determine the coincidence limit of the Green's function, regularise it and renormalise it to obtain the tadpole contribution, and use eq.~\eqref{eq:q_correction} to determine the leading quantum correction to the classical field profile.
\section{Classical symmetron model}
\label{sec:2}
The first step in the calculation is to compute the static profile of the classical symmetron field in the vicinity of an extended, spherical source. The full equation of motion cannot be solved in closed form, but we can obtain an analytical estimate of the true solution in the limit of large sources.

\subsection{Theoretical background}
The action for the symmetron field takes the form \cite{khoury_2010}
\begin{equation}
    S[\phi] = \int\text{d}^4x\,\sqrt{-g}\left[\frac{M_\text{Pl}^2}{2}R + \frac{1}{2}g^{\mu\nu}\partial_\mu\phi\partial_\nu\phi-V(\phi)\right] + \int\text{d}^4x\,\sqrt{-\tilde{g}}\mathcal L_\text{m}(\tilde{g}_{\mu\nu})\:,
\end{equation}
where \(g\) is the determinant of the Einstein-frame metric \(g_{\mu\nu}\), \(R\) is the Ricci scalar, \(M_\text{Pl}\) is the reduced Planck mass and \(\mathcal L_\text{m}\) denotes the Lagrangian density of matter fields. These fields are minimally coupled to the Jordan-frame metric,
\begin{equation}
    \tilde{g}_{\mu\nu} = A(\phi)^2g_{\mu\nu}\:,
\end{equation}
which is related to the Einstein-frame metric by a positive rescaling \(A(\phi)^2\). By the stationary action principle, one obtains the equation of motion of the scalar field \(\phi\),
\begin{equation}
    \square\phi + \od{V}{\phi} - A(\phi)^3\od{A}{\phi}\,\tilde T = \square \phi + \frac{\text{d}V_\text{eff}}{\text{d}\phi} = 0\:,
\end{equation}
\(\tilde T\) is the trace of the matter energy-momentum tensor in the Jordan frame, and \(V_\text{eff}\) is the effective potential. Our matter distribution is non-relativistic and pressureless, so we may write \(\tilde T \approx -\tilde\rho\), with \(\rho = A(\phi)^3\tilde\rho\), as its energy density.

The functions \(V(\phi)\) and \(A(\phi)\) are chosen so that the effective potential \(V_\text{eff}\) has a spontaneously broken \(\mathbb Z_2\) symmetry \(\phi\rightarrow -\phi\) in the limit \(\rho\rightarrow 0\) \cite{Hinterbichler:2011ca}. The potential \(V(\phi)\) takes the form \cite{burrage_2021} 
\begin{equation}
    V(\phi) = -\frac{1}{2}\mu^2\phi^2 + \frac{1}{4}\lambda\phi^4 + \frac{\mu^4}{4\lambda}\:,
\end{equation}
where \(\mu^2>0\) is a tachyonic mass term and \(\lambda>0\) is a dimensionless self-coupling. We have included an additive constant which does not affect the dynamics of the classical field, but is convenient for ensuring that the effective potential vanishes when the field obtains its expectation value \(\pm v\) in the Minkowski vacuum, where
\begin{equation}
    \label{eq:vev}
    v = \frac{\mu}{\sqrt \lambda}\:.
\end{equation}
This amounts to a tuning of the cosmological constant, which is not relevant to the analysis that follows.

For the universal coupling \(A(\phi)\) we have
\begin{equation}
    A(\phi) = 1 + \frac{1}{2M^2}\phi^2\:,
\end{equation}
where \(M\) is a mass scale that controls the strength of the coupling to matter.
\begin{figure}[t]
    \centering
    \includegraphics{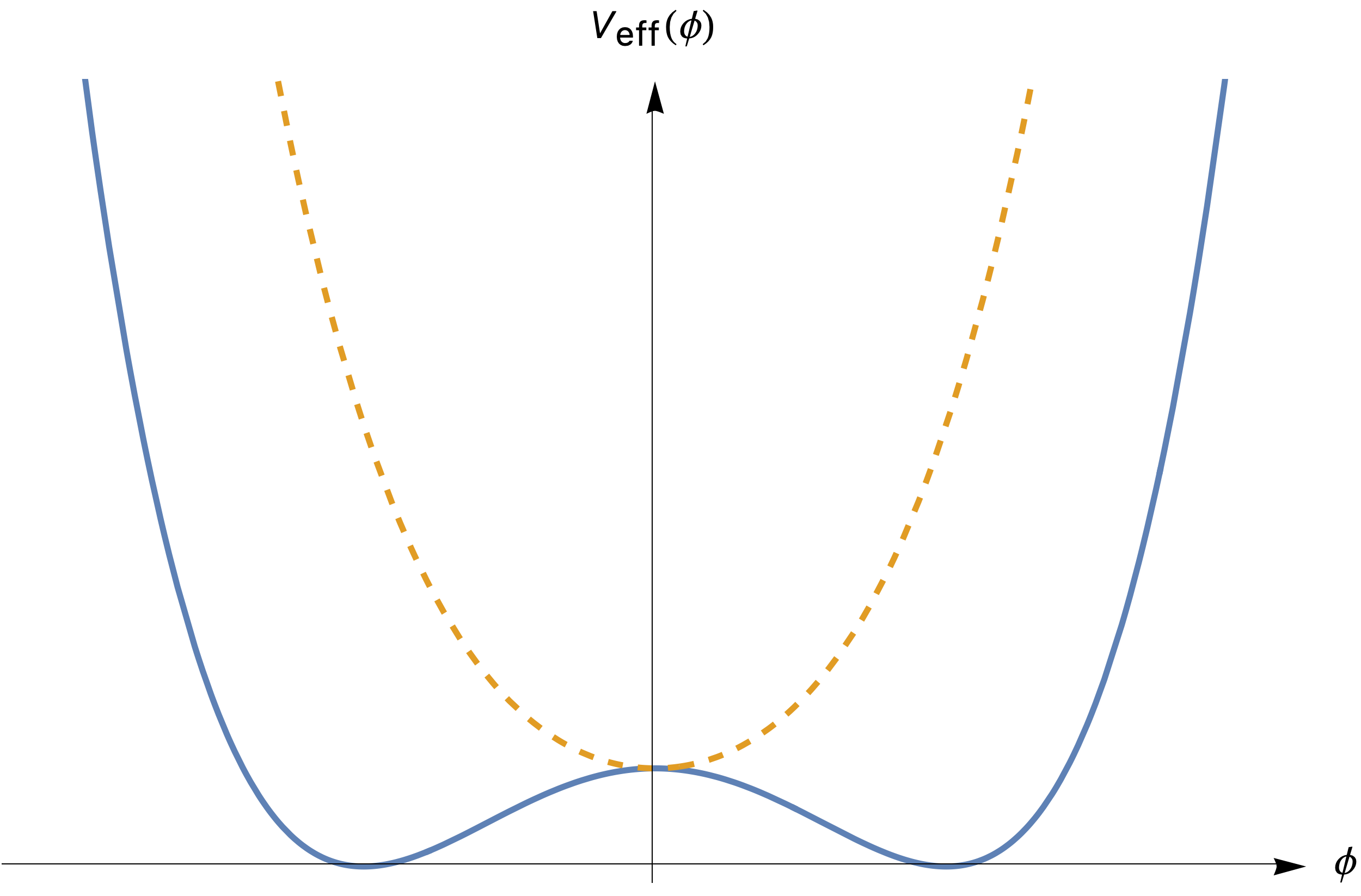}
    \caption{An illustrative plot of the symmetron effective potential with background densities \(\rho>\mu^2M^2\) (orange, dashed) and \(\rho<\mu^2M^2\) (blue, solid).}
    \label{fig:effective_potential}
\end{figure}
In summary, the effective potential for the symmetron field is
\begin{equation}
    \label{eq:effective_potential}
    V_\text{eff}\left(\phi\right) \equiv V(\phi) + A(\phi)\rho =  \frac{1}{2}\left(\frac{\rho}{M^2} - \mu^2\right)\phi^2 + \frac{\lambda}{4}\phi^4 + \frac{\mu^4}{4\lambda}\:.
\end{equation}
See figure \ref{fig:effective_potential} for a plot of this function for both screened and unscreened sources. From the figure, we can directly observe that for \(\rho > \mu^2 M^2\) the field is screened and sits around the minimum at \(\phi = 0\), whereas for \(\rho<\mu^2 M^2\) the field is unscreened and sits in one of the two nonzero minima.

We want to solve for the static classical field configuration, which obeys the equation
\begin{equation}
    \label{eq:static_eom}
    \nabla^2\phi = \od{V_\text{eff}}{\phi}\:.
\end{equation}
We take the extended matter source to be a sphere of uniform density \(\rho_0\) and radius \(R\), described by the function
\begin{equation}
    \label{eq:density}
    \rho(r) = \rho_0\Theta(R - r)\:,
\end{equation}
where \(\Theta\) is the Heaviside step function, or unit step function. The field is therefore spherically symmetric, and the static equation of motion reduces to a radial equation,
\begin{equation}
    \label{eq:static-eom}
    \od{^2\phi}{r^2} + \frac{2}{r}\od{\phi}{r} = \od{V_\text{eff}}{\phi}\:,
\end{equation}
with regularity condition 
\begin{equation}
    \lim_{r\rightarrow 0}\od{\phi}{r} = 0\:.
\end{equation}
The two vacua are physically equivalent, so we are free to set
\begin{equation}
    \label{eq:boundary}
    \lim_{r\rightarrow\infty}\phi(r) = v\:,
\end{equation}
making the classical field a positive function, as well as its derivative. The symmetron model also admits solutions which interpolate between the two degenerate minima, leading to the formation of domain walls. Such solutions are of higher energy than the configuration that we are solving for, and so will have sub-leading contributions to the quantum corrections. See refs.~\cite{PhysRevD.106.043528, PhysRevD.90.124041, clements2023detectingdarkdomainwalls} for discussions of domain walls in scalar-tensor theories.

\subsection{Thin-wall approximation}
We want an analytical understanding of the quantum corrections, but the above system of equations cannot be solved in closed form. One could obtain an approximate solution by linearising the equation of motion \cite{khoury_2010}, but then the Green's function would be independent of the field. Instead, we consider a source whose radius \(R\) is large compared to the characteristic scale of variations of the classical field. That is, \(\mu R \gg 1\). Then, \(\phi\) remains small and grows slowly in the interval \(0< r < R\), experiences a rapid jump around \(r = R\) and then quickly flattens out to \(v\) as \(r\rightarrow \infty\). Thus, we may drop the damping term in eq.~\eqref{eq:static-eom} and turn what was once a radial problem into a purely one-dimensional problem. It is perhaps more conceptually clear to think of this approximation as an asymptotic approximation in the limit \(\mu R \gg 1\). For notational clarity, we write \(s = r - R\) so that the equation of motion reads
\begin{equation}
    \label{eq:symmetron_thin_wall}
    \od{^2\phi }{s^2} = \od{V_\text{eff}}{\phi}\:.
\end{equation}
We identify the origin of the source with \(s\rightarrow -\infty\), and its boundary with \(s = 0\). Spatial infinity is intuitively \(s\rightarrow \infty\). The regularity condition no longer holds by assumption, but does so as a corollary to the condition that the field is completely screened at the origin of the ``infinite'' source,
\begin{equation}
    \lim_{s\rightarrow -\infty}\phi(s) = 0\:.
\end{equation}
This approximation is conceptually similar to Coleman's thin-wall approximation for vacuum decay (see, for example, ref. \cite{coleman_1977}), and shall be referred to as such hereafter.

Equation \eqref{eq:symmetron_thin_wall} may equivalently be written as
\begin{equation}
    \label{eq:order_reduction}
    \frac{1}{2}\od{}{s}\left(\od{\phi}{s}\right)^2 = \od{V_\text{eff}}{s}\:,
\end{equation}
which can be directly integrated on the domain \((s_1, s_2)\subseteq \R\) to yield
\begin{equation}
    \left(\left.\od{\phi}{s}\right\vert_{s = s_2}\right)^2 - \left(\left.\od{\phi}{s}\right\vert_{s = s_1}\right)^2 = 2\left[V_\text{eff}(\phi(s_2)) - V_\text{eff}(\phi(s_1))\right]\:.
\end{equation}
We denote by \(\phi_-(s)\) and \(\phi_+(s)\) the interior and exterior fields respectively. Separately solving for them is simply a matter of picking appropriate values for \(s_1\) and \(s_2\). The full profile
\begin{equation}
    \phi(s) = \Theta(-s)\phi_-(s) + \Theta(s)\phi_+(s)
\end{equation}
is then uniquely determined by imposing continuity of the solution and its first derivative at \(s = 0\). 

For the exterior solution, we choose \(s_1 = s > 0\) and \(s_2 = \infty\) to arrive at the equation
\begin{equation}
    \left(\od{\phi_+}{s}\right)^2 = - \mu^2\phi_+^2 + \frac{\lambda}{2}\phi_+^4 + \frac{\mu^4}{2\lambda}\:.
\end{equation}
Using eq.~\eqref{eq:vev} to eliminate \(\lambda\) leads to a rather simple equation for the normalised field \(\chi_\pm = \phi_\pm/v\),
\begin{equation}
    \od{\chi_+}{s} = \gamma\left(1 - \chi_+^2\right)\:,
\end{equation}
where
\begin{equation}
    \gamma = \frac{\mu}{\sqrt 2}
\end{equation}
is half the effective mass of exterior scalar fluctuations \(m_+ = \sqrt 2\mu\). By comparison with the identity
\begin{equation}
    \od{}{x}\tanh(x) = \sech^2(x) = 1 - \tanh^2(x)\:,
\end{equation}
it is clear that the general solution is
\begin{equation}
    \chi_+(s) = \tanh\left(\gamma s + c_+\right)\:,
\end{equation}
for some real \(c_+\).

For the interior contribution, choosing \(s_1 = -\infty\) and \(s_2 = s < 0\) implies
\begin{equation}
    \left(\od{\phi_-}{s}\right)^2 = \left(\frac{\rho_0}{M^2} - \mu^2\right)\phi_-^2 + \frac{\lambda}{2}\phi_-^4\:.
\end{equation}
Performing similar manipulations as before leads to a slightly more complicated equation for the normalised interior field,
\begin{equation}
    \od{\chi_-}{s} = \gamma\chi_-\sqrt{g^2 + \chi_-^2}\:,
\end{equation}
where 
\begin{equation}
    g = \sqrt{2\left(\frac{\rho_0}{\mu^2 M^2} - 1\right)}
\end{equation}
is the effective mass of interior scalar fluctuations in units of \(\gamma\). By the identity
\begin{figure}[t]
    \centering
    \includegraphics[scale=1]{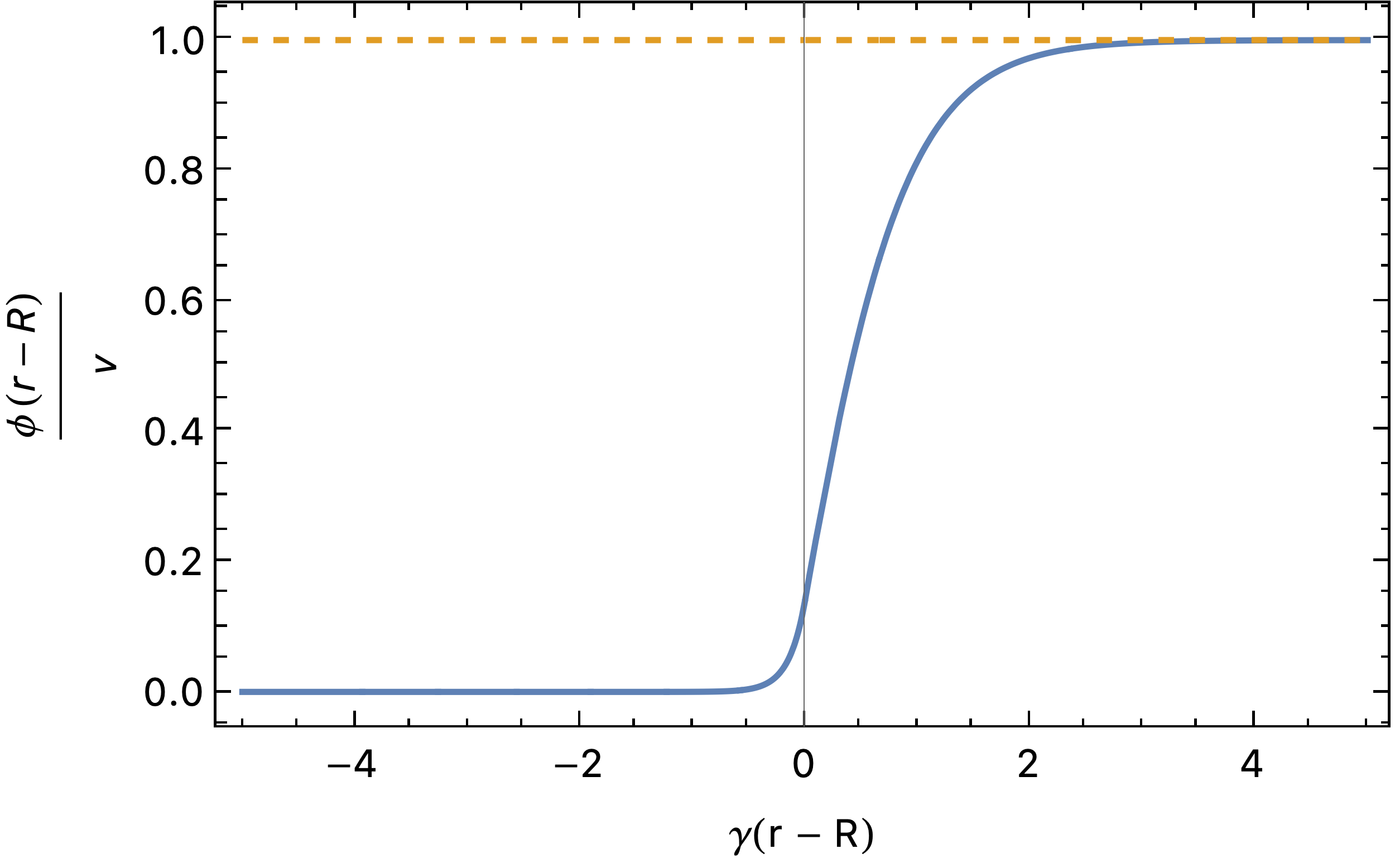}
    \caption{The normalised symmetron field profile for \(\rho_0/\mu^2M^2 = 25\), a value which sits in the region of phase space accessible to hydrogen and muonium spectroscopy \cite{PhysRevD.107.044008}.}
    \label{fig:classical_profile}
\end{figure}
\begin{equation}
    \od{}{x}\csch(x) = -\csch(x)\coth(x) = -\csch(x)\sqrt{1 + \csch^2(x)}\:,
\end{equation}
we see that the general solution is
\begin{equation}
    \chi_-(s) = g\,\csch\left(c_-- g\gamma s\right)\:,
\end{equation}
with \(c_- \in \R\).

Continuity with \(\chi_+\) is easily achieved by defining the constants \(c_\pm\) in terms of the value of the field at the boundary of the source \(\chi_0\), i.e.
\begin{equation}
    c_+ = \artanh\chi_0 \text{ and } c_- = \arcsch\left(\frac{\chi_0}{g}\right)\:.
\end{equation}
Continuity of the first derivative then fixes this boundary value to be
\begin{equation}
    \label{eq:boundary_val}
    \chi_0 = \frac{1}{\sqrt{2+g^2}}\:.
\end{equation}
In conclusion, the static classical field profile is
\begin{equation}
    \label{eq:exact_solution}
    \varphi_\text{cl}(s) = v\left\{\Theta(-s)g\,\csch\left[\arcsch\left(\frac{\chi_0}{g}\right) - g\gamma s\right] + \Theta(s)\tanh\left(\gamma s + \artanh\chi_0\right)\right\}
\end{equation}
in a thin-wall approximation. The solution is stable, as we prove in appendix \ref{sec:eigenmodes}, and a plot is given in figure \ref{fig:classical_profile}. Our solution agrees with those found elsewhere in the literature \cite{PhysRevD.97.064015}.

\section{Green's function}
\label{sec:4}
We now turn to the calculation of the Green's function. By means of the thin-wall approximation, we again reduce the spherical problem to a one-dimensional one. As one would expect from the problem of eigenmodes of a system with a piecewise continuous mass (see, for example, appendix \ref{sec:piecewise_con}), the linearly independent solutions which compose the Green's function behave differently for different frequency intervals. For frequencies below the mass of exterior scalar fluctuations, the solutions asymptotically vanish. Between the masses of interior and exterior fluctuations, the interior modes vanish asymptotically whereas the exterior modes oscillate. Above the interior mass, both interior and exterior modes oscillate unattenuated. However, the expressions for each frequency can be rendered equivalent by a Feynman contour prescription (see appendix \ref{sec:green_technical} for details).

The differential operator from which the Green's function derives is related to the second functional derivative evaluated on \(\varphi_\text{cl}\), slightly altered from the more general eq.~\eqref{eq:inverse_two-point} to include the effective potential,
\begin{equation}
    \label{eq:sfd}
    \left.\fd{^2S[\Phi]}{\Phi(x)\delta\Phi(y)}\right\vert_{\Phi = \varphi_\text{cl}} = \left(-\square - \od{^2V_\text{eff}}{\varphi_\text{cl}^2}\right)\delta^{(4)}(x - y)\:.
\end{equation}
We again refer to the operator in parentheses as the fluctuation operator,
\begin{equation}
    \label{eq:fluc}
    \text{L} = \square + \od{^2V_\text{eff}}{\varphi_\text{cl}^2}\:,
\end{equation}
with the sign convention chosen such that the Green's function satisfies
\begin{equation}
    \text{L}G(x, x') = -\delta^{(4)}(x - x')\:.
\end{equation}
A natural next step is to transform to the frequency-domain, in which the spatial part of the Green's function \(G(\mathbf x, \mathbf x';E)\), defined by the Fourier transform
\begin{equation}
    G(\mathbf x, \mathbf x';t - t') = \int\frac{\text{d}E}{2\pi}e^{-iE(t - t')}G(\mathbf x,\mathbf x';E)\:,
\end{equation}
satisfies the equation
\begin{equation}
    \label{eq:spatial}
    \left[- \nabla^2 - E^2  + \od{^2V_\text{eff}}{\varphi_\text{cl}^2}\right]G(\mathbf x, \mathbf x';E) = -\delta^{(3)}(\mathbf x - \mathbf x')\:.
\end{equation}
From here, we use a partial-wave decomposition,
\begin{equation}
    \label{eq:partial_wave}
    G(\mathbf x, \mathbf x'; E) = \sum_{l=0}^\infty\sum_{m=-l}^l G_l(r, r'; E)Y_l^m\left(\boldsymbol\Omega\right)\bar Y_l^m\left(\boldsymbol \Omega'\right)\:,
\end{equation}
where the \(Y_l^m\) are spherical harmonics, \(l \in \mathbb N\), \(m\in [-l, l]\) is an integer and \(\boldsymbol \Omega\) is a vector containing angular coordinates.
The radial part \(G_l(r, r';E)\) of the Green's function therefore satisfies
\begin{equation}
    \label{eq:radial}
    \left[-\od{^2}{r^2} - \frac{2}{r}\frac{\text{d}}{\text{d}r} + \frac{l(l+1)}{r^2} - E^2  + \od{^2V_\text{eff}}{\varphi_\text{cl}^2}\right]G_l(r, r'; E) = -\frac{\delta(r - r')}{r^2}\:,
\end{equation}
with \(r>0\). The sum over \(m\) follows from the spherical harmonic addition theorem,
\begin{equation}
    \frac{1}{4\pi}\left(2l + 1\right) P_l\left(\cos\theta\right) = \sum_{m=-l}^lY_l^m\left(\boldsymbol\Omega\right)\bar Y_l^m\left(\boldsymbol \Omega'\right)\:,
\end{equation}
where \(\theta\) is the angle between \(\mathbf x\) and \(\mathbf x'\) and \(P_l\) is the Legendre polynomial of order \(l\), allowing us to write 
\begin{equation}
    \label{eq:sum_of_radial_tw}
    G(\mathbf x, \mathbf x'; E) = \frac{1}{4\pi}\sum_{l = 0}^\infty\left(2l + 1\right)P_l\left(\cos\theta\right)G_l(r, r'; E)\:.
\end{equation}
Substituting this into eq.~\eqref{eq:radial} and taking the thin-wall approximation (which amounts to dropping the damping term and replacing \(r\rightarrow R\) in the centrifugal potential and discontinuity \cite{garbrecht_2015_a}) yields
\begin{equation}
    \label{eq:radial_green_general}
    \left[-\od{^2}{s^2} + \frac{l(l+1)}{R^2} - E^2  + \od{^2V_\text{eff}}{\varphi_\text{cl}^2}\right]G_l(s, s'; E) = -\frac{\delta(s - s')}{R^2}\:.
\end{equation}

Details for how we solve for \(G_l(s, s'; E)\) are given in appendix \ref{sec:green_technical}. We summarise the results here. Let \(u(s) = \tanh(\gamma s + c_+)\) and \(z(s) = \csch^2(c_- - g\gamma s)\). Furthermore, define the parameters 
\begin{equation}
    \label{eq:n_iep}
    n = \frac{1}{\gamma}\sqrt{\frac{l(l+1)}{R^2} + 4\gamma^2 - \left(E^2 + i\epsilon\right)}
\end{equation}
and
\begin{equation}
    \label{eq:a_iep}
    a = \frac{1}{2g\gamma}\sqrt{\frac{l(l+1)}{R^2} + g^2\gamma^2 - \left(E^2 + i\epsilon\right)}\:,
\end{equation}
with \(\epsilon > 0\), where the \(i\epsilon\) prescription ensures the following expressions are valid for all \(E\). When both \(s\) and \(s'\) are positive, the solution is
\begin{equation}
    \label{eq:pp}
    G_{l, +}^>(s, s';E) = -\frac{\pi}{2\gamma  R^2}\csc\left(n\pi\right)P_2^{-n}(u)\left(P_2^n(u') + \frac{W\left(P_2^n(u), F_a(z)\right)}{W\left(F_a(z), P_2^{-n}(u)\right)}P_2^{-n}(u')\right)\:,
\end{equation}
where \(W\) denotes the Wro{\'n}skian with respect to \(s\), evaluated at \(s = 0\). \(P_2^n\) is the associated Legendre function of the first kind and \(F_a\) is related to the hypergeometric function \({}_2F_1\) by
\begin{equation}
    F_a(z) = z^a {}_2F_1\left(a - 1, a + \frac{3}{2};2a + 1;-z\right)\:.
\end{equation}
When both \(s\) and \(s'\) are negative, we denote the solution by
\begin{equation}
    \label{eq:mm}
    G_{l, -}^>(s, s';E) = -\frac{1}{4 a g\gamma  R^2}F_a(z')\left(F_{-a}(z) + \frac{W\left(P_2^{-n}(u), F_{-a}(z)\right)}{W\left(F_a(z), P_2^{-n}(u)\right)}F_a(z)\right)\:.
\end{equation}
Finally, when the signs of \(s\) and \(s'\) differ, the Green's function \(G_l(s, s'; E)\) is given by
\begin{equation}
    \label{eq:pm}
    G^>_{l, \pm}(s, s';E) = \frac{P_2^{-n}(u)F_a(z')}{W(F_a(z), P_2^{-n}(u))R^2}\:.
\end{equation}
The superscript \(>\) corresponds to \(s> s'\). Solutions for \(s<s'\) are obtained from the relation \(G_l^<(s, s';E) = G_l^>(s', s;E)\).

All that remains is the sum over \(l\), a task which cannot be completed in closed form but for which we determine an analytical approximation. In the so-called planar-wall approximation \cite{garbrecht_2015_a}, which is compatible with the thin-wall approximation, we consider the source to be so large that its surface can be taken to be a plane. Let \(z_\perp\) be the coordinate perpendicular to the surface and \(\mathbf z_\parallel\) be a vector containing the coordinates which lie within it. We may Fourier transform with respect to the parallel coordinates, introducing a 2-momentum \(\mathbf k\) in the process,
\begin{equation}
    \label{eq:continuum_app}
    G(\mathbf x, \mathbf x';E) = \int\frac{\text{d}^2\mathbf k}{\left(2\pi\right)^2}e^{i\mathbf k\cdot\left(\mathbf z_\parallel - \mathbf z'_\parallel\right)}G(z, z';\mathbf k, E)\:,
\end{equation}
writing \(z_\perp = z\) for ease. The function \(G(z, z';\mathbf k, E)\) satisfies the equation
\begin{equation}
    \left[-\pd{^2}{z^2} + k^2 - E^2  + \od{^2V_\text{eff}}{\varphi_\text{cl}^2}\right]G(z, z';\mathbf k, E) = -\delta(z - z')\:,
\end{equation}
which is just the equation for \(R^2G_l(s, s';E)\) but with the discrete parameter \(l(l+1)/R^2\) replaced by the continuous parameter \(k^2\). We will adopt the notation \(G(s, s';E, k)\) to reflect this change.

\section{Tadpole contribution and quantum corrections}
\label{sec:5}
We are now able to compute the tadpole contribution, which is essentially the renormalised coincidence limit of the Green's function that we found in the last section. To do so, we need to integrate over the modes from the continuum approximation, resulting in a divergent loop integral that we regularise with a simple momentum cutoff. While the counterterms can be found exactly, the renormalised tadpole contribution can only be found via a numerical method, by subtracting divergences at the level of the integrand in the loop integral \cite{garbrecht_2015_a, garbrecht_2015_b}.

\subsection{Coincidence limit}
\label{sec:cont-app}
Following from eq.~\eqref{eq:continuum_app}, the coincidence limit of the Green's function in the planar-wall approximation takes the form
\begin{equation}
    \label{eq:cont-modes}
    G(\mathbf x;E) = R^2\int_0^\infty\frac{\text{d}k}{2\pi}\,k\,G(s; E, k)\:,
\end{equation}
where \(G(\mathbf x;E)\equiv G(\mathbf x, \mathbf x;E)\) and \(G(s;E, k)\equiv G(s, s;E, k)\). Recall that the \(R^2\) prefactor follows from taking the thin-wall approximation (see eq.~\eqref{eq:radial_green_general}). Adding back the time dependence and taking the coincidence limit in time yields
\begin{equation}
    \label{eq:coinc-cont}
    G(x) = 2R^2\int_0^\infty\frac{\text{d}E}{2\pi}\,\int_0^\infty\frac{\text{d}k}{2\pi}\,k\,G(s;E, k)\:,
\end{equation}
where \(G(x)\equiv G(x, x)\), and we have exploited the fact that \(G(s;E, k)\) is an even function in \(E\). This loop integral is well defined after making a Wick rotation (at least exterior to the source), the action of which is equivalent to replacing every occurrence of \(E^2\) with \(-E^2\) at the cost of a factor of \(i\), giving
\begin{equation}
    G(x) = 2iR^2\int_0^\infty\frac{\text{d}E}{2\pi}\,\int_0^\infty\frac{\text{d}k}{2\pi}\,k\,G(s;E^2 + k^2)\:.
\end{equation}
This integral is UV-divergent, so we regularise it with a simple UV cutoff \(\Lambda\) and write
\begin{figure}
    \centering
    \includegraphics{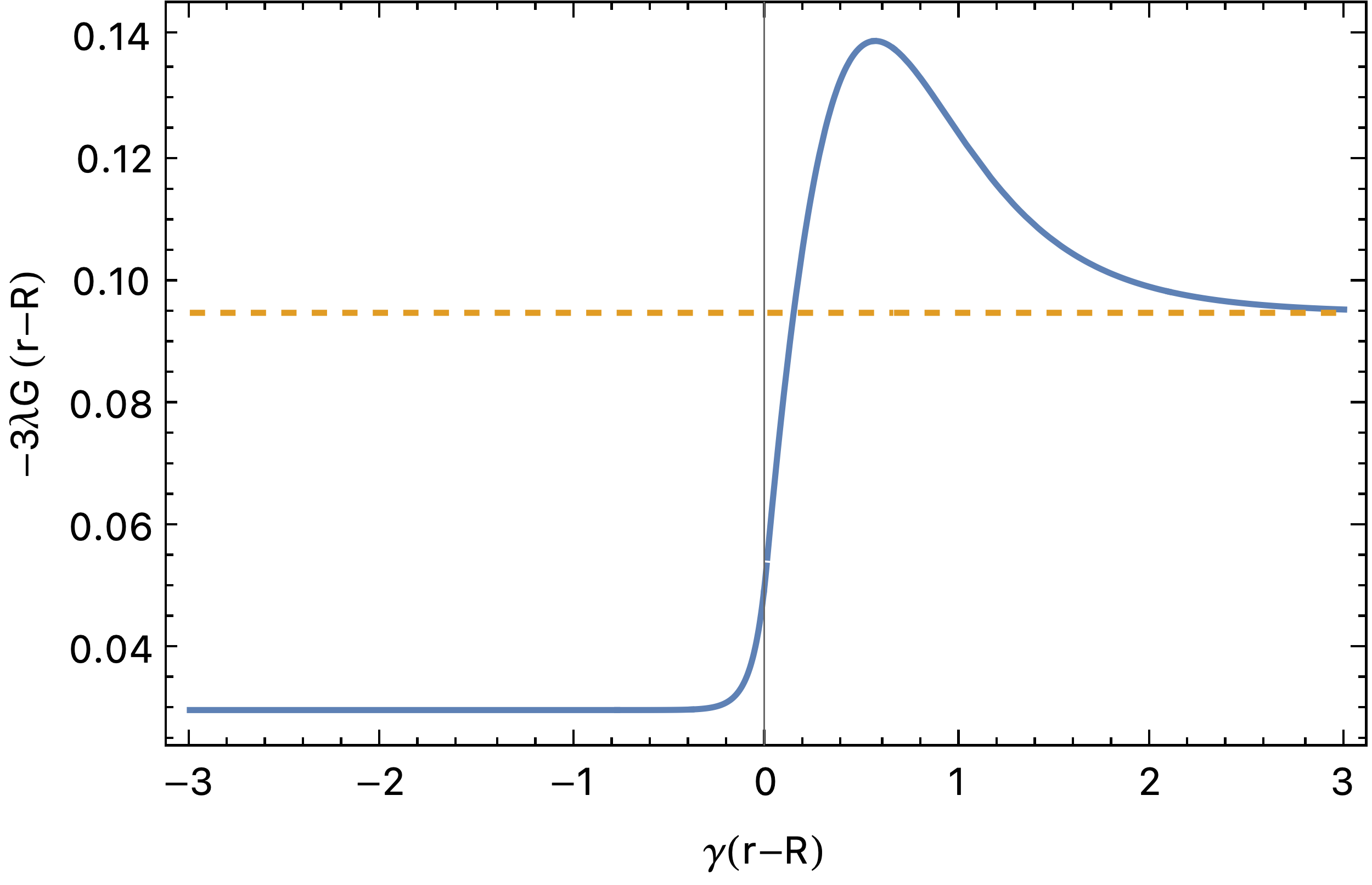}
    \caption{The coincidence limit of the Green's function, where we have chosen \(p = \gamma\). We have set \(\lambda = 0.1\) and \(\rho/\mu^2 M^2 = 25\).}
    \label{fig:green}
\end{figure}
\begin{equation}
    \label{eq:loop}
    G(x) = \frac{iR^2}{2\pi^2}\int_0^\Lambda\text{d}p\,p^2 G(s;p)\:,
\end{equation}
where \(p^2 = E^2 + k^2\) is the Euclidean momentum. One would expect \(G(s;p)\) to be flat for \(s\rightarrow \pm\infty\), where the classical configuration is also essentially constant, with some nontrivial behaviour near the surface of the source. The plot in figure \ref{fig:green}, showing \(G(s;p=\gamma)\), exhibits this behaviour. If we also take the form of the quantum correction in eq.~\eqref{eq:q_correction} into account, one could already guess that the most striking difference between the tree-level and one-loop fields probably appears between \(s = 1/2\gamma\) and \(s = \gamma\).

To evaluate eq.~\eqref{eq:loop}, we first analyse the exterior part, for which we again use a subscript \(+\). Before Wick rotating, it is
\begin{equation}
    G_+(s;E, k) = -\frac{\pi}{2\gamma R^2}\csc\left(n\pi\right)P_2^{-n}(u)\left(P_2^n(u) + \frac{W\left(P_2^n, F_a\right)}{W\left(F_a, P_2^{-n}\right)}P_2^{-n}(u)\right)\:.
\end{equation}
We split this expression into a part \(G_+^{(2)}\), which explicitly depends on the value of the field at the surface of the source, and a part \(G_+^{(1)}\), which does not (at least when taken as a function of the field itself). The latter may be treated exactly and is given by
\begin{align}
    \nonumber G_+^{(1)}(s;E, k) &= -\frac{\pi}{2\gamma R^2}\csc\left(n\pi\right)P_2^{-n}(u)P_2^n(u)\\
    &= \frac{-\left(k^2-E^2\right)^2+3 \gamma^2 \left(u^2-2\right) \left(k^2-E^2\right)-9 \gamma ^4 \left(u^2-1\right)^2}{2R^2 \left(k^2-E^2\right) \left(k^2 + 3 \gamma ^2-E^2\right) \sqrt{k^2 + 4 \gamma ^2-E^2}}\:.
\end{align}
Note the singularities: simple poles at \(E = \pm k\) and \(E = \pm\sqrt{k^2 + 3\gamma^2}\) and a branch point at \(E = \pm \sqrt{k^2 + 4\gamma^2}\), all of which lie on the real energy axis. The result of the \(i\epsilon\) prescription is a deflection of the positive and negative singularities below and above the real line, respectively, as shown in figure \ref{fig:gplus_contour}. The integral over the contour shown in this figure vanishes by Cauchy's integral theorem, and the integrals over \(C_1\) and \(C_2\) vanish by Jordan's lemma. Thus, the integrals over real and imaginary \(E\) in the given directions are equal and opposite, as expected for a Wick rotation to be valid. One can now compute the regularised loop integral over \(G_+^{(1)}\) to arrive at the expression
\begin{figure}
    \centering
    \begin{tikzpicture}
        \draw[->] (-4,0) -- (4,0) node[right] {\(\operatorname{Re}(E)\)};
        \draw[->] (0,-4) -- (0,4) node[above] {\(\operatorname{Im}(E)\)};
        
        \draw[very thick,->-] (-3.5,0) -- (0,0) node[midway, below] {\(L_1\)};
        \draw[very thick,->-] (0,0) -- (3.5,0) node[midway, below] {};
        \draw[very thick,->-] (0,3.5) -- (0,0) node[midway, right] {\(L_2\)};
        \draw[very thick,->-] (0,0) -- (0,-3.5) node[midway, right] {};
        
        \draw[very thick,->-] (3.5,0) arc[start angle=0, end angle=90, radius=3.5] node[midway, right] {\(C_1\)};
        \draw[very thick,->-] (0,-3.5) arc[start angle=270, end angle=180, radius=3.5] node[midway, right] {\(C_2\)};

        \filldraw[black] (0.5,-0.5) circle (2pt) node[below] {\(s_1\)};
        \filldraw[black] (1.5,-0.5) circle (2pt) node[below] {\(s_2\)};
        \filldraw[black] (2.5,-0.5) circle (2pt) node[below] {\(s_3\)};
        \draw[thick,decorate,decoration=zigzag] (2.5, -0.5) -- (4,-0.5);

        \filldraw[black] (-0.5,0.5) circle (2pt) node[above] {\(\bar s_1\)};
        \filldraw[black] (-1.5,0.5) circle (2pt) node[above] {\(\bar s_2\)};
        \filldraw[black] (-2.5,0.5) circle (2pt) node[above] {\(\bar s_3\)};
        \draw[thick,decorate,decoration=zigzag] (-2.5, 0.5) -- (-4,0.5);
    
    \end{tikzpicture}
    \caption{The Wick contour for the energy integral of the contribution \(G^>_1\), where \(s_1 = k - i\epsilon\), \(s_2 = \sqrt{k^2 + 3\gamma^2} - i\epsilon\) and \(s_3 = \sqrt{k^2 + 4\gamma^2} - i\epsilon\).\\}
    \label{fig:gplus_contour}
\end{figure}
\begin{equation}
    G_+^{(1)}(u) = -\frac{i\gamma^2}{8 \pi^2}\left(\frac{\Lambda ^2}{\gamma^2}+2- \left(1-3 u^2\right) \log \left(\frac{\gamma^2 }{\Lambda^2 }\right)-\pi  \sqrt{3} u^2\left(1-u^2\right)\right)\:.
\end{equation}
This is exactly \(i\) times the result obtained from the Euclidean calculation \cite{garbrecht_2015_a}, as we might expect given the similarity of the external part of the classical solution and the resulting differential problem to that of determining the quantum corrected bounce configurations in the Coleman description of thin-wall vacuum decay. The second term \(G_+^{(2)}\) is too complicated to be amenable to a similar analysis. However, the loop integral of \(G_+^{(2)}\) converges, and thus all we need to know is that it will contribute some finite shift, which can be calculated using numerical methods.

The interior contribution is
\begin{equation}
    G_-(s; k, E) = - \frac{1}{4a g\gamma  R^2} F_a(z)\left(F_{-a}(z) + \frac{W(P_2^{-2}, F_{-a})}{W(F_a, P_2^{-n})}F_a(z)\right)\:.
\end{equation}
It is tempting to split this expression up along similar lines as the exterior contribution.\begin{figure}[t]
    \centering
    \begin{tikzpicture}
        \draw[->] (-4,0) -- (4,0) node[right] {\(\operatorname{Re}(E)\)};
        \draw[->] (0,-4) -- (0,4) node[above] {\(\operatorname{Im}(E)\)};
        
        \draw[very thick,->-] (-3.5,0) -- (0,0) node[midway, below] {\(L_1\)};
        \draw[very thick,->-] (0,0) -- (3.5,0) node[midway, below] {};
        \draw[very thick,->-] (0,3.5) -- (0,0) node[midway, right] {\(L_2\)};
        \draw[very thick,->-] (0,0) -- (0,-3.5) node[midway, right] {};

        \draw[very thick,->-] (3.5,0) arc[start angle=0, end angle=90, radius=3.5] node[midway, right] {\(C_1\)};
        \draw[very thick,->-] (0,-3.5) arc[start angle=270, end angle=180, radius=3.5] node[midway, right] {\(C_2\)};

        \filldraw[black] (0.5,-2.5) circle (2pt) node[below] {\(s_1\)};
        \filldraw[black] (1.5,-0.5) circle (2pt) node[below] {\(s_2\)};
        \filldraw[black] (2.5,-0.5) circle (2pt) node[below] {\(s_3\)};
        \draw[thick,decorate,decoration=zigzag] (2.5, -0.5) -- (4,-0.5);

        \filldraw[black] (-0.5,2.5) circle (2pt) node[above] {\(\bar s_1\)};
        \filldraw[black] (-1.5,0.5) circle (2pt) node[above] {\(\bar s_2\)};
        \filldraw[black] (-2.5,0.5) circle (2pt) node[above] {\(\bar s_3\)};
        \draw[thick,decorate,decoration=zigzag] (-2.5, 0.5) -- (-4,0.5);
    \end{tikzpicture}
    \caption{The Wick contour for the attempted energy integral of the contribution \(G^{(1)}_-\), where \(s_1 = \sqrt{k^2 - 3 g^2\gamma^2} - i\epsilon\) (displayed for \(k^2<3g^2\gamma^2\)), \(s_2 = k-i\epsilon\) and \(s_3 = \sqrt{k^2 + g^2\gamma^2} - i\epsilon\).\\}
    \label{fig:gminus_contour}
\end{figure}
Upon doing so, the term which does not contain Wro{\'n}skians of Legendre and hypergeometric functions is
\begin{align}
    \nonumber G_-^{(1)}(s;p) &= - \frac{1}{4a g\gamma R^2} F_a(z)F_{-a}(z)\\
    &= -\frac{9 \gamma ^4 g^4 z (z+1)+3 \gamma ^2 g^2 (z+1) \left(E^2-k^2\right)+\left(k^2-E^2\right)^2}{2 \left(k^2-E^2\right) \left(k^2-3 g^2\gamma ^2 -E^2\right) \sqrt{k^2+g^2\gamma ^2 -E^2}}\:.
\end{align}
Note the singularities: simple poles at \(E = \pm k\) and \(E = \pm \sqrt{k^2 - 3 g^2\gamma^2}\) and branch points at \(E = \pm \sqrt{k^2 + g^2\gamma^2}\). Note further that, for all \(k\), the pair of branch points and first pair of poles lie on the real energy axis, whereas the second pair of poles lie on the imaginary energy axis if \(k^2 < 3 g^2\gamma^2\). In this case, the \(i\epsilon\) prescription deflects the real singularities as before (positive below and negative above the real line), but the imaginary poles undergo an infinitesimal anticlockwise rotation, as shown in figure \ref{fig:gminus_contour}. We seem to have a propagator whose mass takes the wrong sign, i.e., a tachyon, which contributes a residue with an additional factor of \(i\). This tachyonic mode turns out to be spurious --- a fact corroborated by our considerations in appendix \ref{sec:eigenmodes} and continuity with the exterior contribution --- but prevents us from calculating \(G_-^{(2)}\) in closed form, at least by any method of which we are aware. Consequently, the renormalisation will have to be performed numerically. We elaborate on this in the next section.

\subsection{Renormalisation}
\label{sec:renorm}
For a constant field \(\varphi\), the one-loop contribution to the equation of motion is contained within the first derivative of the Coleman--Weinberg effective potential \cite{PhysRevD.7.1888}, given by
\begin{equation}
    V_\text{CW}(\phi) = V(\phi) + \frac{1}{2}\delta m^2\phi^2 + \frac{1}{4}\delta\lambda\phi^4 - \frac{i}{2}\int\,\frac{\text{d}^4p}{\left(2\pi\right)^4}\,\ln\left(1 - \frac{1}{p^2}\od{^2V}{\phi^2}\right)\:,
\end{equation}
where \(\delta m^2\) and \(\delta \lambda\) denote the mass and coupling counterterms, respectively. We choose the label \(V_\text{CW}\) to avoid confusion with the previously defined classical effective potential \(V_\text{eff}\). The above integral reduces to a one-dimensional integral in much the same way as the loop integral for the Green's function, and it is straightforward to show that
\begin{align}
    \nonumber \int\,\frac{\text{d}^4p}{\left(2\pi\right)^4}\,\ln\left(1 - \frac{1}{p^2}\od{^2V}{\phi^2}\right) &= \frac{i}{2\pi^2}\int_0^\Lambda\text{d}p\,p^2\left(\sqrt{p^2 + \od{^2V}{\phi^2}} - p\right)\:,
\end{align}
assuming spherical symmetry. Imposing the renormalisation conditions \cite{garbrecht_2015_a},
\begin{equation}
    \left.\pd{^2V_\text{CW}}{\phi^2}\right\vert_{\phi = v} = 4 \gamma^2
\end{equation}
and
\begin{equation}
    \left.\pd{^4V_\text{CW}}{\phi^4}\right\vert_{\phi = v} = 6\lambda\:,
\end{equation}
uniquely determines the counterterms to be
\begin{equation}
    \delta \lambda = -\frac{9\lambda^2 }{16 \pi ^2}\left(\ln \left(\frac{\gamma^2}{\Lambda ^2}\right)+5 \right)
\end{equation}
and
\begin{equation}
    \delta m^2 = -\frac{3 \lambda\gamma^2}{8 \pi ^2}\left(\frac{\Lambda ^2}{\gamma^2}-\log \left(\frac{\gamma ^2}{\Lambda ^2}\right)-31 \right)\:.
\end{equation}
The renormalisation is independent of the source density, as expected.

Since we are unable to determine the UV-divergent part of the interior Green's function in closed form, we will need to integrate the Green's function numerically. We do this using standard functions in Mathematica. To effect the renormalisation, we introduce pseudo-counterterms that allow us to subtract the divergences at the level of the integrand \cite{garbrecht_2015_a}. These are functions \(\Delta m^2(p)\) and \(\Delta\lambda(p)\) that satisfy the integral equations
\begin{equation}
    \frac{1}{2\pi^2}\int_0^\Lambda\text{d}p\,p^2\Delta m^2(p) = \delta m^2\:,
\end{equation}
and
\begin{equation}
    \frac{1}{2\pi^2}\int_0^\Lambda\text{d}p\,p^2\Delta \lambda(p) = \delta\lambda\:.
\end{equation}
Clearly \(\Delta\lambda\) must vary like
\begin{equation}
    \Delta\lambda(p) = \frac{1}{p^2}\left(\frac{1}{\sqrt{p^2 + 4\gamma^2}}A + \frac{\sqrt{p^2 + 4\gamma^2}}{p^2 + 3\gamma^2}B\right)\:,
\end{equation}
since it is terms like these that give rise to logarithmic divergences, but not quadratic divergences. Integrating and matching like terms fixes the constants to be
\begin{equation}
    A = \frac{9 \left(15 \sqrt{3}+\pi \right) \lambda ^2}{4 \pi}
\end{equation}
and
\begin{equation}
    B = -\frac{135 \sqrt{3} \lambda ^2}{4 \pi }\:.
\end{equation}
By similar logic, \(\Delta m^2\) must vary like
\begin{equation}
    \Delta m^2(p) = \frac{1}{p^2}\left(\frac{C}{\sqrt{p^2 + 4\gamma^2}}+\frac{D \sqrt{p^2 + 4\gamma^2}}{p^2 + 3\gamma^2}+E \sqrt{p^2 + 4\gamma^2}\right)\:,
\end{equation}
with constants determined to be
\begin{equation}
    C = \frac{3 \gamma^2\left(\pi -99 \sqrt{3}\right) \lambda }{2 \pi }\:,
\end{equation}
\begin{equation}
    D = \frac{297 \sqrt{3} \gamma ^2 \lambda }{2 \pi }\:,
\end{equation}
and
\begin{equation}
    E = -\frac{3 \lambda }{2}\:.
\end{equation}
\begin{figure}
    \centering
    \includegraphics{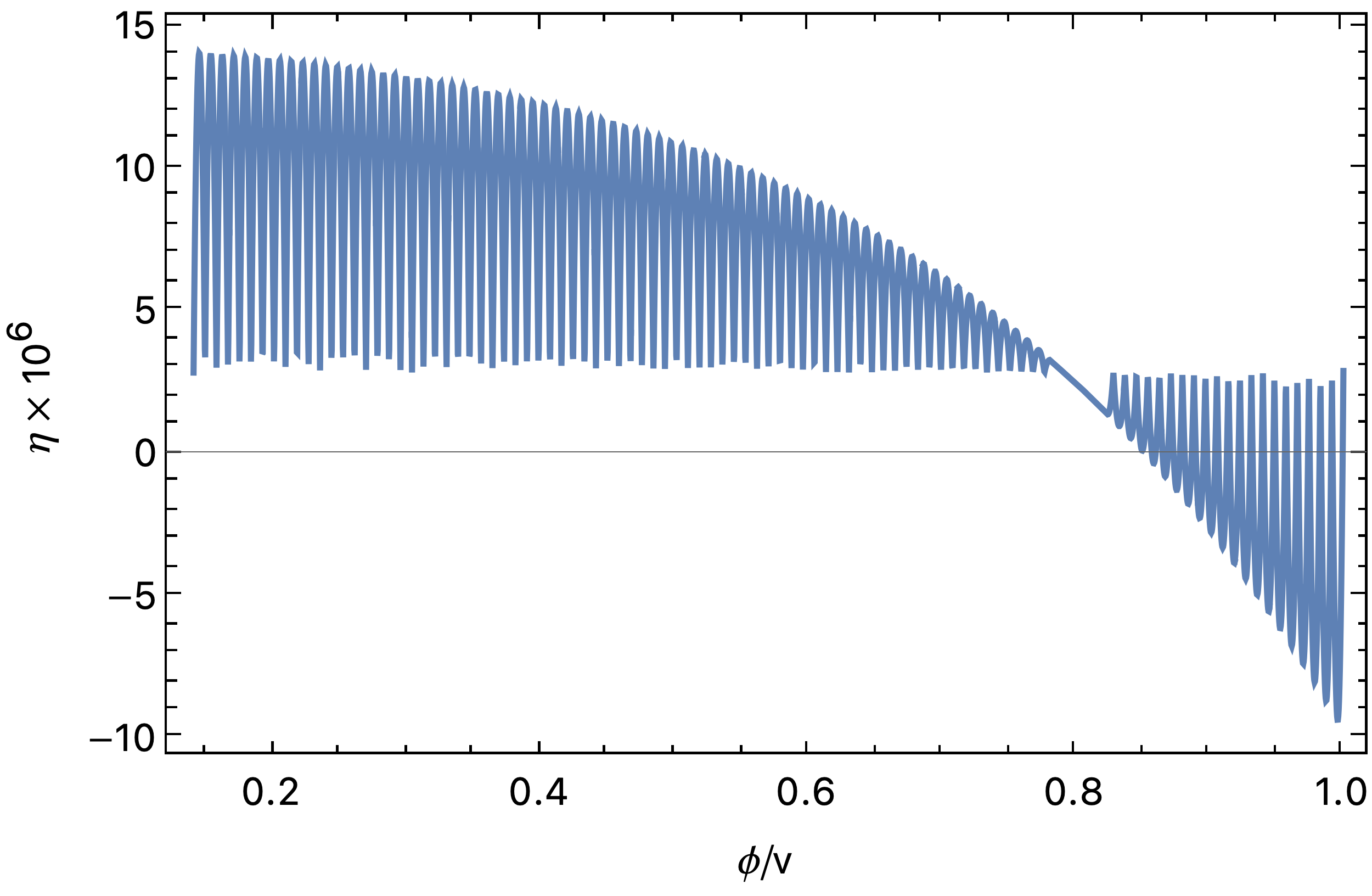}
    \caption{The relative error \(\eta\) between the exact and numerical calculations of the integral of \(G_+^{(1)}\). Note that the horizontal axis begins at \(\phi_0/v\), where \(\phi_0\) is the value of the field at the surface of the source.}
    \label{fig:tadpole_error}
\end{figure}

The renormalised tadpole contribution is given by
\begin{align}
    \Pi^R(\phi) = 3\lambda i G(\phi) + \delta m^2 + \delta\lambda \phi^2\:.
\end{align}
As a consistency check, we note that the contribution from \(G_+^{(1)}\) is
\begin{equation}
    \label{eq:euc_tad}
    \Pi_+^{(1), R}(u) = \frac{9\lambda\gamma^2}{8\pi^2}\left[6 + \left(1 - u^2\right)\left(5 - \frac{\pi}{\sqrt 3} u^2\right)\right]\:,
\end{equation}
which is in line with the result from ref.~\cite{garbrecht_2015_a} (six times larger if the quartic coefficient is set to \(\lambda/4!\) instead of \(\lambda/4\)). We shall use this as a benchmark of the accuracy of the numerical method used to compute the integral
\begin{equation}
    \Pi^R(\phi)= \frac{1}{2\pi^2}\int_0^\Lambda\text{d}p\,p^2\left[-3\lambda G(s;p) + \Delta m^2(p) + \Delta\lambda(p)\phi^2\right]\:.
\end{equation}
If we replace \(G\) by \(G_+^{(1)}\), the result should closely match eq.~\eqref{eq:euc_tad}. Indeed, in figure \ref{fig:tadpole_error}, we see that the relative error \(\eta\) between the exact and approximate expressions is, at worst, of order \(10^{-5}\). The shape and order of magnitude of this curve are largely independent of parameter values, suggesting an overall systematic percentage error of about \(0.0015\%\) on the numerical calculation of \(\Pi^R\). We remark that this result derives from numerical estimates which are linearly interpolated. Higher-order interpolation reduces the relative error by an order of magnitude, to about \(0.00028\%\), but results in artefacts near domain boundaries.

\subsection{One-loop correction}
With the tadpole contribution successfully renormalised, we can now solve the equation of motion for the quantum correction \(\delta\phi\). The renormalised correction to the classical field configuration \(\delta\phi\) is time dependent in general and satisfies
\begin{equation}
    \left(\pd{^2}{t^2} - \pd{^2}{s^2} +\frac{\rho(s)}{M^2}- \mu^2 + 3\lambda\varphi_\text{cl}(s)^2\right)\delta\phi = -\Pi^R(\varphi_\text{cl})\varphi_\text{cl}(s)\:.
\end{equation}
Notice that the differential operator is that which defines the Green's function for \(k = E\), equivalently \(n=2\), where \(n\) is the order of the associated Legendre function. Therefore, this equation is solved by the convolution
\begin{align}
    \nonumber \delta\phi &= \int\text{d}s'\int\text{d}t'\,R^2G(s, s';t-t';k=E)\Pi^R\left(\varphi_\text{cl}(s')\right)\varphi_\text{cl}(s')\\
    \nonumber &= \int\text{d}s'\int\text{d}t'\int\frac{\text{d}E}{2\pi}e^{-iE(t-t')}\,R^2G(s, s';k=E)\Pi^R\left(\varphi_\text{cl}(s')\right)\varphi_\text{cl}(s')\\
    &= \int\text{d}s'\,R^2G(s, s';n=2)\Pi^R\left(\varphi_\text{cl}(s')\right)\varphi_\text{cl}(s')\:,
\end{align}
where the time dependence falls out because \(G(s, s';k=E) \equiv G(s, s';n=2)\) does not depend on \(E\). The Green's function has a pole at this point in momentum space and thus, hereafter, it should be understood that we are working in the limit \(n\rightarrow 2\). The exponential is thus the only part which depends on \(E\), and integrating over it will yield a delta function, the integral of which is unity.
\begin{figure}
    \centering
    \includegraphics{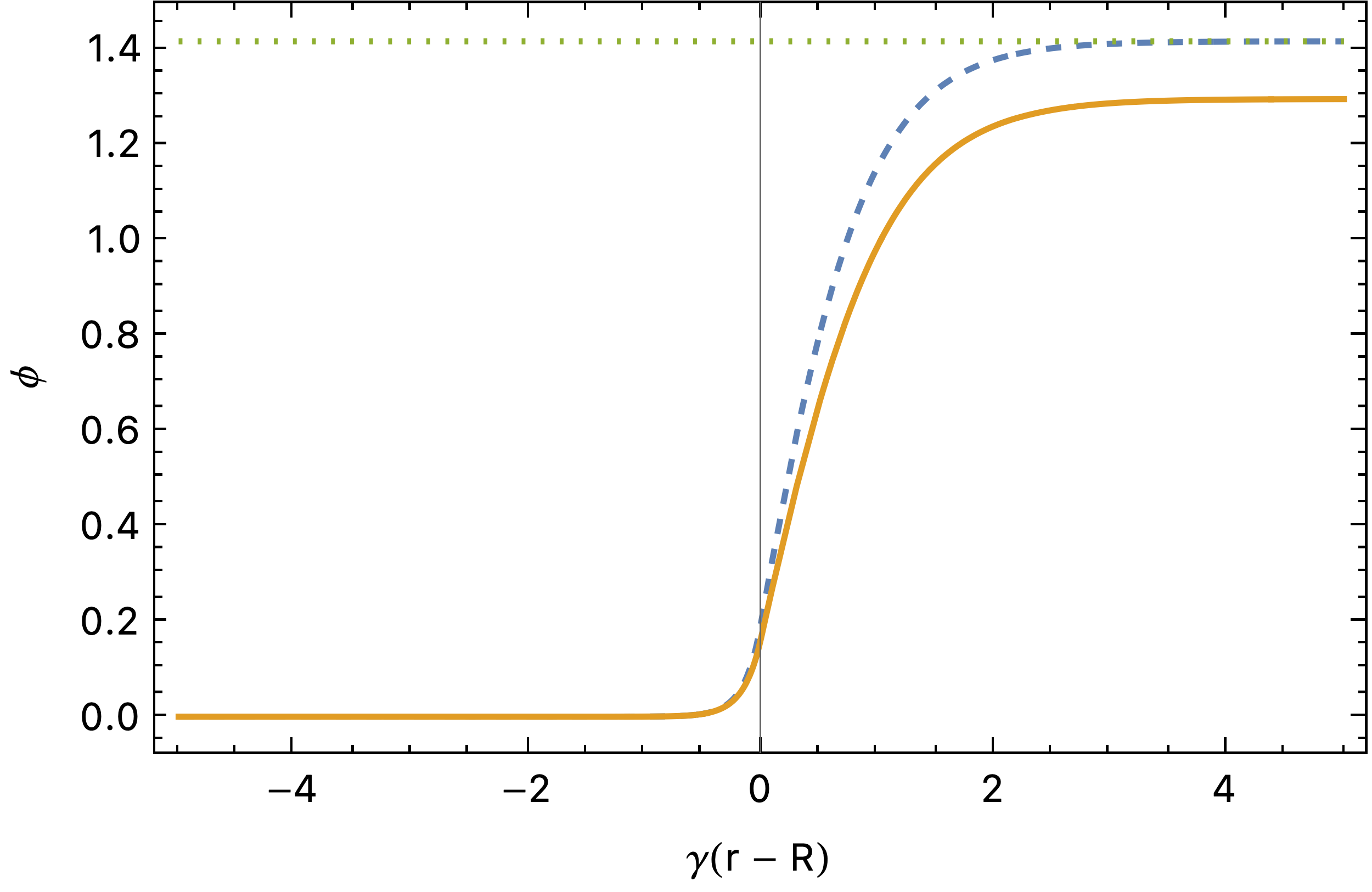}
    \caption{A plot of the symmetron field with (orange, dotted) and without (blue, solid) the one-loop correction. The parameter values are as they were in figure \ref{fig:classical_profile}, and we have set \(\lambda = 0.5\).\\}
    \label{fig:clvsqu}
\end{figure}

\begin{figure}
    \centering
    \includegraphics{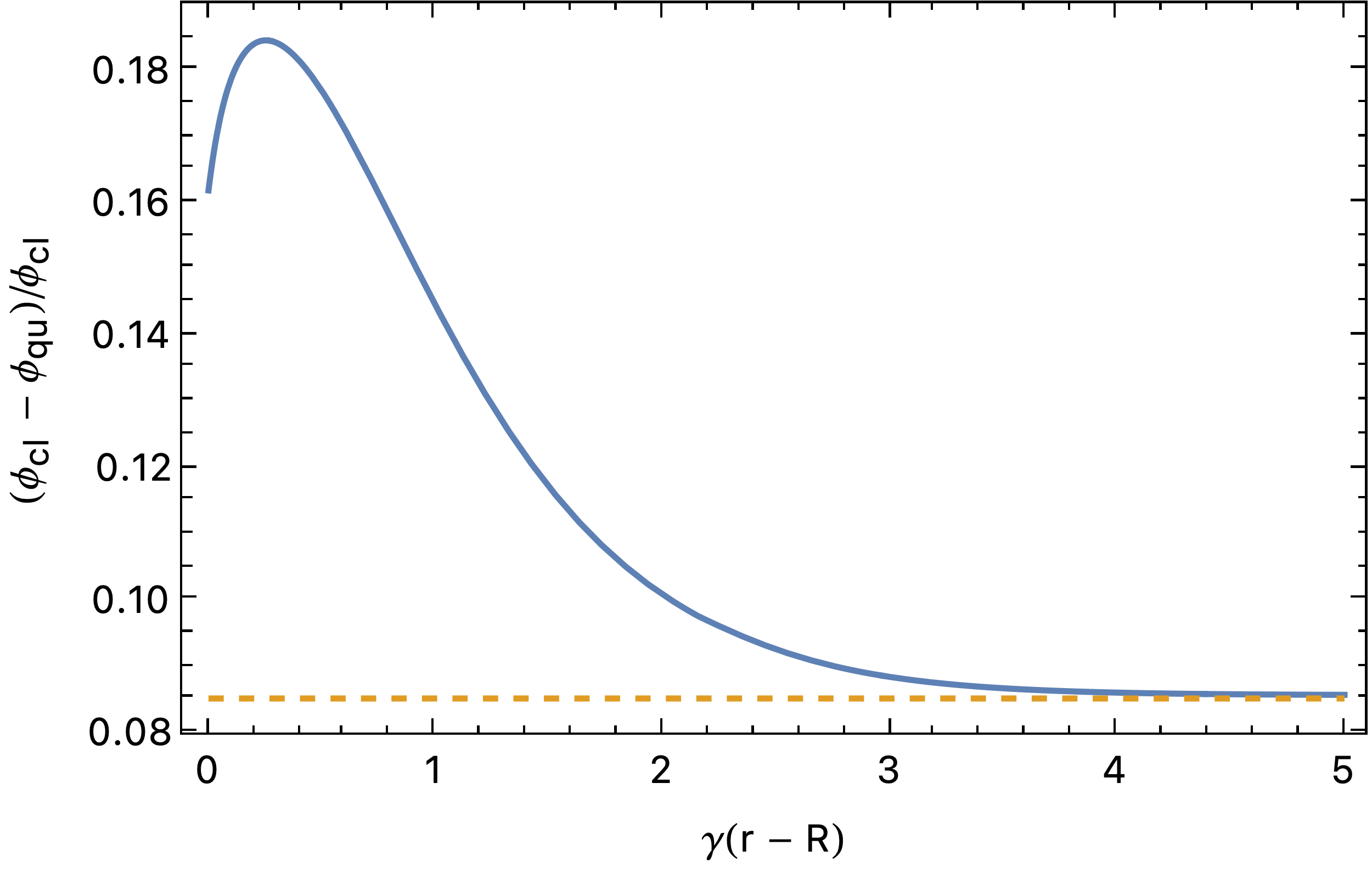}
    \caption{A plot of the relative difference between the tree-level and one-loop symmetron fields. Parameter values match figure \ref{fig:clvsqu}.\\}
    \label{fig:field_diff}
\end{figure}

\begin{figure}
    \centering
    \includegraphics{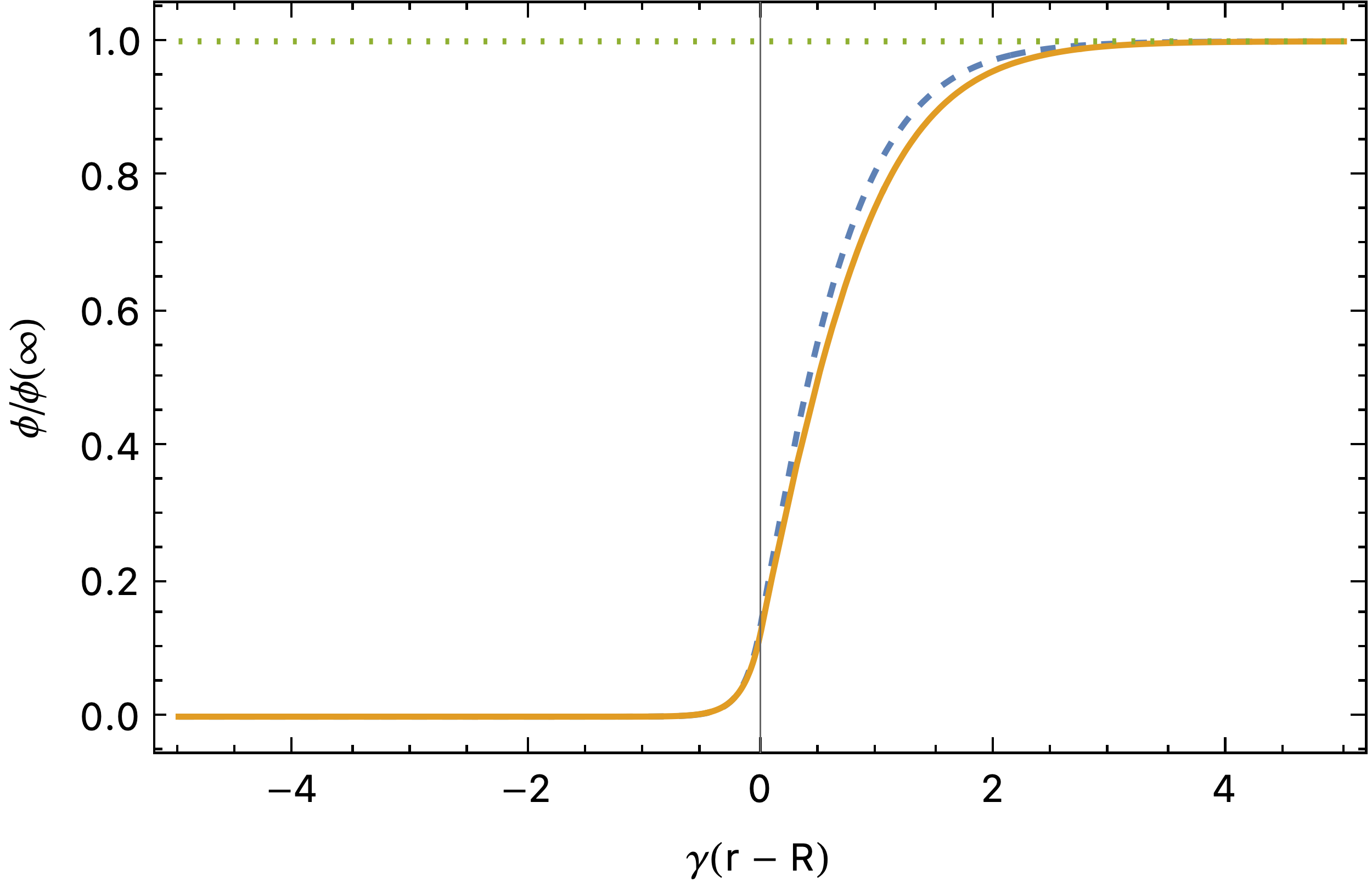}
    \caption{A plot with the tree-level (blue, dashed) and one-loop (orange, solid) symmetron fields normalised to their respective vacuum expectation values.}
    \label{fig:clvsqu_norm}
\end{figure}

\begin{figure}
    \centering
    \includegraphics{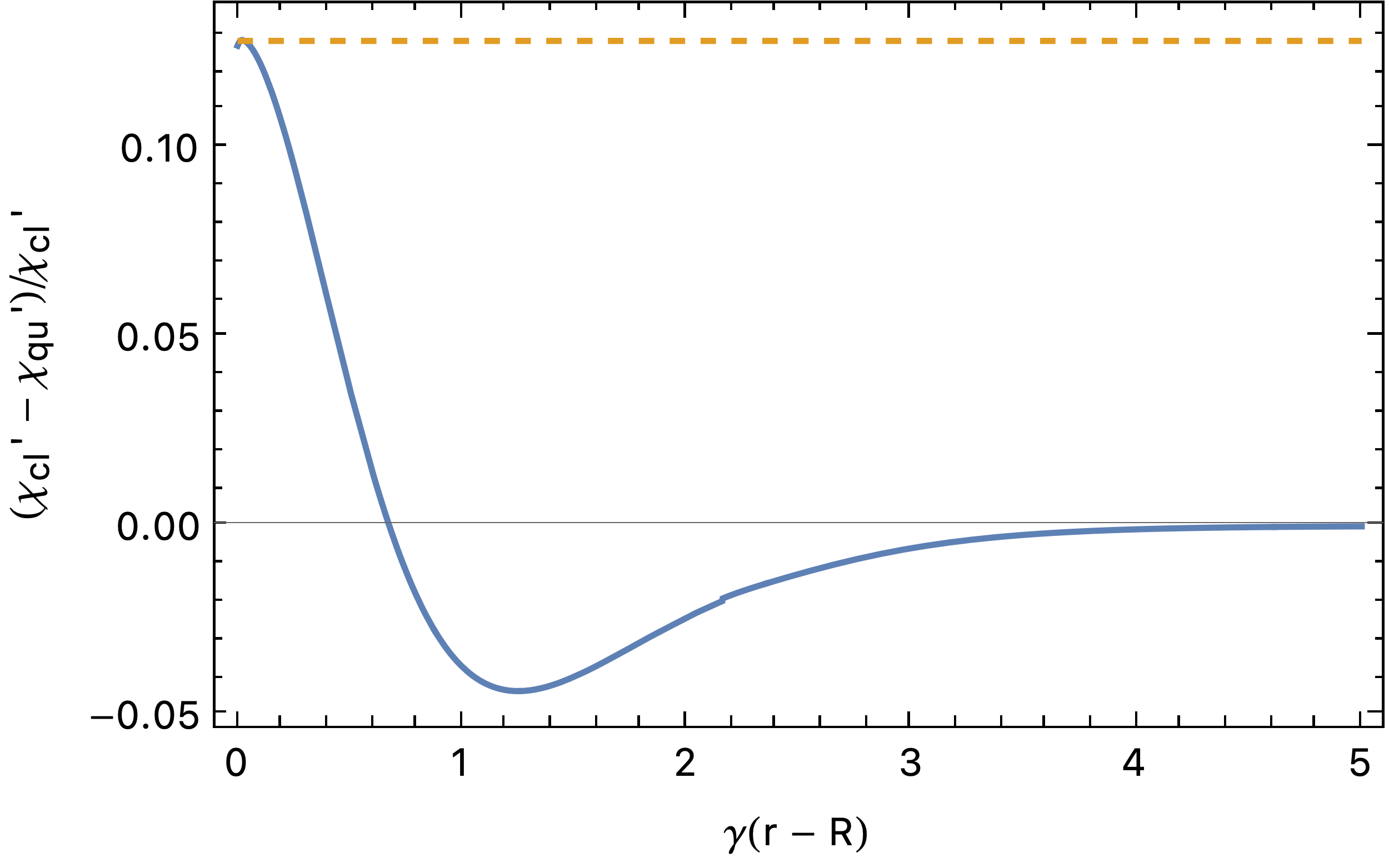}
    \caption{A plot showing the relative difference in the derivative of the normalised fields outside the source. For brevity, we write \(\chi = \phi/\phi(\infty)\). The small discontinuity is a numerical artefact.}
    \label{fig:slope_diff}
\end{figure}

A plot comparing the classical field profile with the one-loop field profile is given in figure \ref{fig:clvsqu}. There are two features of interest in this plot. The easiest thing to see is that the one-loop vacuum \(v_\text{qu}\) sits at a lower value than the tree-level vacuum \(v\). Suppose we may write \(v_\text{qu} = v + \Delta v\). Then the slope of the Coleman--Weinberg effective potential near \(v\) can be written as a series in \(\Delta v\). To first order in the Taylor expansion, we have
\begin{equation}
    \left.\frac{\partial V_\text{CW}}{\partial \phi}\right\vert_{v + \Delta v} = \left.\frac{\partial V_\text{CW}}{\partial \phi}\right\vert_v +  \left.\frac{\partial^2 V_\text{CW}}{\partial \phi^2}\right\vert_v \Delta v + O\left((\Delta v)^2\right)\:.
\end{equation}
Stipulating that the shifted vacuum be the true minimum of the one-loop potential gives
\begin{equation}
    \Delta v = -\frac{27\mu\sqrt{\lambda}}{16\pi^2} = -\frac{27\lambda}{16\pi^2}v\:.
\end{equation}
For the parameters used to generate figure \ref{fig:clvsqu}, this amounts to a shift of about \(8.5\%\) in the vacuum expectation value, which is precisely what we observe in figure \ref{fig:field_diff}. The more subtle feature hiding in figure \ref{fig:clvsqu} is that the quantum field is not simply a rescaled version of the classical field. The derivative of the field profile, when normalised relative to its VEV, has also shifted. Figure \ref{fig:clvsqu_norm} shows this. The mass roughly determines the rate with which the field interpolates between its two asymptotes. Thus, it follows that the shift in the derivative is due to the shift in the relationship between the Lagrangian mass parameter \(\mu\) and the physical mass \(m\). At tree level, the physical mass is just the previously derived mass of scalar fluctuations,
\begin{equation}
    m_\text{tree}^2 = 2\mu^2\:.
\end{equation}
The one-loop physical mass may be obtained by determining the second derivative of the one-loop potential at the true vacuum, which is
\begin{equation}
    m_\text{1-loop}^2 := \left.\frac{\partial^2V_\text{CW}}{\partial \phi^2}\right\vert_{v+\Delta v} = 2\mu^2\left(1 - \frac{81\lambda}{16\pi^2}\right) + O\left(\lambda^2\right)\:.
\end{equation}
For the parameters used to generate figure \ref{fig:clvsqu}, this amounts to about a \(12.8\%\) shift in the mass, which we reason accounts for the comparable shift in the value of the derivative at the origin. This closely matches figure \ref{fig:slope_diff}, where we have plotted the difference in the derivatives of the normalised fields, relative to the slope of the normalised classical field at the origin, in an attempt to decouple the intrinsic change in the derivative from the change due to the shift in the VEV.

\begin{figure}[t]
    \centering
    \begin{subfigure}[h]{\textwidth}
        \centering
        \includegraphics{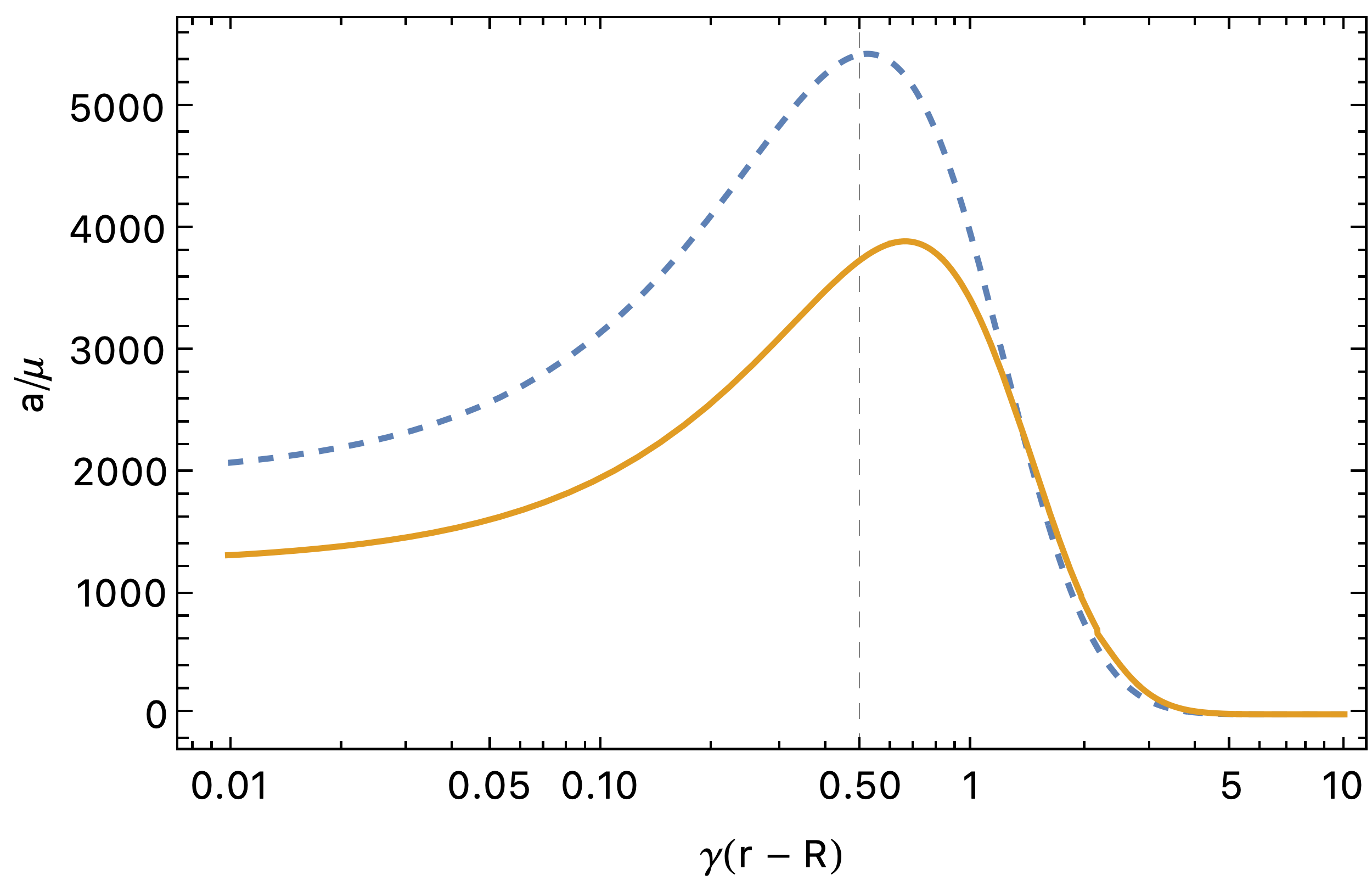}
        \caption{Parameters appropriate for hydrogen spectroscopy \cite{PhysRevD.107.044008} (\(\mu = 1\,\)GeV, \(M = 10\,\)MeV, \(\lambda = 0.5\) and \(\rho_0 = 2.54\times 10^{-3}\,\)GeV\(^4\)).\\}
        \label{fig:qforce_spec}
    \end{subfigure}
    \hfill
    \begin{subfigure}[h]{\textwidth}
        \centering
        \hspace*{-1cm}\includegraphics[scale=1.1]{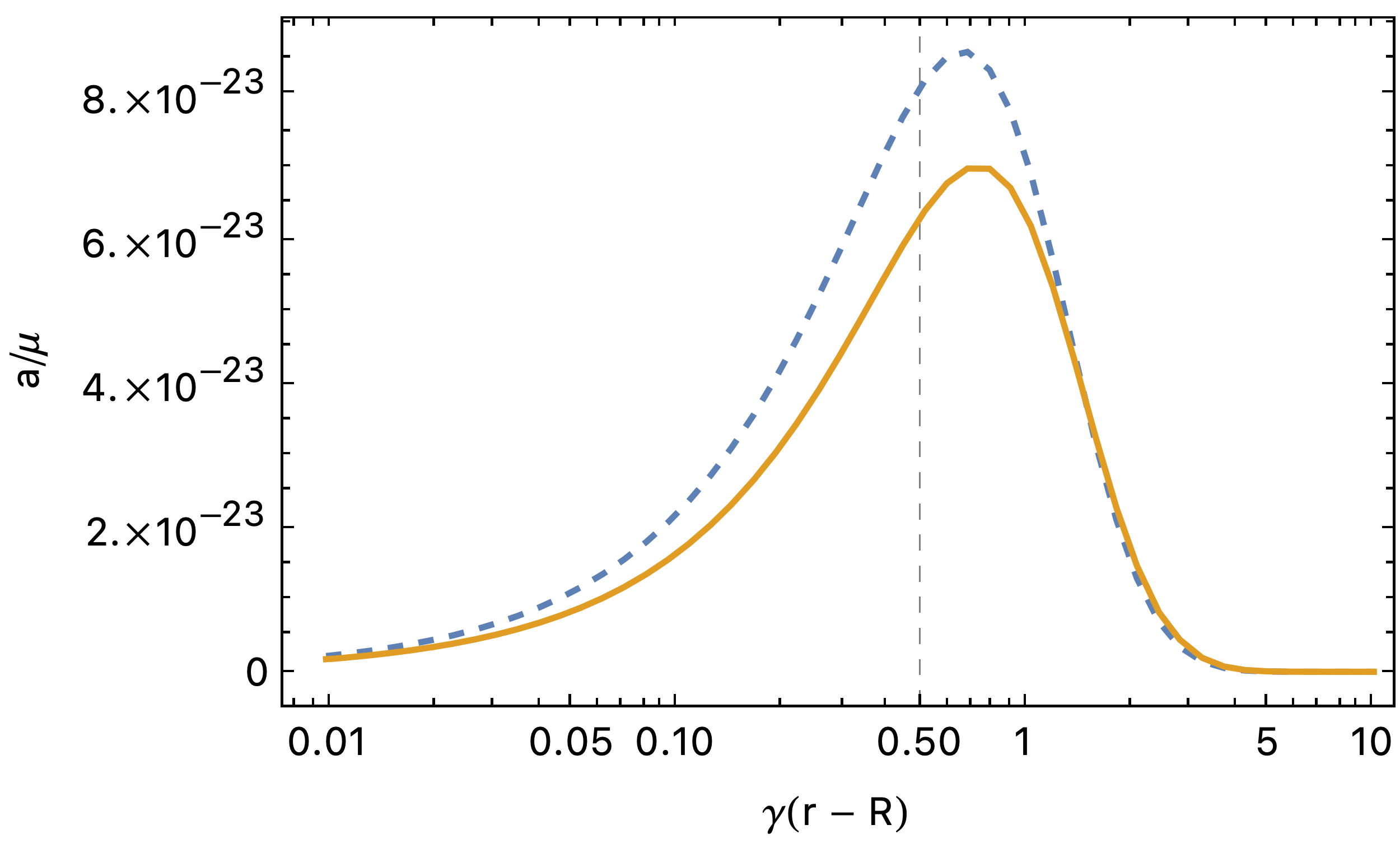}
        \caption{Parameters appropriate for atom interferometry \cite{panda} (\(\mu = 10^{-1}\)meV, \(M = 10^{-2}\,\)GeV, \(\lambda = 10^{-0.5}\) and \(\rho_0 = 8.178\times 10^{-5}\,\)MeV\(^4\)).\\}
        \label{fig:qforce_int}
    \end{subfigure}
    \caption{Plots of the acceleration experienced by a test particle, in units of \(\mu\), assuming tree-level (blue, dashed) and one-loop (orange, solid) forces. The vertical line represents one Compton wavelength from the surface of the source.}
    \label{fig:qforce}
\end{figure}

\begin{figure}[t]
    \centering
    \begin{subfigure}{0.45\linewidth}
        \centering
        \includegraphics[scale=0.55]{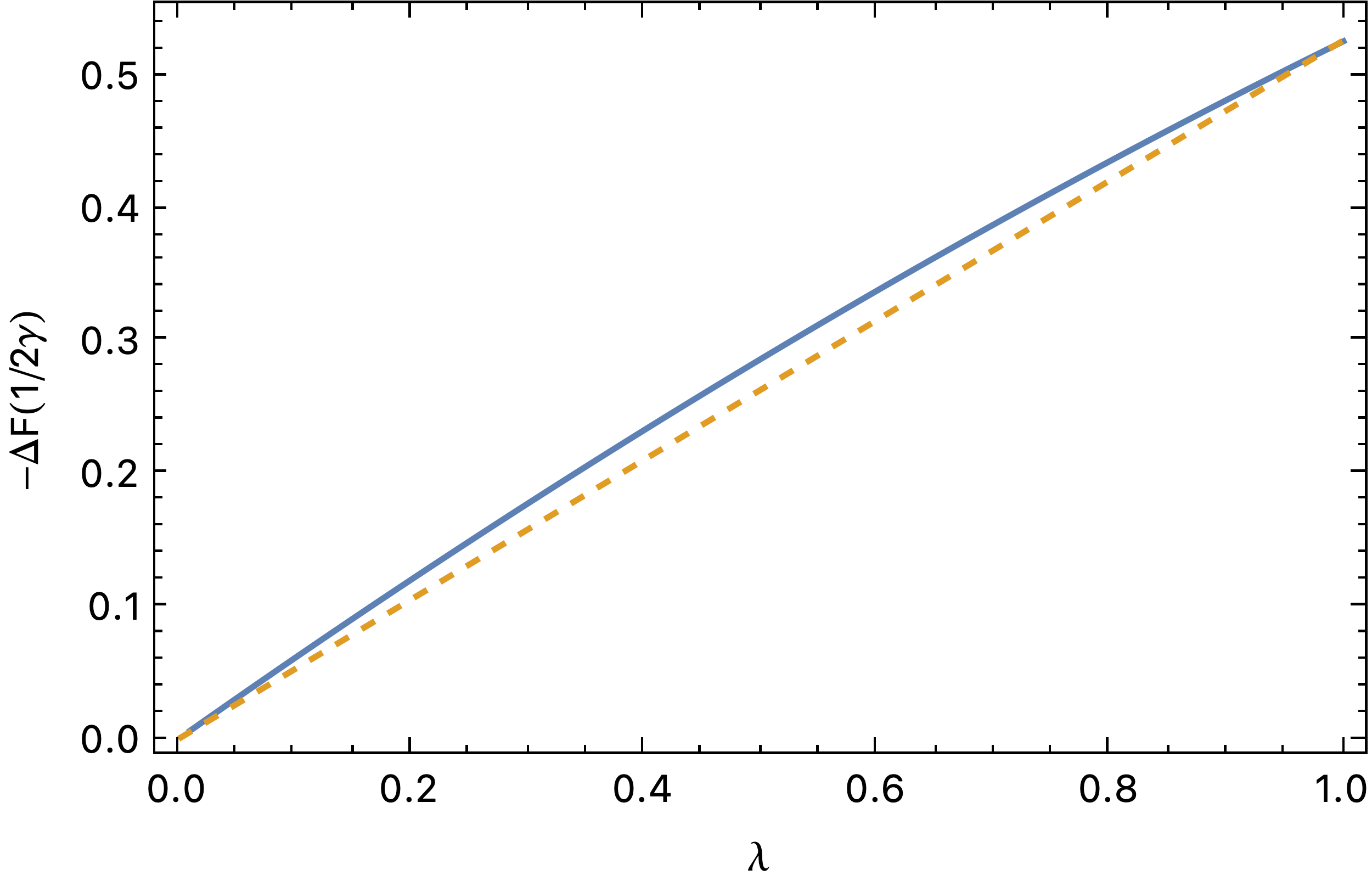}
        \caption{Parameter values, apart from \(\lambda\), are as they were in Figure \ref{fig:qforce_spec}, appropriate for hydrogen spectroscopy.\\}
        \label{fig:fchange_lambda_spec}
    \end{subfigure}
    \hfill
    \begin{subfigure}{0.5\linewidth}
        \centering
        \includegraphics[scale=0.55]{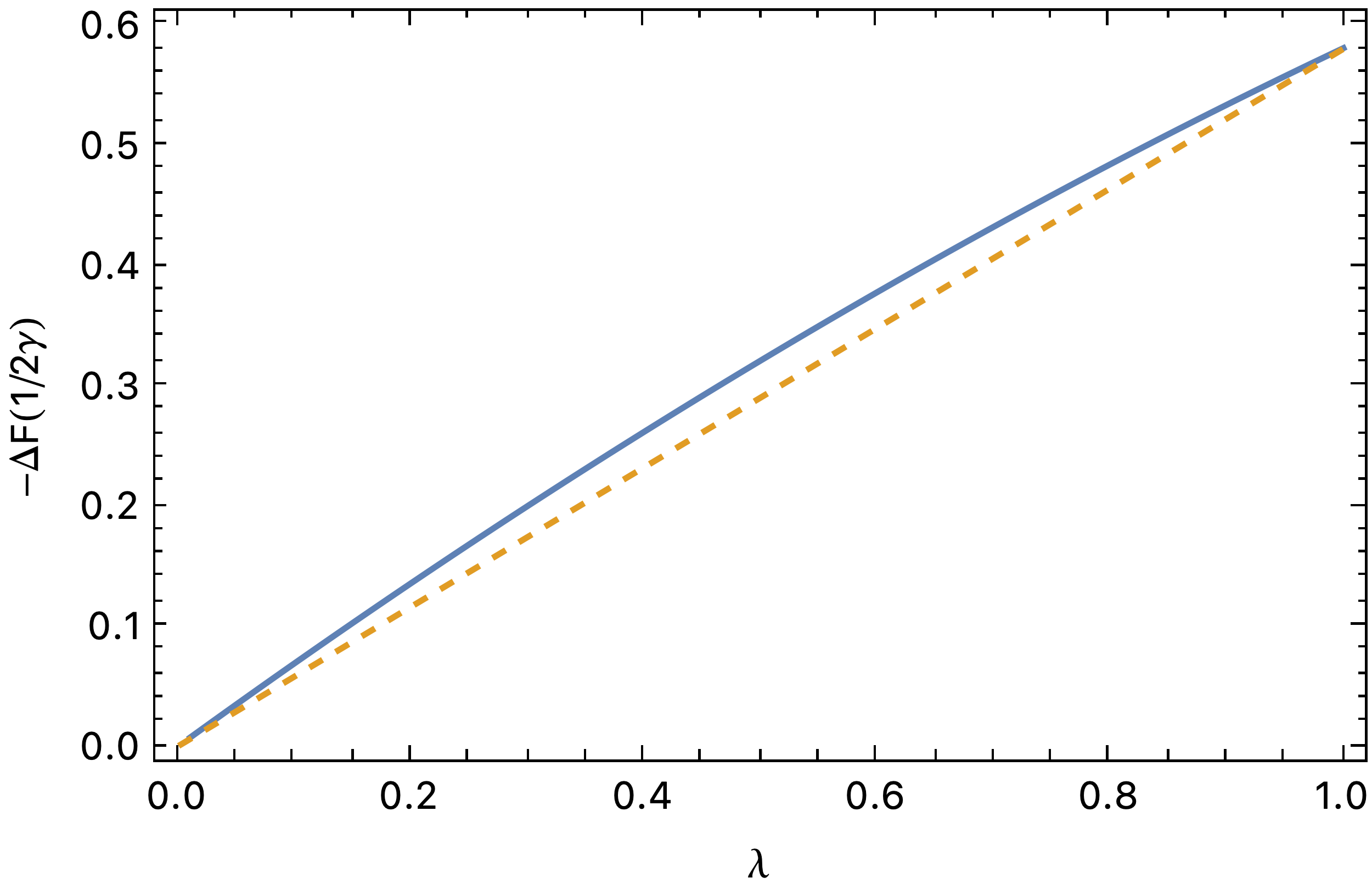}
        \caption{Parameter values, apart from \(\lambda\), are as they were in Figure \ref{fig:qforce_int}, appropriate for atom interferometry.\\}
        \label{fig:fchange_lambda_int}
    \end{subfigure}
    \hfill
    \begin{subfigure}{0.45\linewidth}
        \centering
        \includegraphics[scale=0.57]{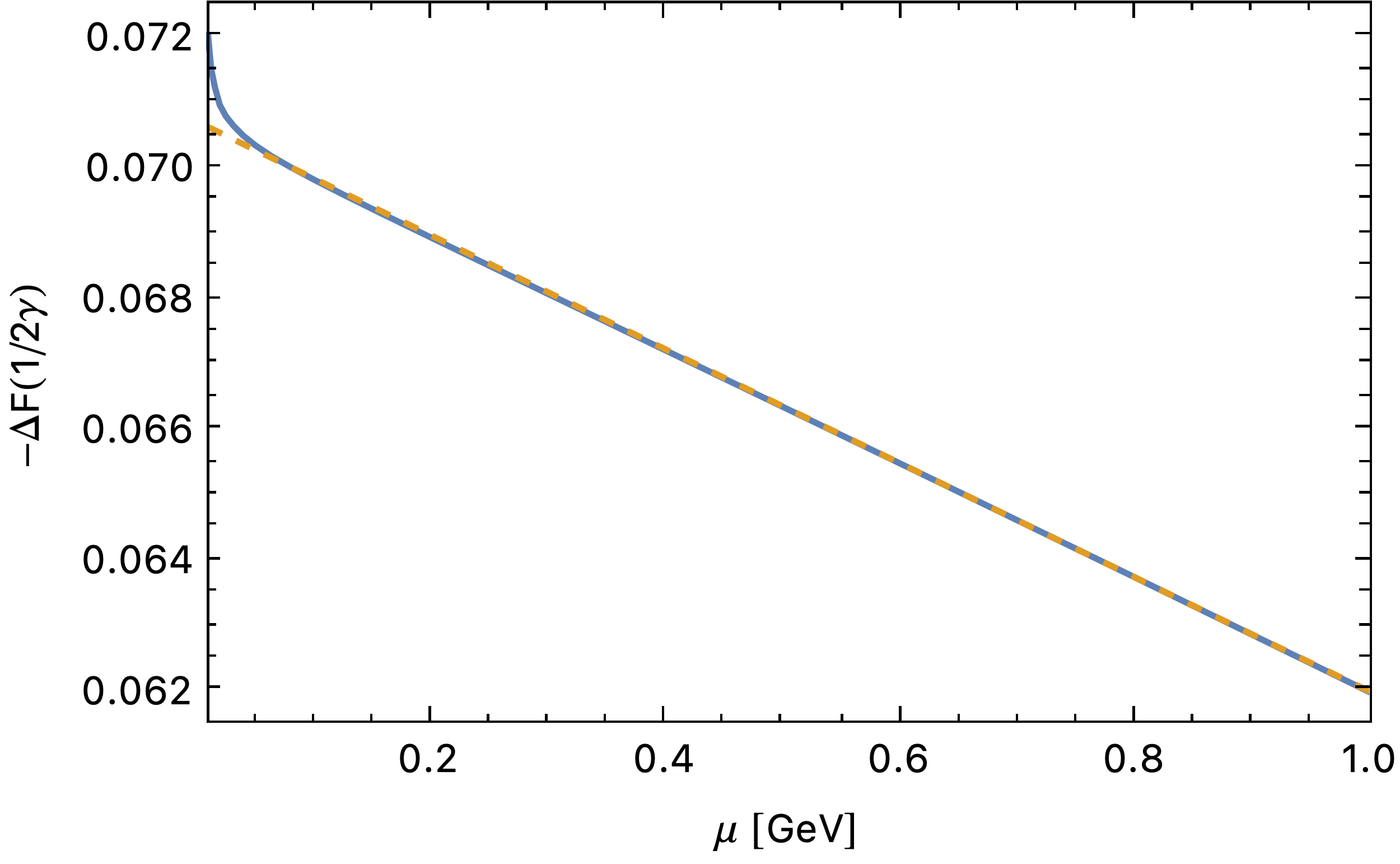}
        \caption{Parameter values, apart from \(\mu\), are as they were in Figure \ref{fig:qforce_spec}, appropriate for hydrogen spectroscopy. \(\lambda = 0.1\).\\}
        \label{fig:fchange_mu_spec}
    \end{subfigure}
    \hfill
    \begin{subfigure}{0.5\linewidth}
        \centering
        \includegraphics[scale=0.57]{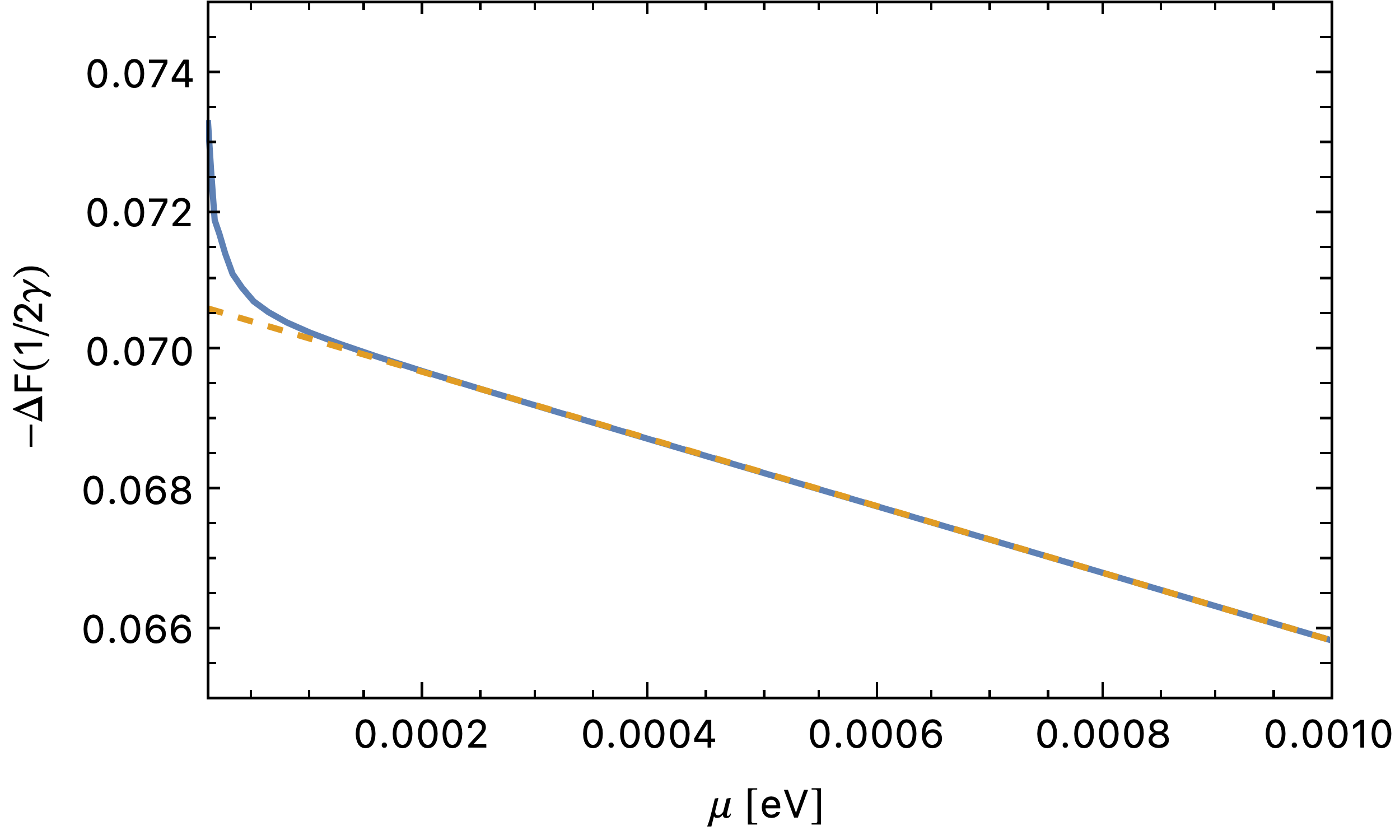}
        \caption{Parameter values, apart from \(\mu\), are as they were in Figure \ref{fig:qforce_int}, appropriate for atom interferometry. \(\lambda = 0.1\).\\}
        \label{fig:fchange_mu_int}
    \end{subfigure}
    \caption{A summary of the parameter dependence of the fractional change in the specific symmetron force due to quantum corrections (blue, solid), along with comparisons to linear dependence (orange, dotted).}
    \label{fig:fchange_lambda}
\end{figure}

Both the correction to the mass and the VEV will combine to enhance the correction to the force. One can equivalently talk just of the acceleration a test particle would feel. In the vicinity of a source which produces a symmetron field profile \(\phi\), the acceleration \(a\) is given by
\begin{equation}
    \label{eq:force}
    a = -\frac{1}{M^2}\phi\nabla\phi\:.
\end{equation}
Since all functions here are spherically symmetric, it is sufficient to consider only the radial component of the force hereafter. We do not attempt to determine the exact way in which the mass and VEV shifts contribute to \(\delta\phi\). However, from eq.~\eqref{eq:force}, one might naively assume that the VEV shift contributes ``twice'', in a sense (since the field is squared), while the mass shift (which applies because of the derivative) contributes ``once''. This very rough argument implies that the relative shift in the force or acceleration \(\Delta F\) is of order
\begin{equation}
    \Delta F = \frac{a_\text{cl} - a_\text{qu}}{a_\text{cl}}\sim 2\times \frac{27 \lambda}{16\pi^2} + \frac{81\lambda}{32\pi^2}\:.
\end{equation}
From a naive perturbative expansion, one might thus expect the shift in the fifth force to scale as \(6\lambda/\pi^2\), that is remarkably close to the scaling which find numerically. This expression suggests that the shift in the force due to the self-interaction is, to a good approximation, linear in the coupling --- at least in the perturbative regime  --- despite the non-perturbative dependence of the VEV on the coupling. Figure \ref{fig:fchange_lambda} corroborates this claim. For the parameters used to generate figure \ref{fig:clvsqu}, the naive calculation predicts a \(30\%\) shift in the strength of the force, which is precisely what we observe, around a distance of one Compton wavelength, in figure \ref{fig:qforce_spec}. This approximation holds well throughout the parameter space; the values used to generate figure \ref{fig:qforce_int} would imply a correction of about \(20\%\), which the plot agrees with. Naturally, the approximation breaks down very close to or far from the surface of the source, where one would expect quantum effects to become less relevant. Notice, however, that, for large masses, we observe strong deviations from tree-level even close to the source.

The quantum-corrected force is generally weaker than its classical counterpart, except after about four Compton wavelengths from the surface of the source, where the quantum force becomes stronger. While this strengthening represents a relatively large fractional change, in absolute terms, the difference is negligible. As one would expect, the shift in the force vanishes as \(\lambda \rightarrow 0\). Note, however, that, for sufficiently small \(\lambda\), quantum corrections due to interactions with Standard Model fields are expected to become the dominant contribution, which we leave for future work.

It is worth considering how the magnitude of the correction to the force depends on the other free parameters. Since changes in the mass scale \(M\) are degenerate with changes in the density \(\rho_0\) of the source, the only remaining free parameter of interest is the field mass \(\mu\). Our rough arguments from earlier in this subsection would suggest that changing the mass parameter should have next to no effect on the force. This is approximately true. We observe a slight negative correlation between \(\mu\) and \(-\Delta F\), but one that amounts to a change of about \(0.4\%\) in the magnitude of the relative correction to the force for all \(\mu \in (0, 1]\,\)GeV (the widest mass range where our numerical methods yield trustworthy results) in figure \ref{fig:fchange_mu_spec}, with a similar relationship shown in figure \ref{fig:fchange_mu_int}. Note that we have extrapolated the linear relationship, ignoring the apparent divergence for small masses. This divergence may be a numerical artefact, a consequence of underflow errors. Alternatively, this could be a signal of the breakdown of the thin-wall approximation, since for \(\mu\rightarrow 0\) we have \(1/\mu \gg R\).

Although these corrections seem comparatively large in places, the strength of the implications for experiments is not necessarily so. For symmetron fields with masses in, say, the \(\mu\)eV range, not only are the forces very small, but the distance where we predict the most significant disparity between the tree-level and one-loop forces is far outside the range of tabletop experiments designed to detect them. Our results have the strongest implications for experiments which probe heavier symmetron fields.

\section{Conclusions}
\label{sec:6}
We set out to estimate the magnitude of the leading-order quantum corrections to the symmetron field around extended sources. To this end, we have solved for the exact field profile around spherical sources whose radii are much larger than the Compton wavelength of the field, meaning our results are also relevant for systems with planar geometry \cite{cannex, universe7070234, PhysRevD.88.083004, PhysRevLett.107.111301, cronenberg_acoustic_2018, PhysRevLett.110.031301, PhysRevD.86.102003}. We have computed the inverse of the fluctuation operator (the Green's function) and the renormalised tadpole contribution. With that, we derived and numerically solved the equation of motion satisfied by the one-loop correction to the classical field profile. We have found that the action of quantum corrections is to flatten the gradients of the classical field profiles and reduce their vacuum expectation values, which agrees with similar analyses of the impact of quantum corrections on tunnelling configurations \cite{garbrecht_2015_a}. As a result, forces derived from one-loop field profiles are generally weaker than those computed from tree-level profiles, for the same input parameters.

The aim of this work was to obtain an analytical estimate of the relevance of quantum corrections to calculations of symmetron fifth forces. This is to enable a first quantification of the theoretical uncertainty on fifth-force calculations that results from ignoring radiative effects, as is commonly done in the literature. To make the problem analytically tractable, we employed the thin-wall and planar approximations, allowing us to treat spherical and planar geometries on the same footing. Of course, the variation of the fifth force far away from the surface of the source depends strongly on the dimensionality of the system, whether spherical, cylindrical or planar. However, as we have seen, the quantum corrections are maximal over a finite range close to the surface of the source where the gradients in the field profile and the nonlinearities in the field equations are largest. We therefore do not expect the dimensionality of the problem to have a significant impact on the percentage shift in the field due to quantum corrections. Quantifying the systematic error in this shift resulting from these approximations in the spherical case would require a numerical analysis of both the classical field profile and the resulting Green's function. Such an analysis is beyond the scope of this work and may be presented elsewhere.

In certain areas of parameter space, the classical field remains virtually identical to the quantum field. This is true in particular for very small self-coupling, as one might expect. One should note, however, that for \(\lambda\lesssim10^{-40}\), it is expected that couplings to the Standard Model dominate, and thus an additional source of quantum corrections would have to be taken into account. For sufficiently high but still perturbative couplings, we observe that the quantum-corrected force can be considerably weaker than the classical prediction. Indeed, for regions of parameter space that sit within current constraints \cite{PhysRevD.107.044008, yin_experimental_2025, universe10070297}, the symmetron force is as much as 30\% weaker than classically predicted, for the same input parameters.

While our results speak mostly to perturbative self-couplings, there is a clear trend that shows that the strength of the correction scales almost linearly with the strength of the self-interaction. We suspect that this trend continues into the non-perturbative regime and may even amount to a symmetron force that is practically undetectable for sufficiently high self-couplings. Phenomenologically, it would also be difficult to use such a field to account for observations related to dark matter, since even environments with low ambient density may have vanishingly small fifth forces. Our results also apply directly to systems with large, screened and spherical sources or systems with planar geometry. However, such an idealisation does not mean that we should expect quantum corrections to be minor for systems with, say, cylindrical symmetry \cite{panda}, or relatively small sources. Field profiles still flatten away from the thin-wall approximation \cite{garbrecht_2018}.

We emphasise two different conclusions to draw from our work. The first is that, due to renormalisation, the relationship between Lagrangian parameters and physical observables shifts from the tree-level one. This conclusion alone would imply that current constraints apply to slightly different regions of parameter space. The second conclusion is that the spatial variation of the fifth force also changes, as indicated by figure \ref{fig:qforce}, generally growing more slowly and peaking at a different point in space in the one-loop approximation. This could impact how we optimise geometries of future tabletop experiments \cite{universe7070234}. The spatial variation of the quantum correction is a crucial point. Since \(\delta\phi\) changes with position, there is no point in space that we could choose to define the physical mass and self-coupling such that the quantum correction vanishes everywhere. Put differently, there does not exist a convenient choice of renormalisation scheme that makes the quantum corrections disappear. They cannot be fine-tuned away, as it were. The spatially varying quantum fluctuations about the classical background are an intrinsic and unavoidable feature of the theory.

We propose several directions for future work. First, the size of the one-loop corrections suggests that higher-order corrections may still have measurable contributions. To help with this endeavour, by use of the Schwinger-Keldysh closed-time path formalism \cite{schwinger_brownian_1961, Keldysh:1964ud}, it may be possible to simplify the portion of our analysis concerned with deriving an expression for quantum-corrected observables and the quantum field, and perhaps even obtain its equation of motion, all without reference to the effective action. This may be presented in future work. Second, the method presented in this paper can likely produce estimates of quantum corrections for symmetron-like models, such as the complex symmetron \cite{universe11050158} or generalised symmetron \cite{astronomy2020009}. Third, it has been shown that one can get around some experimental constraints by choosing models in which screening arises at the one-loop level \cite{PhysRevLett.117.211102}, provided the background field is constant \cite{garbrecht_2015_b}. The methods developed in this paper would allow us to revisit this result and take into account the spatial variation of the field profile. In addition, a future investigation could consider the effect of quantum corrections for a broader range of observables, including shifts in frequency spectra \cite{KADING2025101788}, properties of white dwarfs (mass, radius, luminosity, etc.) \cite{universe11050158} and gravitational lensing \cite{astronomy2020009}. Finally, we hope that this or a similar analysis could be used to further explore the quantum nature of fifth-force theories in general.
\section*{Acknowledgements}
The authors thank Clare Burrage, Ben Elder, Christian Käding and Björn Garbrecht for helpful discussions and comments on this draft. This work was supported by the University of Manchester,  the Science and Technology Facilities Council (STFC) [Grant No.\ ST/X00077X/1], and a United Kingdom Research and Innovation (UKRI) Future Leaders Fellowship [Grant Nos.\ MR/V021974/1 and\ MR/V021974/2]. The Mathematica notebook supporting the results presented in this paper can be found at: \url{DOI:10.5281/zenodo.18431681}.
\appendix
\section{Spectrum of the fluctuation operator}
\label{sec:eigenmodes}
As mentioned at various points in this paper, our method of quantisation is valid as long as the discrete spectrum of the fluctuation operator is positive definite. Every indication suggests that this is indeed the case. For completeness, we compute the spectrum of the fluctuation operator in this section.

\subsection{Eigenvalue problem}
\begin{figure}
    \centering
    \begin{subfigure}{\textwidth}
        \centering
        \includegraphics{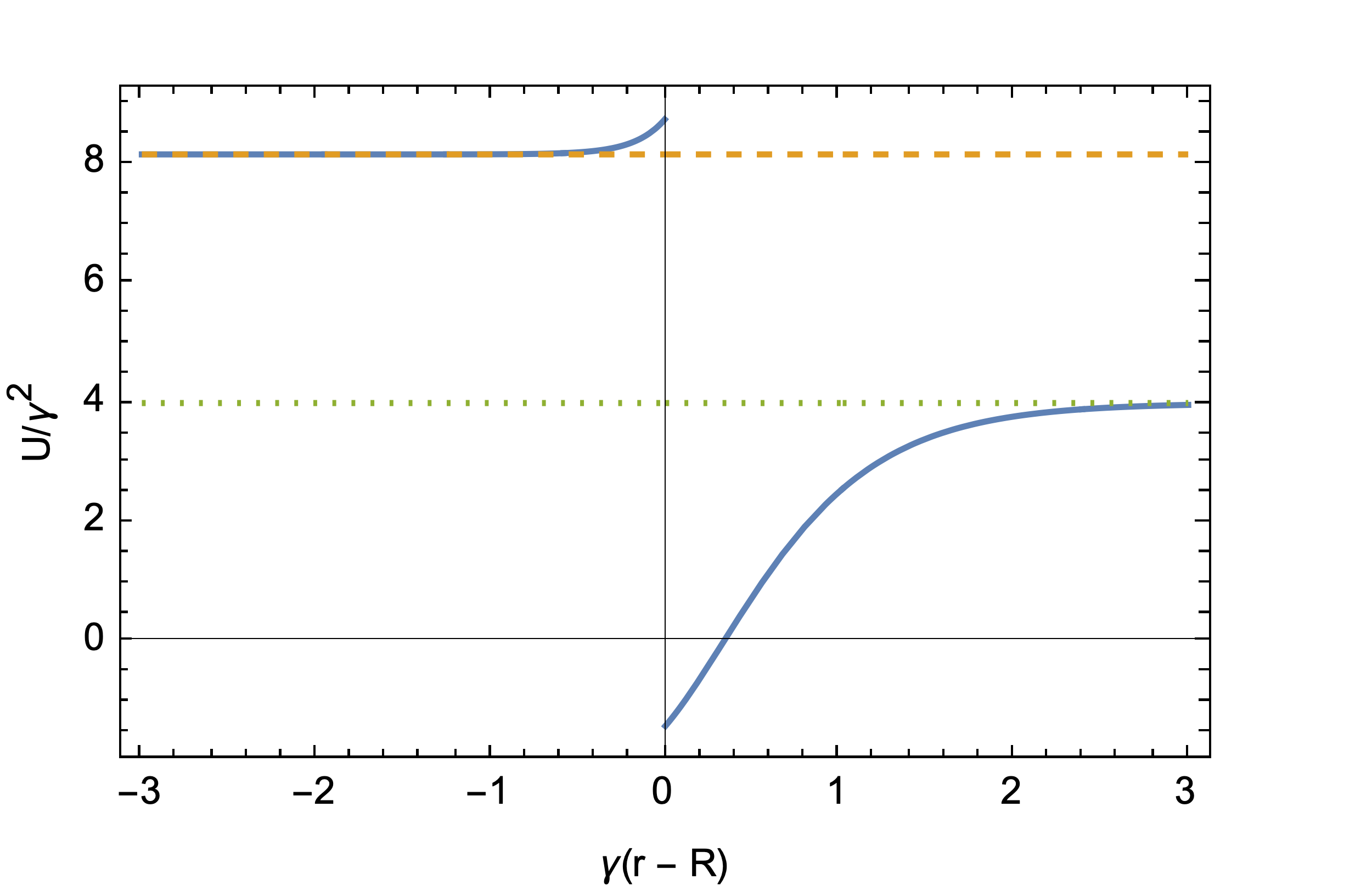}
        \subcaption{The effective potential of the classical configuration as a function of position. The dashed green and orange lines correspond to the values \(4 \gamma^2\) and \(g^2\gamma^2\) respectively.}
        \label{fig:eigen_potential_a}
    \end{subfigure}
    \hfill
    \begin{subfigure}{\textwidth}
        \centering
        \includegraphics{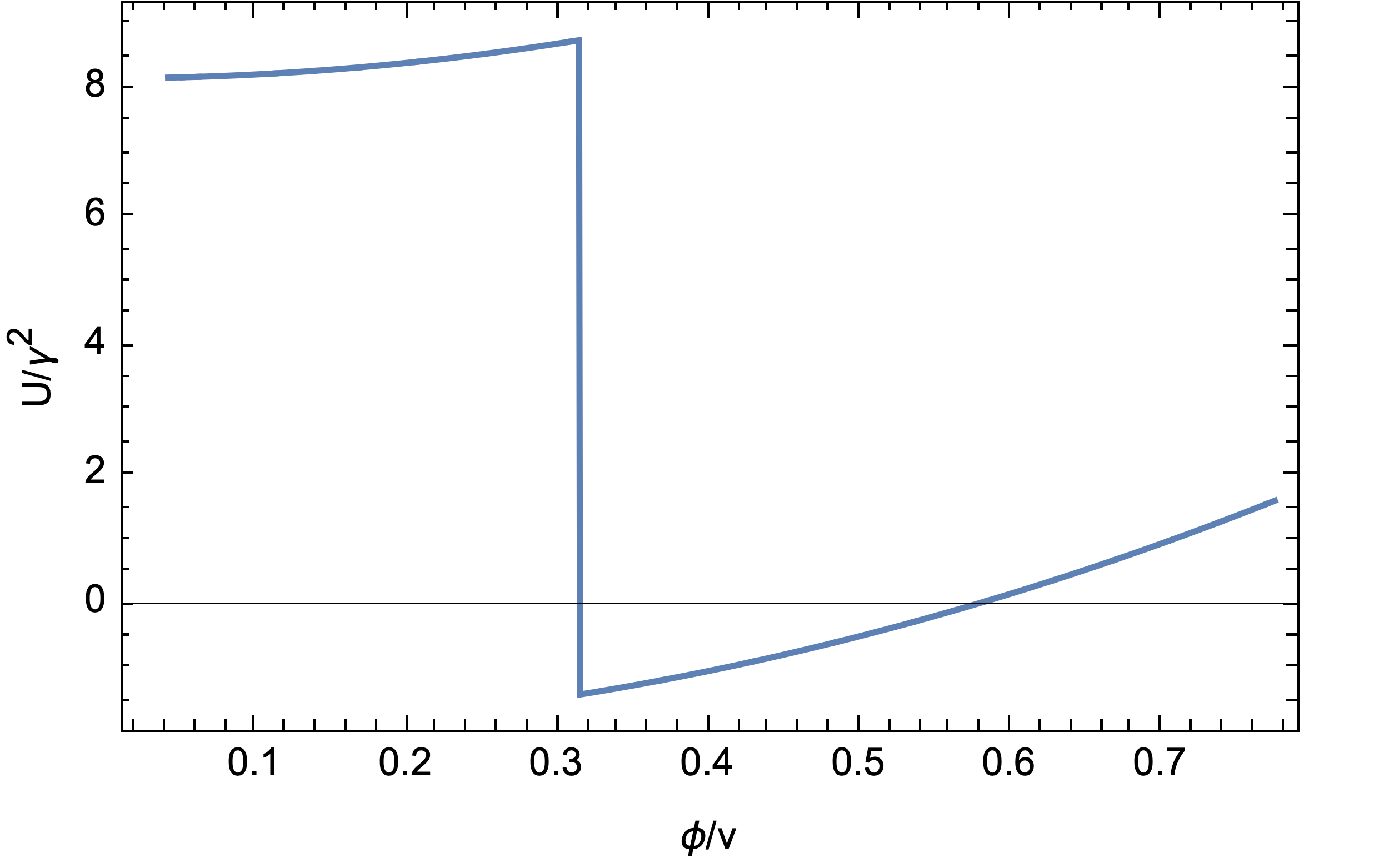}
        \subcaption{The effective potential of the classical configuration as a function of the normalised field. The effective potential is piecewise-convex in this representation.}
    \label{fig:eigen_potential_b}
    \end{subfigure}
    \caption{The effective potential of the classical configuration.}
\end{figure}

We denote by \(\Psi_\mathbf{n}(t, \mathbf x)\) the discrete eigenfunctions of the fluctuation operator \(\text{L}\), as defined in eq.~\eqref{eq:fluc}. They satisfy
\begin{equation}
    \text{L}\Psi_\mathbf{n}(t, \mathbf x) = -\lambda_\mathbf{n}\Psi_\mathbf{n}(t, \mathbf x)\:,
\end{equation}
where \(\mathbf{n}\) contains an analogue of the principal quantum number and angular momentum quantum number. With the separation of variables \(\Psi(t, \mathbf x) = f(t)\Psi(\mathbf x)\), we find that the time-dependent part is a pure phase
\begin{equation}
    f(t) = e^{\pm i\omega t}\:,
\end{equation}
and the spatial part of the solution satisfies
\begin{equation}
    \left[-\nabla^2 + \od{^2V_\text{eff}}{\varphi_\text{cl}^2} -\omega^2\right]\Psi_\mathbf{n}(\mathbf x) = -\lambda_\mathbf{n}\Psi_\mathbf{n}(\mathbf x)\:,
\end{equation}
along with the orthogonality condition
\begin{equation}
    \int_{\R^3}\text{d}^3\mathbf x\,\Psi_\mathbf{n}(\mathbf x)\Psi^*_{n'}(\mathbf x) = \delta_{\mathbf{nn}'}\:.
\end{equation}
If we define the parameter \(E_\mathbf{n} = \omega^2 - \lambda_\mathbf{n}\) and the function \(U(r) = \text{d}^2V_\text{eff}/\text{d}\varphi_\text{cl}^2\), then the eigenvalue equation can be written in the suggestive form
\begin{equation}
    \left[-\nabla^2 + U(r)\right]\Psi_\mathbf{n}(\mathbf x) = E_\mathbf{n}\Psi_\mathbf{n}(\mathbf x)\:.
\end{equation}
In other words, the spatial part of the eigenfunctions of \(\text{L}\) are simply the eigenfunctions of the ``Hamiltonian'' \(\hat H = -\nabla^2 + U(r)\). Plots of the ``central potential'' \(U\) are given in figures \ref{fig:eigen_potential_a} and \ref{fig:eigen_potential_b}.

Modes with zero or negative eigenvalues are problematic in the Euclidean version of this calculation. The former lead to divergences while the latter indicate instabilities. There is a subtle difference in the nature of the zero/negative mode problem when considered in Minkowski spacetime vs. Euclidean space. Our dispersion relation does not fix a single value of \(\omega\), so it would appear that the Minkowski space fluctuation operator can only have a continuous spectrum. However, since the spectral representation of the Minkowski space Green's function has the schematic form
\begin{equation}
    G(x, x') = \int\frac{\text{d}\omega}{2\pi}e^{-i\omega(t - t')}\left[\sum_\mathbf{n}\frac{\Psi_{\lambda_\mathbf{n}}(\mathbf x)\Psi_{\lambda_\mathbf{n}}(\mathbf x')^*}{\lambda_\mathbf{n}(\omega)} + \int\text{d}\lambda\,\frac{\Psi_\lambda(\mathbf x)\Psi_\lambda(\mathbf x')^*}{\lambda(\omega)}\right]\:,
\end{equation}
a divergence could still arise if the spatial part of the fluctuation operator admits a discrete zero mode. In other words, the problem of zero/negative modes only applies to the spatial part of the Minkowski space operator.

In the literature on vacuum decay \cite{coleman_1977, Devoto_2022,garbrecht_2015_a}, one tends to show the existence of zero modes by taking the gradient of the equation of motion \eqref{eq:static_eom}. It would appear that we can do the same here by writing
\begin{equation}
    \label{eq:zero_mode_eq}
    0 = \nabla\left(\nabla^2\varphi_\text{cl} - \od{V_\text{eff}}{\varphi_\text{cl}}\right) \stackrel{?}{=} \left(\nabla^2 + \od{^2V_\text{eff}}{\varphi_\text{cl}^2}\right)\nabla\varphi_\text{cl}\:,
\end{equation}
which looks like the equation of a zero mode for \(\omega = 0\). There are two problems with this. First, even if the second equation were correct, \(\nabla\varphi_\text{cl}\) is not a differentiable function (that is, it does not have a continuous first derivative), and thus is not an eigenfunction of the differential operator \(\text{L}\). Secondly, the effective potential depends explicitly on position because of the matter density 
\(\rho(\mathbf x)\), so taking its gradient results in an additional contribution proportional to \(\delta(\mathbf x)\rho_0\). Consequently, the operator in the first equation in eq.~\eqref{eq:zero_mode_eq} is not truly the fluctuation operator. Similar arguments tell us that taking the radial derivative will not yield the equation of motion of the negative eigenmode, as it does for vacuum decay. In principle, such modes may still exist, so we continue the calculation.

Since \(\text{L}\) is spherically symmetric, the \(\Psi_\mathbf{n}(\mathbf x)\) admit a spherical harmonic expansion
\begin{equation}
    \label{eq:expansion}
    \Psi_\mathbf{n}(\mathbf x) = \Psi_{nl}(r, \boldsymbol \Omega) = \frac{\psi_{nl}(r)}{r}Y_l^m(\boldsymbol \Omega)\:,
\end{equation}
where \(\boldsymbol \Omega\) is a vector containing the angular coordinates, \(l\in\Z\) is the angular momentum number, \(m\in[-l, l]\) and \(n\) is a discrete label which is not necessarily an integer. Naturally, any discrete labelling may just as well be expressed with integer indices, but this will prove to be unnecessary. The \(\psi_{nl}\) satisfy
\begin{equation}
    \label{eq:evp}
    \left[-\od{^2}{r^2} + \frac{l(l+1)}{r^2} + U(r)\right]\psi_{nl} = E_{nl}\psi_{nl}\:.
\end{equation}
Hereafter, we will work in the thin-wall approximation. Translating to it involves the replacements that we made in the earlier section, along with the assignment \(r \rightarrow R\) in the centripetal term \cite{garbrecht_2015_a}. This changes our interpretation of the Schr{\"o}dinger equation. No longer does it describe a central potential but an effective one-dimensional potential
\begin{equation}
    U_\text{eff}(s) = \frac{l(l+1)}{R^2} + \od{^2V_\text{eff}}{\varphi_\text{cl}^2}\:.
\end{equation}
For completeness, we note the main features of this potential. For \(s<0\), it is given by
\begin{equation}
    U_\text{eff}(s) = \frac{l(l+1)}{R^2}+g^2 \gamma^2\left(1 + 6 \text{csch}^2\left(c_- -g \gamma s\right)\right)\:.
\end{equation}
As \(s\rightarrow -\infty\), \(\csch\left(c_- - m_- s\right)\rightarrow 0\). Therefore,
\begin{equation}
    \lim_{s\rightarrow -\infty}U(s) = \frac{l(l+1)}{R^2} + g^2\gamma^2 \:.
\end{equation}
The potential initially grows slowly before experiencing rapid growth near the boundary of the source and reaching a global maximum
\begin{equation}
    \operatorname{max}\left(U_\text{eff}\right) = \frac{l(l+1)}{R^2} + \gamma ^2 \left(\frac{6}{2 + g^2} + g^2\right)\:,
\end{equation}
and then discontinuously dropping to a global minimum
\begin{equation}
    \operatorname{min}\left(U_\text{eff}\right) = \frac{l(l+1)}{R^2} + \gamma ^2 \left(\frac{6}{2 + g^2}-2\right)\:.
\end{equation}
For \(s>0\), the potential is
\begin{equation}
    U_\text{eff}(s) = \frac{l(l+1)}{R^2} + 2\gamma^2 \left(3 \tanh ^2\left(\gamma  s+c_+\right)-1\right)\:.
\end{equation}
Since \(\tanh\left(\gamma s + c_1\right)\rightarrow 1\) as \(s\rightarrow \infty\), we find that the lower asymptote is
\begin{equation}
    \lim_{s\rightarrow \infty} U(s) = \frac{l(l+1)}{R^2} + 4\gamma^2\:.
\end{equation}
In conclusion, if the fluctuation operator admits discrete modes, they can only exist for \(\operatorname{min}\left(U_\text{eff}\right) < E_{nl} < l(l+1)/R^2 + 4\gamma^2\). These correspond to bound states of the ``Hamiltonian''.

\subsection{General solution}
We now determine the canonical solutions for the eigenvalue problem. To this end, we split around the discontinuity at \(s = 0\) and consider two separate equations for the interior \(\psi^{nl}_-\) and exterior \(\psi^{nl}_+\) parts. The full \(l^\text{th}\) component of the eigenmode with index \(n\) is then given by
\begin{equation}
    \psi_{nl}(s) = \Theta(-s)\psi^{nl}_-(s) + \Theta(s)\psi^{nl}_+(s)\:,
\end{equation}
and it is uniquely determined by imposing continuity, continuity of the first derivative and normalisability. The exterior modes satisfy
\begin{equation}
    \label{eq:ext}
    \left[\od{^2}{s^2} - \frac{l(l+1)}{R^2} + E_{nl} + 2\gamma^2 - 6\gamma^2\chi_+^2\right]\psi^{nl}_+ = 0\:.
\end{equation}
By changing the independent variable to the normalised external field
\begin{equation}
    \label{eq:u}
    \chi_+ = \tanh\left(\gamma s + c_1\right)\Rightarrow \od{}{s} = \gamma\left(1-\chi_+^2\right)\od{}{\chi_+}\:,
\end{equation}
we transform eq.~\eqref{eq:ext} to
\begin{equation}
    \label{eq:legendre}
    \od{}{\chi_+}\left[\left(1-\chi_+^2\right)\od{\psi^{nl}_+}{\chi_+}\right] + \left[j(j+1) - \frac{n^2}{1-\chi_+^2}\right]\psi^{nl}_+ = 0\:,
\end{equation}
and recover the general Legendre equation with degree \(j = 2\) and order 
\begin{equation}
    \label{eq:order}
    n = \pm\frac{1}{\gamma}\left(\frac{l(l+1)}{R^2} - E_{nl} + 4\gamma^2\right)^{1/2}\:.
\end{equation}
The canonical solutions are \(P_2^n\) and \(Q_2^n\), the associated Legendre functions of the first and second kind, respectively. Thus, the general solution is
\begin{equation}
    \label{eq:general_ext_mode}
    \psi^{nl}_+(s) = A_l P_2^n\left(\chi(s)\right) + B_l Q_2^n\left(\chi(s)\right)\:.
\end{equation}
Since \(l(l+1)/R^2 - E_{nl}>-4\gamma^2\), \(n\) is real. If \(n<0\), normalisability requires that we set \(B_l = 0\) for all \(l\). If \(n\) is positive, we may swap \(P_2^n\) for \(P_2^{-n}\) (see section \ref{sec:green_technical}) and are still required to set \(B_l = 0\). Furthermore, since \(P_2^n\) for negative \(n\) and \(P_2^{-n}\) for positive \(n\) coincide, we may set \(n > 0\) without loss of generality and represent the exterior part of the eigenfunctions as
\begin{equation}
    \psi^{nl}_+(s) = A_lP_2^{-n}\left(\chi_+(s)\right)\:,
\end{equation}
with \(n>0\).

Rearranging eq.~\eqref{eq:order} gives the spectrum
\begin{equation}
    \label{eq:spectrum}
    \lambda_{nl} = \omega^2 + \gamma^2(n^2 - 4) - \frac{l(l+1)}{R^2}\:.
\end{equation}
Requiring that the interior eigenvalue problem gives rise to the same spectrum yields
\begin{equation}
    \label{eq:int}
    \left[\od{^2}{s^2} + \gamma^2\left(4 - n^2 - g^2 - 6\chi_-^2\right)\right]\psi^{nl}_- = 0\:.
\end{equation}
We encounter this equation in section \ref{sec:green_technical} and show that it is solved by functions resembling hypergeometric functions,
\begin{equation}
    \psi_-(z) = C_lF_a(z) + D_lF_{-a}(z)\:,
\end{equation}
where \(z = \chi_-^2/g^2\). See eq.~\eqref{eq:hyper-like} for the definition of \(F_{\pm a}\). Since \(F_{-a}\) blows up at the origin of the source, normalisability demands \(D_l = 0\) for all \(l\), which leaves
\begin{equation}
    \psi_-(s) = C_lF_a\left(\frac{\chi_-(s)^2}{g^2}\right)\:.
\end{equation}

\subsection{Continuity, differentiability and normalisation}
\begin{figure}[t]
    \centering
    \includegraphics{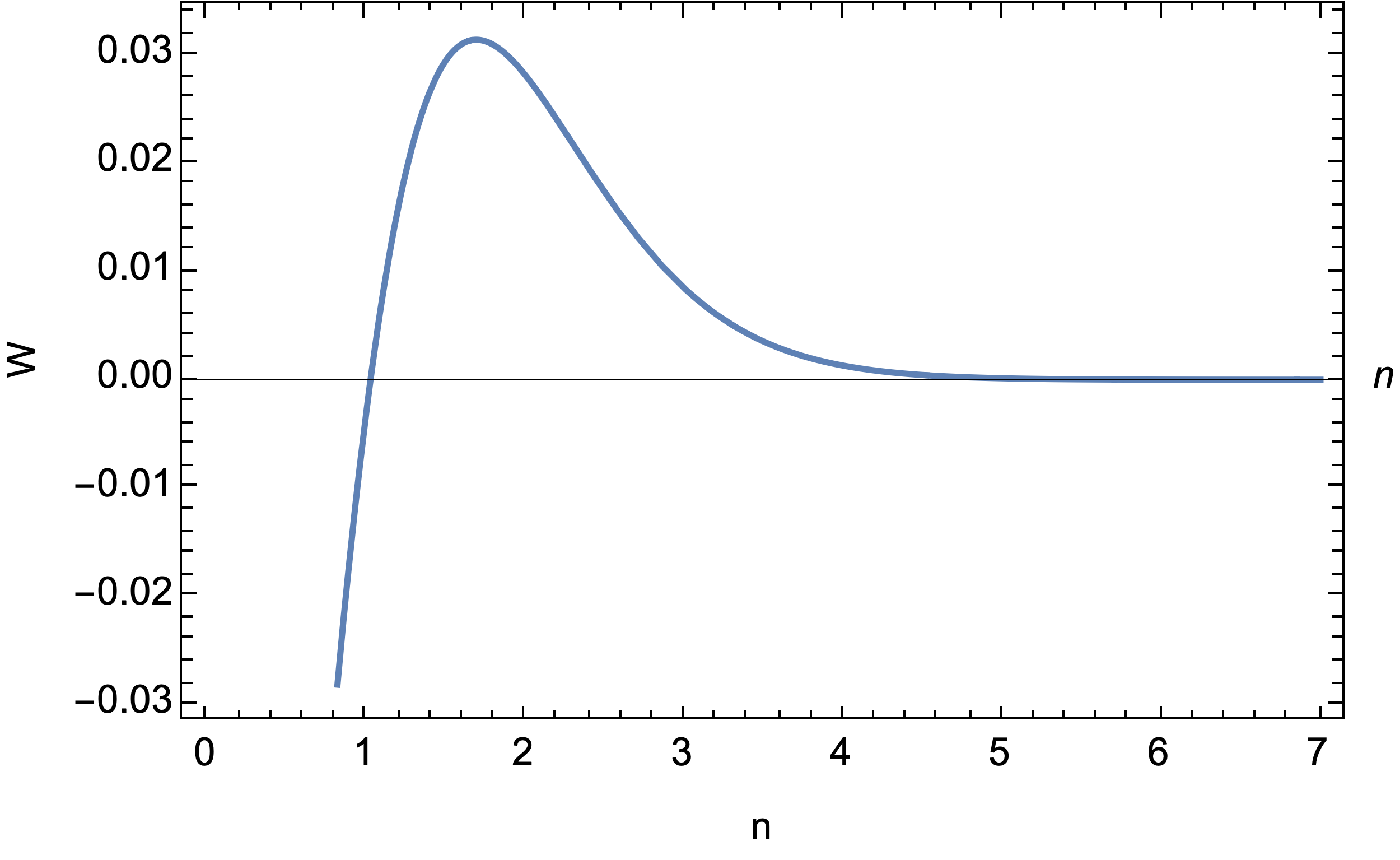}
        \caption{A plot of the Wro{\'n}skian \(\left.W\left[F_a\left(z(s)\right),P_2^{-n}(u(s))\right]\right\vert_{s = 0}\) as a function of \(n\). The root corresponds to the value of \(n\) which satisfies the ``quantisation condition'' \ref{eq:eigenmode_Wronskian}. Note its proximity to \(n=1\).}
        \label{fig:quantisation}
\end{figure}

\begin{figure}[t]
    \centering
    \includegraphics{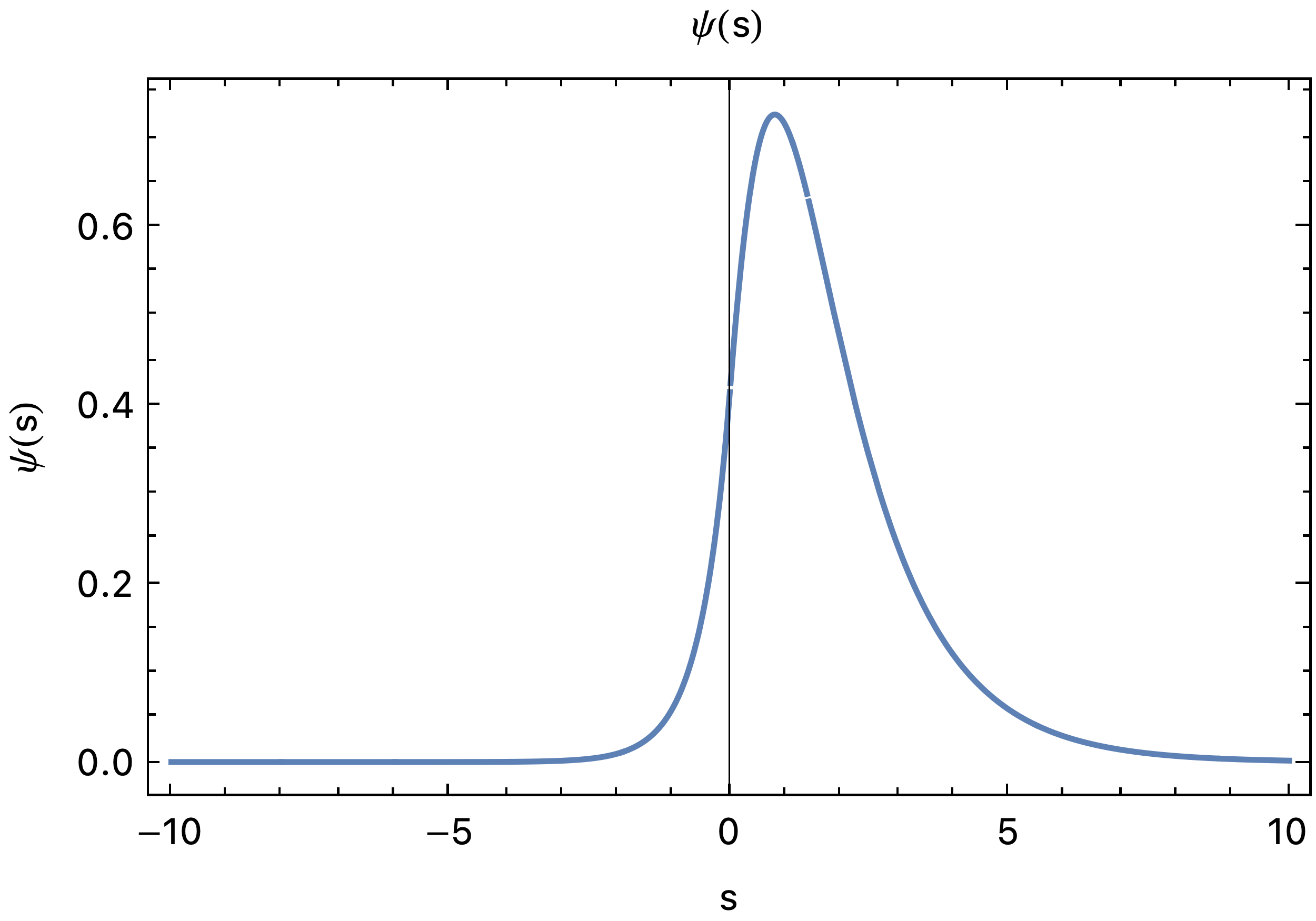}
    \caption{A qualitative plot of the \(n = 1 + \epsilon\) eigenfunction. Note that the mode is bound to the source.}
    \label{fig:eigenmode}
\end{figure}

The continuity condition reads
\begin{equation}
    A_l P_2^{-n}(u_0) = C_l F_a(z_0)\Leftrightarrow C_l = A_l\frac{P_2^{-n}(u_0)}{F_a\left(z_0\right)}\:,
\end{equation}
fixing one of the constants. A choice of normalisation fixes the other. Differentiability requires
\begin{equation}
    \left.A_l \od{}{s}P_2^{-n}(u(s))\right\vert_{s = 0} = \left.C_l \od{}{s}F_a\left(z(s)\right)\right\vert_{s = 0}
\end{equation}
which combines with the continuity condition to produce the constraint
\begin{equation}
    \label{eq:eigenmode_Wronskian}
    \left.W\left[F_a\left(z(s)\right),P_2^{-n}(u(s))\right]\right\vert_{s = 0} = 0\:,
    \end{equation}
where \(W\) denotes the Wro{\'n}skian. This is the ``quantisation condition'', which constrains the allowed values of \(n\). It cannot be solved in closed form but is easily approximated numerically. A plot of the Wro{\'n}skian at \(s = 0\) as a function of \(n\) is given in figure \ref{fig:quantisation} for a set of test values of the Lagrangian parameters. We observe one root near \(n = 1\), omitting the infinite number of roots for negative \(n\) that normalisability rules out. Indeed, for a wide range of physical parameter values, the root is always close to unity. A simple Newton-Raphson iteration is sufficient to arrive at a decent approximation to \(n\), and we have plotted the resulting eigenfunction in figure \ref{fig:eigenmode}.

In conclusion, the spatial part of the fluctuation operator admits one tower of discrete modes with positive eigenvalues. There are no zero/negative modes.

\section{Green's function}
This appendix provides a full derivation of the Green's function quoted at the end of section \ref{sec:4}. As a pedagogical introduction to the manipulations involved, we first analyse the simpler problem of a free field with piecewise constant mass in appendix \ref{sec:piecewise_con}. Then the calculation relevant for the symmetron field is given in detail in appendix \ref{sec:green_technical}. 
\subsection{Example: piecewise constant mass}
\label{sec:piecewise_con}
First, we consider a 1+1 dimensional system, which could correspond to a field with constant mass in the positive and negative domain but experiences a discontinuous jump at the origin. The Green's function for such a system satisfies the equation
\begin{equation}
    \label{eq:green_step}
    \left(\pd{^2}{t^2}-\pd{^2}{x^2} + V(x)\right)G(t, t';x, x') = -\delta(t - t')\delta(x - x')\:,
\end{equation}
where the potential \(V(x)\) is given by 
\begin{equation}
    V(x) = \begin{cases}
        m_+^2\:, & x > 0\\
        m_-^2\:, & x < 0\:,
    \end{cases}
\end{equation}
with \(m_+<m_-\). This equation describes a step-wise free system with an effective mass that depends on the sign of the coordinate. We suppose that the coordinate-space Green's function \(G(t, t';x, x')\) may be written as a weighted superposition of temporal plane waves
\begin{equation}
    G(t, t';x, x') = \int\frac{\text{d}E}{2\pi}\,e^{-iE(t - t')}G(x, x';E)\:,
\end{equation}
so that the frequency-domain Green's function \(G(x, x';E)\) satisfies
\begin{equation}
    \left[-\pd{^2}{x^2} + V(x) - E^2\right]G(x, x';E) = -\delta(x - x')\:.
\end{equation}
We denote by \(G^>(x, x';E)\) and \(G^<(x, x';E)\) the contributions to \(G(x, x';E)\) in the regions \(x > x'\) and \(x < x'\) respectively. The latter two are completely specified by the former, since
\begin{equation}
    G(x, x';E) = \Theta(x - x')G^>(x, x';E) + \Theta(x' - x)G^<(x, x';E)\:,
\end{equation}
and \(G^<(x, x';E) = G^>(x', x;E)\). The two functions satisfy the homogeneous equation
\begin{equation}
    \left[-\od{^2}{x^2} + V(x) - E^2\right]G^\gtrless(x, x';E) = 0\:.
\end{equation}

Without loss of generality, we may consider the case in which \(E\) is positive. The behaviour of the Green's function depends on the value of \(E\) relative to \(V(x)\). We identify three distinct regimes: the bound regime \(0<E<m_+\), the tunnelling regime \(m_+ < E < m_-\) and the scattering regime \(m_- < E\). Additionally, the discontinuities at \(x = 0\), \(x' = 0\) and \(x = x'\) split the \(xx'\)-plane into six distinct regions. We denote this with a subscript \((\operatorname{sgn} x, \operatorname{sgn} x')\), as summarised in figure \ref{fig:regions}.

\subsubsection*{Bound regime}
We shall use the symbol \(B\) to denote the bound contribution to the Green's function. While \(0<E<m_+\), the spectral sum runs over normalisable eigenfunctions of the differential operator in eq.~\eqref{eq:green_step}. The equation
\begin{equation}
    \left[-\od{^2}{x^2} + m_+^2- E^2\right]y_+(x) = 0
\end{equation}
describes the \(x\)- and \(x'\)-dependence of \(B^\gtrless_{l, +}(x, x';E)\), the \(x\)-dependence of \(B^>_{l, \pm}(x, x';E)\) and the \(x'\)-dependence of \(B^<_{l, \pm}(x, x';E)\). The equation
\begin{equation}
    \left[-\od{^2}{x^2} + m_-^2- E^2\right]y_-(x) = 0
\end{equation}
describes the \(x\)- and \(x'\)-dependence of \(B^\gtrless_{l, -}\), the \(x'\) dependence of \(B^>_{l, \pm}(x, x';E)\) and the \(x\)-dependence of \(B^<_{l, \pm}(x, x';E)\). Hyperbolic functions solve both equations:
\begin{equation}
    y_+(x) \sim \exp\left(\pm k_+x\right)\;,
\end{equation}
where \(k_+^2 = m_+^2 - E^2\) and 
\begin{equation}
    y_-(x) \sim \exp\left(\pm k_-x\right)\:,
\end{equation}
where \(k_-^2 = m_-^2 - E^2\). The usual homogeneous boundary conditions will require \(B^>_{l, +}(x, x';E)\) to tend to zero as \(x\rightarrow \infty\) but leave it unconstrained in \(x'\). Thus, 
\begin{equation}
    B^>_{l, +}(x, x';E) = e^{-k_+x}\left(a_1 e^{k_+x'}+ a_2 e^{-k_+x'}\right) = B^<_{l, +}(x', x;E)\:.
\end{equation}
Similarly, \(B^>_{l, \pm}\) tends to zero as \(x\rightarrow \infty\) and as \(x'\rightarrow -\infty\), which suggests
\begin{equation}
    B^>_{l, \pm}(x, x';E) = be^{k_-x'-k_+x} = B^<_{l, \pm}(x', x;E)\:.
\end{equation}
Finally, \(B^>_{l, -}(x, x';E)\) tends to zero as \(x'\rightarrow -\infty\) and is unconstrained in \(x\). Hence, 
\begin{equation}
    B^>_{l, -}(x, x';E) = e^{k_-x'}\left(c_1 e^{-k_-x}+ c_2 e^{k_-x}\right) = B^<_{l, -}(x', x;E)\:.
\end{equation}

The Green's function is continuous in \(x\) and \(x'\), but it's derivative experiences a jump discontinuity along \(x = x'\), which is given by
\begin{equation}
    \left.\od{}{x}G^>(x, x';E)\right\vert_{x=x'} - \left.\od{}{x}G^<(x, x';E)\right\vert_{x = x'} = 1\:.
\end{equation}
Each pair of bound functions with the same combination of signs in their subscript must satisfy this condition. They may each be solved to give the following parameter values:
\begin{equation}
    a_1 = -\frac{1}{2 k_+}\:,\, b = -\frac{1}{k_- + k_+}\:,\,c_1 = -\frac{1}{2 k_-}\:.
\end{equation}

Note that the derivative jump condition is imposed at \(x = x' = 0\) for the mixed-sign contributions. Continuity is imposed along the \(x\) and \(x'\) axes: \(B^>_{l, +}(x, 0;E) = B^>_{l, \pm}(x, 0;E)\) and \(B^>_{l, -}(0, x';E) = B^>_{l, \pm}(0, x';E)\). This gives the remaining parameters
\begin{equation}
    a_2 = -\frac{1}{2 k_+}\frac{k_+ - k_-}{k_+ + k_-}\:,\,c_2 = -\frac{1}{2 k_-}\frac{k_- - k_+}{k_+ + k_-}\:.
\end{equation}
Note also the resemblance between these parameters and the transmission and reflection amplitudes. The bound Green's function \(B(x, x';E)\) is thus completely specified by
\begin{equation}
    B^>_{l, +}(x, x';E) = -\frac{1}{2 k_+}e^{-k_+x}\left(e^{k_+x'} + \frac{k_+ - k_-}{k_+ + k_-} e^{-k_+x'}\right) = B^<_{l, +}(x', x;E)\:,
\end{equation}
\begin{equation}
    B^>_{l, \pm}(x, x';E) = -\frac{e^{-k_+x}e^{k_-x'}}{k_- + k_+} = B^<_{l, \pm}(x', x;E)\:,
\end{equation}
and
\begin{equation}
    B^>_{l, -}(x, x';E) = -\frac{1}{2 k_-}e^{k_-x'}\left(e^{-k_-x} + \frac{k_- - k_+}{k_+ + k_-} e^{k_-x}\right) = B^<_{l, +}(x', x;E)\:.
\end{equation}

\subsubsection*{Tunneling regime}
While \(m_+ < E < m_-\), exterior eigenfunctions look like plane waves, while interior eigenfunctions are still attenuated. Consequently, homogeneous boundary conditions remain for equations defined for \(x,x'<0\), while the boundary behaviour of solutions defined for \(x, x'>0\) will, for now, remain unspecified.

The equation for \(y_+(x)\) is solved by trigonometric functions,
\begin{equation}
    y_+(x) \sim \exp\left(\pm i p_+x\right)\:,
\end{equation}
where \(p_+^2 = E^2 - m_+^2\). We will use the symbol \(T\) to denote tunnelling contributions to the Green's function. That \(T^\gtrless_{l, +}(x, x';E)\) appears unconstrained suggests it takes the form
\begin{equation}
    T^>_{l, +}(x, x';E) = a_1e^{ip_+(x - x')} + a_2 e^{-ip_+(x - x')} + a_3e^{ip_+(x + x')} + a_4e^{-ip_+(x + x')}\:.
\end{equation}
The falloff condition does still apply, but in accordance with the Feynman contour prescription. This is equivalent to adding a small imaginary part to the momentum, and thus, a falloff is achieved only if \(a_4 = 0\). What's more, since \(x > x'\), we must also have \(a_2 = 0\), so the constrained form of \(T^>_{l, +}\) is
\begin{equation}
    T^>_{l, +}(x, x';E) = a_1e^{ip_+(x - x')} + a_3e^{ip_+(x + x')} = T^<_{l, +}(x', x;E)\:.
\end{equation}
\(T^>_{l, \pm}(x, x';E)\) tends to zero as \(x'\rightarrow-\infty\) and as \(x\rightarrow \infty\) (when the momentum acquires a small imaginary part), suggesting the form
\begin{equation}
    T^>_{l, \pm}(x, x';E) = b e^{ip_+x + k_-x'} = T^<_{l, \pm}(x', x;E)\:.
\end{equation}
Finally, \(T^>_{l, -}(x, x';E)\) tends to zero as \(x'\rightarrow -\infty\) but is unconstrained in \(x\), suggesting 
\begin{equation}
    T^>_{l, -}(x, x';E) = c_1 e^{k_-x'}\left(e^{-k_-x} + c_2 e^{k_-x}\right) = T^<_{l, -}(x', x;E)\:.
\end{equation}

The discontinuity in the derivative at \(x = x'\) gives us three of the five constants,
\begin{equation}
    a_1 = -\frac{i}{2p_+}\:,\,b = \frac{1}{ip_+ - k_-}\:,\,c_1 = -\frac{1}{2k_-}\:.
\end{equation}
Continuity gives us the remaining two,
\begin{equation}
    a_3 = -\frac{i}{2p_+}\left(\frac{ip_+ + k_-}{ip_+ - k_-}\right)\:,\,c_2 = -\frac{1}{2 k_-}\left(\frac{k_- + ip_+}{k_- - ip_+}\right)\:.
\end{equation}
In conclusion, \(T(x, x';E)\) is completely specified by
\begin{equation}
    T^>_{l, +}(x, x';E) = -\frac{i}{2p_+}\left[e^{i p_+ (x-x')} + \frac{ip_+ + k_-}{ip_+ - k_-} e^{i p_+ (x+x')}\right] = T^<_{l, +}(x', x;E)\:,
\end{equation}
\begin{equation}
    T^>_{l, \pm}(x, x';E) = \frac{e^{ip_+x + k_-x'}}{ip_+ - k_-} = T^<_{l, \pm}(x', x;E)\:,
\end{equation}
and
\begin{equation}
    T^>_{l, -}(x, x';E) = -\frac{1}{2k_-}e^{k_-x'}\left(e^{-k_-x} + \frac{k_- + ip_+}{k_- - ip_+}e^{k_-x}\right) = T^<_{l, -}(x', x;E)\:.
\end{equation}

\subsubsection*{Scattering regime}
While \(m_- < E\), both exterior and interior eigenfunctions look like plane waves. The equation for \(y_-(x)\) is again solved by trigonometric functions
\begin{equation}
    y_-(x) \sim \exp(\pm ip_- x)\:,
\end{equation}
where \(p_-^2 = E^2 - m_-^2\). We adopt the symbol \(S\) to denote the scattering contributions to the Green's function. The form of the solution in the top-right quadrant of the \(x-x'\) plane does not change,
\begin{equation}
    S^>_{l, +}(x, x';E) = a_1 e^{i p_+(x - x')} + a_2 e^{ip_+(x + x')} = S^<_{l, +}(x', x;E)\:.
\end{equation}
The mixed sign solutions are unconstrained in general, but the falloff condition means the only term that contributes is the outgoing travelling wave,
\begin{equation}
    S^>_{l, \pm}(x, x';E) = b e^{i\left(p_+x - p_-x'\right)} = S^<_{l, \pm}(x', x;E)\:.
\end{equation}
Finally, the variation of \(S^>_{l, -}(x, x';E)\) mirrors that of the positive-sign solution, with the appropriate sign changes,
\begin{equation}
    S^>_{l, -}(x, x';E) = c_1 e^{i p_-(x - x')} + c_2 e^{-ip_-(x + x')} = S^<_{l, -}(x', x;E)\:.
\end{equation}
The jump discontinuity in the derivative once again gives us three of the five constants,
\begin{equation}
    a_1 = -\frac{i}{2p_+}\:,\,b = \frac{1}{ip_+ + ip_-}\:,\,c_1 = -\frac{i}{2p_-}\:.
\end{equation}
Continuity gives us the rest,
\begin{equation}
    a_2 = -\frac{i}{2p_+}\left(\frac{p_+ - p_-}{p_+ + p_-}\right)\:,\,c_2 = -\frac{i}{2p_-}\left(\frac{p_- - p_+}{p_- + p_+}\right)\:.
\end{equation}
Therefore, the scattering part of the Green's function is completely specified by
\begin{equation}
    S^>_{l, +}(x, x';E) =  -\frac{i}{2p_+}\left[ e^{i p_+(x - x')} + \frac{p_+ - p_-}{p_+ + p_-} e^{ip_+(x + x')} \right]= S^<_{l, +}(x', x;E)\:,
\end{equation}
\begin{equation}
    S^>_{l, \pm}(x, x';E) = \frac{e^{i\left(p_+x - p_-x'\right)}}{ip_+ + ip_-} = S^<_{l, \pm}(x', x;E)\:,
\end{equation}
and
\begin{equation}
    S^>_{l, -}(x, x';E) = -\frac{i}{2p_-} \left[e^{i p_-(x - x')} + \frac{p_- - p_+}{p_- + p_+} e^{-ip_-(x + x')} \right]= S^<_{l, -}(x', x;E)\:.
\end{equation}

\subsubsection*{All frequencies}
The similarities between the oscillatory and attenuated expressions strongly hint towards a universal expression that is valid for all frequencies. Since the solutions in each frequency interval are exponential functions, it is a trivial matter to see that this is true in the case of piecewise constant mass. One simply chooses the branch which sends \(k_\pm\) to \(ip_\pm\) when \(E\) goes from being less than to greater than \(m_\pm\). We save an explicit demonstration of this equivalence for the next section, where it is slightly easier to see immediately.

\subsection{Symmetron field}
\label{sec:green_technical}
We now give a detailed derivation of the Green's function for the symmetron fluctuation operator in the thin-wall approximation. Recall from section \ref{sec:4} the equation for the thin-wall Green's function,
\begin{equation}
    \left[-\od{^2}{s^2} + \frac{l(l+1)}{R^2} - E^2  + \od{^2V_\text{eff}}{\varphi_\text{cl}^2}\right]G_l(s, s'; E) = -\frac{\delta(s - s')}{R^2}\:.
\end{equation}
We denote by \(G^>_l(s, s';E)\) and \(G^<_l(s, s';E)\) the contributions to \(G_l(s, s';E)\) defined for \(s > s'\) and \(s<s'\) respectively. By the reciprocity condition,
\begin{equation}
    G^>_l(s, s';E) = G^<_l(s', s;E)\:,
\end{equation}
it follows that \(G_l(s, s';E)\), which is given by
\begin{equation}
    G_l(s, s';E) = \Theta(s' - s)G^<_l(s, s';E) + \Theta(s - s')G^>_l(s, s';E)\:,
\end{equation}
is completely characterised by either one of the \(G^>_l(s, s';E)\) or \(G^<_l(s, s';E)\). The step function in the effective potential introduces discontinuities on the lines \(s = 0\) and \(s'=0\), in addition to the standard \(s=s'\) discontinuity due to the delta function. Hence, the \(G^\gtrless_l(s, s';E)\) are further split into three contributions depending on the signs of \(s\) and \(s'\), taking the form
\begin{equation}
    \label{eq:lot_of_theta}
    G^\gtrless_l(s, s';E) = \Theta_s\Theta_{s'}G_{l,+}^\gtrless(s, s';E) + \Theta_s\Theta_{-s'}G_{l,\pm}^\gtrless(s, s';E) + \Theta_{-s}\Theta_{-s'}G_{l,-}^\gtrless(s, s';E)\:,
\end{equation}
where we append a sign to the subscript to identify these contributions. For brevity, we write \(\Theta_s \equiv \Theta(s)\). In order of appearance, we refer to each term as the exterior, boundary and interior contributions. A graphical summary of this decomposition is given in figure \ref{fig:regions}. We will solve for contributions to the \(G_l^\gtrless\) for different values of \(E\), adopting a naming scheme analogous to that in quantum mechanics: contributions for \(0<E<2\gamma\) are called bound, \(2\gamma < E < g\gamma\) tunnelling and \(E>g\gamma\) scattering (cf. appendix \ref{sec:piecewise_con}).

\begin{figure}
    \centering
    \begin{tikzpicture}[scale=1.25]
        \begin{axis}[
            axis lines = middle,
            xlabel = \(s\),
            ylabel = \(s'\),
            ticks = none,
            xmax = 2,
            xmin = -2,
            ymax = 2,
            ymin = -2]
        \addplot [color = black, dashed, domain=-1.45:1.45, samples=100] {x};
        \addplot [mark = none] coordinates {(1, 0.5)} node[label = 0:\(G_{l,+}^>\)]{};
        \addplot [mark = none] coordinates {(0.5, 1)} node[label = 90:\(G_{l,+}^<\)]{};
        \addplot [mark = none] coordinates {(-0.75, 0.75)} node[label = 135:\(G_{l,\pm}^<\)]{};
        \addplot [mark = none] coordinates {(-1, -0.5)} node[label = 180:\(G_{l,-}^<\)]{};
        \addplot [mark = none] coordinates {(-0.5, -1)} node[label = 270:\(G_{l,-}^>\)]{};
        \addplot [mark = none] coordinates {(0.75, -0.75)} node[label = 315:\(G_{l,\pm}^>\)]{};
        \node[draw, fill=white] at (1.5, 1.5) {exterior};
        \node[draw, fill=white] at (-1.5, -1.5) {interior};
        \node[draw, fill=white] at (0.75, -0.5) {boundary};
        \node[draw, fill=white] at (-0.75, 0.5) {boundary};
        \end{axis}
    \end{tikzpicture}
    \caption{A depiction of the contributions to the Green's function depending on the sign of \(s\) and \(s'\). \(G_{l, +}^\gtrless\), \(G_{l, -}^\gtrless\) and \(G_{l, \pm}^\gtrless\) are collectively referred to as the exterior, interior and boundary contributions, respectively.}
    \label{fig:regions}
\end{figure}

\subsubsection*{Bound regime}
When \(0<E<2\gamma\), the differential equation
\begin{equation}
    \label{eq:++}
    \left[-\od{^2}{s^2} + \frac{l(l+1)}{R^2} - E^2 - 2\gamma^2  + 6\gamma^2\chi_+(s)^2\right]y_+(s) = 0
\end{equation}
describes the \(s\)- and \(s'\)-dependence of \(G^\gtrless_{l, +}(s, s';E)\), the \(s\)-dependence of \(G^>_{l, \pm}(s, s';E)\) and the \(s'\)-dependence of \(G^<_{l, \pm}(s, s';E)\).
By changing the independent variable to the normalised external field,
\begin{equation}
    u = \chi_+ = \tanh\left(\gamma s + c_+\right)\Rightarrow \od{}{s} = \gamma\left(1-\chi_+^2\right)\od{}{\chi_+}\:,
\end{equation}
we transform this equation to
\begin{equation}
    \od{}{u}\left[\left(1-u^2\right)\od{y_+}{u}\right] + \left[j(j+1) - \frac{n^2}{1-u^2}\right]y_+(u) = 0\:,
\end{equation}
and recover the general Legendre equation with degree \(j = 2\) and order 
\begin{equation}
    n = \frac{1}{\gamma}\sqrt{\frac{l(l+1)}{R^2} - E^2 + 4\gamma^2}\:,
\end{equation}
which is real while we are in the bound regime. The canonical solutions are \(P_2^n\) and \(Q_2^n\), the associated Legendre functions of the first and second kind, respectively. Any solution set which uses both can be expressed just as well by a combination of \(P_2^n\) and \(P_2^{-n}\) via the identity \cite{olver}
\begin{equation}
    P_j^{n}(x) = \sec(n\pi)\frac{\Gamma(j + n + 1)}{\Gamma(j - n + 1)}P_j^{-n}(x) + \frac{2}{\pi}\tan(n\pi)Q_j^{n}(x)\:,
\end{equation}
so we may write
\begin{equation}
    y_+(s) \sim P_2^{\pm n}\left(u(s)\right)\:.
\end{equation}

The equation
\begin{equation}
    \label{eq:--}
    \left[-\od{^2}{s^2} + \frac{l(l+1)}{R^2} - E^2 + g^2\gamma^2 + 6\gamma^2\chi_-(s)^2\right]y_-(s) = 0
\end{equation}
describes the \(s\)- and \(s'\)-dependence of \(G^\gtrless_{l, -}(s, s';E)\), the  \(s'\)-dependence of \(G^>_{l, \pm}(s, s';E)\) and the \(s\)-dependence of \(G^<_{l, \pm}(s, s';E)\).
This time, we perform the change of variables 
\begin{equation}
    \label{eq:interior_substitution}
    u = \frac{\chi_-}{g} = \csch(c_- - g\gamma s)\Rightarrow \od{}{s} = g\gamma u\sqrt{1 + u^2}\od{}{u}\:,
\end{equation}
transforming the equation into
\begin{equation}
    \left[u^2(1+u^2)\od{^2}{u^2} + u(1 + 2u^2)\od{}{u} + \frac{4 - n^2 - g^2}{g^2} - 6u^2\right]y_-(u) = 0\:.
\end{equation}
This equation bears a strong similarity to the hypergeometric equation in \(-u^2\). To see this more clearly, we use the change of variables \(z = u^2\) to write 
\begin{equation}
    \label{eq:almost_hyper}
    \left[z(1+z)\od{^2}{z^2} + \left(1 + \frac{3}{2}z\right)\od{}{z} - \frac{3}{2} + \frac{4 - n^2 - g^2}{4g^2z}\right]y_-(z) = 0\:,
\end{equation}
then ``peel off'' the large \(z\) behaviour with the substitution
\begin{equation}
    y_-(z) = z^a f(z;a)\:,
\end{equation}
where \(a\) is a free parameter and \(f\) may depend on \(a\). Choosing \(a\) to be
\begin{equation}
    \label{eq:alpha}
    a = \frac{1}{2g\gamma}\sqrt{\frac{l(l+1)}{R^2} - E^2 + g^2\gamma^2}\:,
\end{equation}
allows us to write the solutions in terms of hypergeometric functions
\begin{equation}
    f_1(z,a) = {}_2F_1\left(a - 1, a + \frac{3}{2};2a + 1;-z\right)
\end{equation}
and 
\begin{equation}
    f_2(z;a) = z^{-2a}{}_2F_1\left(-a - 1, -a + \frac{3}{2};-2a + 1;-z\right)\:.
\end{equation}
By a nice coincidence, the solutions \(y_-\) are thus linked by a change of sign in \(a\). For brevity, we define the ``hypergeometric-like'' function
\begin{equation}
    \label{eq:hyper-like}
    F_a(z)= z^a {}_2F_1\left(a - 1, a + \frac{3}{2};2a + 1;-z\right)\:,
\end{equation}
and write
\begin{equation}
    y_-(s) \sim F_{\pm a}\left(z(s)\right)\:.
\end{equation}
Note that \(a\) will remain real until we reach the scattering regime.

Let \(u' = u(s')\) and \(z' = z(s')\):
\paragraph{Exterior:}
\(G^>_{l, +}(s, s';E)\) tends to zero as \(s\rightarrow \infty\) but is unconstrained in \(s'\), suggesting that
\begin{equation}
    G_{l, +}^>(s, s';E) = P_2^{-n}(u)\left(AP_2^n(u') + BP_2^{-n}(u')\right) = G_{l, +}^<(s', s;E)\:,
\end{equation}
where \(A\) and \(B\) are constants.
\paragraph{Boundary:}
\(G^>_{l, \pm}(s. s';E)\) tends to zero as \(s\rightarrow\infty\) and \(s'\rightarrow -\infty\), suggesting that
\begin{equation}
    G^>_{l, \pm}(s, s';E) = \frac{P_2^{-n}(u)F_a(z')}{\mathcal W} = G^<_{l, \pm}(s', s;E)\:,
\end{equation}
where \(\mathcal W\) is a constant. 
\paragraph{Interior:}
finally, \(G^>_{l, -}(s, s')\) tends to zero as \(s'\rightarrow -\infty\) but is unconstrained in \(s\), suggesting that
\begin{equation}
    G_{l, -}^>(s, s';E) = F_a(z')\left(CF_{-a}(z) + DF_a(z)\right) = G_{l, -}^<(s', s;E)\:,
\end{equation}
where \(C\) and \(D\) are both constants.

The Green's function is continuous in \(s\) and \(s'\), but its derivative experiences a jump discontinuity along the line \(s = s'\), given by
\begin{equation}
    \label{eq:d-jump}
    \left.\od{}{s}G^>(s, s')\right\vert_{s = s'} - \left.\od{}{s}G^<(s, s')\right\vert_{s = s'} = \frac{1}{R^2}\:.
\end{equation}
This applies to pairs of bound functions with the same sign subscript. Let \(u_0 = u(0)\) and \(z_0 = z(0)\). Then the parameter values are given by \cite{olver, andrews}
\begin{equation}
    A = \frac{1}{R^2}\left[W\left(P_2^{n}(u), P_2^{-n}(u)\right)\right]^{-1} = -\frac{\pi}{2\gamma R^2}\csc\left(n\pi\right)\:,
\end{equation}
\begin{equation}
    C = \frac{1}{R^2}\left[W\left(F_{a}(z), F_{-a}(z)\right)\right]^{-1} = \frac{1}{4ag\gamma   R^2}\:,
\end{equation}
and
\begin{equation}
    \mathcal{W} = W(F_a(z_0), P_2^{-n}(u_0))R^2\:.
\end{equation}
Note that, since the only place where \(s = s'\) for the mixed-sign contributions is \(s = s' = 0\), the relevant Wro{\'n}skian must be evaluated at the origin. One can find the remaining constants by enforcing continuity of the Green's function and its first derivative on the \(s\)- and \(s'\)-axes. For any \(s\), the values and derivatives of \(G_{l, +}^>(s, s';E)\) and \(G_{l, \pm}^>(s, s';E)\) must match at \(s' = 0\). The first of these conditions implies
\begin{equation}
    \frac{1}{\mathcal{W}R^2} = F_a(z_0)^{-1}\left(-\frac{\pi}{2\gamma R^2}\csc\left(n\pi\right)P_2^n(u_0) + B P_2^{-n}(u_0)\right)
\end{equation}
while the second one implies
\begin{equation}
    \frac{1}{\mathcal{W}R^2} = \left.\left(\od{}{s'}F_a(z')\right)^{-1}\left(-\frac{\pi}{2\gamma R^2}\csc\left(n\pi\right)\od{}{s'}P_2^n(u') + B \od{}{s'}P_2^{-n}(u')\right)\right\vert_{s' = 0}\:.
\end{equation}
These conditions are, in fact, degenerate. One way to prove this is to eliminate \(B\). This superficial difference helps derive a relatively simple expression for \(B\), however, which is
\begin{equation}
    B = -\frac{\pi}{2\gamma R^2}\csc(n\pi) \frac{W(P_2^n(u_0), F_a(z_0))}{W(F_a(z_0), P_2^{-n}(u_0))}\:.
\end{equation}
For any \(s'\), the values and derivatives of the functions \(G_{l, -}^>(s, s';E)\) and \(G_{l, \pm}^>(s, s';E)\) must match at \(s = 0\). The first of these conditions implies
\begin{equation}
    \frac{1}{\mathcal{W}R^2} = \left(P_2^{-n}(u_0)\right)^{-1}\left(-\frac{1}{4 a g\gamma  R^2}F_{-a}(z_0) + D F_a(z_0)\right)\:,
\end{equation}
while the second one implies
\begin{equation}
    \frac{1}{\mathcal{W}R^2} = \left.\left(\od{}{s}P_2^{-n}(u)\right)^{-1}\left(-\frac{1}{4 a g\gamma  R^2}\od{}{s}F_{-a}(z) + D \od{}{s}F_a(z)\right)\right\vert_{s = 0}\:.
\end{equation}
Like before, the difference between these expressions is superficial, but allows us to derive a convenient form for the missing variable,
\begin{equation}
    D = -\frac{1}{4 a g\gamma  R^2}\frac{W\left(P_2^{-n}(u_0), F_{-a}(z_0)\right)}{W\left(F_a(z_0), P_2^{-n}(u_0)\right)}\:.
\end{equation}

Let's summarise our results. The exterior contributions are
\begin{equation}
    \label{eq:pp}
    G_{l, +}^>(s, s';E) = -\frac{\pi}{2\gamma  R^2}\csc\left(n\pi\right)P_2^{-n}(u)\left(P_2^n(u') + \frac{W\left(P_2^n, F_a\right)}{W\left(F_a, P_2^{-n}\right)}P_2^{-n}(u')\right) = G_{l, +}^<(s', s;E)\:,
\end{equation}
the interior contributions are
\begin{equation}
    \label{eq:mm}
    G_{l, -}^>(s, s';E) = -\frac{1}{4 a g\gamma  R^2}F_a(z')\left(F_{-a}(z) + \frac{W\left(P_2^{-n}, F_{-a}\right)}{W\left(F_a, P_2^{-n}\right)}F_a(z)\right) = G_{l, -}^<(s, s';E)\:,
\end{equation}
and the boundary contributions are
\begin{equation}
    \label{eq:pm}
    G^>_{l, \pm}(s, s';E) = \frac{P_2^{-n}(u)F_a(z')}{W(F_a, P_2^{-n})R^2} = G^<_{l, \pm}(s', s;E)\:.
\end{equation}
All Wro{\'n}skians are evaluated at \(s = 0\).

\subsubsection*{Tunnelling regime}
When \(2\gamma <E<g\gamma \), exterior eigenfunctions should experience undamped oscillations while the interior eigenfunctions are still attenuated. The equation for \(y_+(s)\) is now solved by associated Legendre functions of imaginary order
\begin{equation}
    y_+(s) \sim P_2^{\pm i \nu}(u(s))\:,
\end{equation}
where
\begin{equation}
    \nu = \frac{1}{\gamma }\sqrt{E^2 - \frac{l(l+1)}{R^2} - 4\gamma ^2}\:.
\end{equation}
As one would expect, these solutions are oscillatory and have Dirac normalisation \cite{bielski}. When considering the homogeneous boundary conditions, we take the Feynman prescription, \(E^2\rightarrow E^2 + i\epsilon\), equivalent to integrating over a particular contour, shown later. Using similar reasoning as for the bound regime, we find that the contributions are as follows:
\paragraph{Exterior:}
\begin{equation}
    G^>_{l, +}(s, s';E) = A P_2^{i\nu}(u)P_2^{-i\nu}(u') + B P_2^{i\nu}(u)P_2^{i\nu}(u') = G^<_{l, +}(s', s;E)\:.
\end{equation}
\paragraph{Boundary:}
\begin{equation}
    G^>_{l, \pm}(s, s';E) = \frac{P_2^{i\nu}(u) F_{a}(z')}{\mathcal{W}} =G^<_{l, \pm}(s', s;E)\:.
\end{equation}
\paragraph{Interior:}
\begin{equation}
    G_{l, -}^>(s, s';E) = F_a(z')\left(C F_{-a}(z) + D F_a(z)\right) = G_{l, -}^<(s', s;E)\:.
\end{equation}

The derivative jump condition [eq.~\eqref{eq:d-jump}] gives us three of the five constants,
\begin{equation}
    A = -\frac{i \pi}{2\gamma R^2}\operatorname{csch}(\nu\pi)\:,\,\mathcal{W} = W\left[F_a(z_0), P_2^{i\nu}(u_0)\right]\:,\,C = \frac{1}{4ag\gamma R^2}\:,
\end{equation}
while continuity and differentiability give us the remaining two,
\begin{equation}
    B = -\frac{i \pi}{2\gamma R^2}\operatorname{csch}(\nu\pi)\frac{W\left(P_2^{-i\nu}(u_0), F_a(z_0)\right)}{W\left(F_a(z_0), P_2^{i\nu}(u_0)\right)}\:,\, D = \frac{1}{4ag\gamma R^2}\frac{W\left(P_2^{i\nu}(u_0), F_{-a}(z_0)\right)}{W\left(F_a(z_0), P_2^{i\nu}(u_0)\right)}\:.
\end{equation}
In summary, the tunnelling contributions to the Green's function are given by
\begin{equation}
    G^>_{l, +}(s, s';E) =  -\frac{i \pi}{2\gamma R^2}\operatorname{csch}(\nu\pi) P_2^{i\nu}(u)\left(P_2^{-i\nu}(u') + \frac{W\left(P_2^{-i\nu}, F_a\right)}{W\left(F_a, P_2^{i\nu}\right)}P_2^{i\nu}(u')\right)\:,
\end{equation}
\begin{equation}
    G^>_{l, \pm}(s, s';E) = \frac{P_2^{i\nu}(u) F_{a}(z')}{W\left(F_a, P_2^{i\nu}\right)R^2}\:,
\end{equation}
and
\begin{equation}
    G_{l, -}^>(s, s';E) = \frac{1}{4ag\gamma R^2}F_a(z')\left(F_{-a}(z) + \frac{W\left(P_2^{i\nu}, F_{-a}\right)}{W\left(F_a, P_2^{i\nu}\right)} F_a(z)\right)\:.
\end{equation}

\subsubsection*{Scattering regime}
When \(E > g\gamma\), both exterior and interior modes exhibit undamped oscillations. The equation for \(y_-(x)\) is now solved by
\begin{equation}
    y_-(s) \sim  F_{\pm i\alpha}(z(s))\:,
\end{equation}
where
\begin{equation}
    \alpha = \frac{1}{2 g\gamma }\sqrt{E^2 - \frac{l(l+1)}{R^2} - g^2\gamma^2}\:.
\end{equation}
\paragraph{Exterior:}
the form of the exterior solutions does not change.
\begin{equation}
    G^>_{l, +}(s, s';E) = A P_2^{i\nu}(u)P_2^{-i\nu}(u') + B P_2^{i\nu}(u')P_2^{i\nu}(u) = G^<_{l, +}(s', s;E)\:.
\end{equation}
\paragraph{Boundary:} the boundary contributions are
\begin{equation}
    G^>_{l, \pm}(s, s';E) = \frac{P_2^{i\nu}(u) F_{-i\alpha}(z')}{\mathcal{W}} = G^<_{l, \pm}(s', s;E)\:.
\end{equation}
\paragraph{Interior:} the interior contributions are
\begin{equation}
    G^>_{l, -}(s, s';E) = CF_{-i\alpha}(z')F_{i\alpha}(z) + DF_{-i\alpha}(z')F_{-i\alpha}(z) = G^<_{l, -}(s', s;E)\:.
\end{equation}

The jump discontinuity yields parameter values which are similar to those derived in the earlier cases,
\begin{equation}
    A =  -\frac{i \pi}{2\gamma R^2}\operatorname{csch}(\nu\pi)\:,\, \mathcal{W} = W\left(F_{-i\alpha}(z_0), P_2^{i\nu}(u_0)\right)R^2\:,\, C = -\frac{i}{4 \alpha  g\gamma  R^2}\:,
\end{equation}
as does continuity,
\begin{equation}
    B = -\frac{i \pi}{2\gamma R^2}\operatorname{csch}(\nu\pi)\frac{W\left(P_2^{-i\nu}(u_0), F_{-i\alpha}(z_0)\right)}{W\left(F_{-i\alpha}(z_0), P_2^{i\nu}(u_0)\right)}\:,\, D = -\frac{i}{4 \alpha  g\gamma  R^2}\frac{W\left(P_2^{i\nu}(u_0), F_{i\alpha}(z_0)\right)}{W\left(F_{-i\alpha}(z_0), P_2^{i\nu}(u_0)\right)}\:.
\end{equation}
Thus, the scattering contributions to the Green's function are
\begin{equation}
    G_{l, +}^>(s, s';E) = -\frac{i \pi}{2\gamma R^2}\operatorname{csch}(\nu\pi) P_2^{i\nu}(u)\left(P_2^{-i\nu}(u') + \frac{W\left(P_2^{-i\nu}, F_{-i\alpha}\right)}{W\left(F_{-i\alpha}, P_2^{i\nu}\right)} P_2^{i\nu}(u')\right)\:,
\end{equation}
\begin{equation}
    G_{l, \pm}^>(s, s';E) = \frac{P_2^{i\nu}(u) F_{-i\alpha}(z')}{W\left(F_{-i\alpha}, P_2^{i\nu}\right)R^2}\:,
\end{equation}
and
\begin{equation}
    G_{l, -}^>(s, s';E) = -\frac{i}{4 \alpha  g\gamma  R^2}F_{-i\alpha}(z')\left(F_{i\alpha}(z) + \frac{W\left(P_2^{i\nu}, F_{i\alpha}\right)}{W\left(F_{-i\alpha}, P_2^{i\nu}\right)}F_{-i\alpha}(z)\right)\:.
\end{equation}

\subsubsection*{All frequencies}
Naively, one would expect that increasing \(E\) until it is greater than the sum of the angular momentum and mass terms maps \(n\) to \(i\nu\), but the validity of such a statement depends on the branch taken. The branch ambiguity is resolved in the same way we resolved the issue of boundary conditions: by the \(i\epsilon\) prescription.

Consider the expressions
\begin{equation}
    z = \lim_{\epsilon\rightarrow 0}\sqrt{c^2 - (x^2 + i\epsilon)}
\end{equation}
and 
\begin{equation}
    w = \sqrt{x^2 - c^2}\:,
\end{equation}
where \(c > 0\) is fixed and \(x \geq 0\). We claim that \(-z = iw\) for all \(x\geq 0\). The proof is as follows: first, we establish the use of the convention in which complex arguments lie in the interval \((-\pi, \pi]\). With this convention in mind, we note that we may write 
\begin{equation}
    z = \lim_{\epsilon\rightarrow 0}\sqrt{|c^2 - x^2 - i\epsilon|}\exp\left(\frac{i}{2}\operatorname{arg}\left(c^2 - x^2 - i\epsilon\right)\right)
\end{equation}
and 
\begin{equation}
    w =  \sqrt{|x^2 - c^2|}\exp\left(\frac{i}{2}\operatorname{arg}\left(x^2 - c^2\right)\right)\:.
\end{equation}
In the case that \(x=0\), we have
\begin{align}
    \nonumber z &= \lim_{\epsilon\rightarrow 0}\sqrt{|c^2 - i\epsilon|}\exp\left(\frac{i}{2}\operatorname{arg}\left(c^2 - i\epsilon\right)\right)\\
    &= c\:,
\end{align}
and
\begin{align}
    \nonumber w &= \sqrt{|- c^2|}\exp\left(\frac{i}{2}\operatorname{arg}\left(- c^2\right)\right)\\
    &= ce^{i\pi/2}\:.
\end{align}
For the case in which \(0<x<c\), equivalently \(c^2 - x^2 > 0\), we have
\begin{equation}
    z = \sqrt{c^2-x^2}\:,
\end{equation}
while 
\begin{equation}
    w = \sqrt{c^2 - x^2}e^{i\pi/2}\:.
\end{equation}
The proof only becomes nontrivial for \(x>c\), equivalently \(x^2 - c^2 >0\). For these values of \(x\), we have
\begin{equation}
    z = \sqrt{x^2 - c^2}e^{-i\pi/2}\:,
\end{equation}
noting that the exponent comes with a minus sign since the \(i\epsilon\) term pushes \(c^2 - x^2\) below the negative real line, and
\begin{equation}
    w = \sqrt{x^2 - c^2}\:.
\end{equation}
It is clear to see that \(-z = \sqrt{c^2 - x^2}e^{i\pi} = iw\) holds in every case, and so the claim holds for all \(x\geq 0\).

We implement this result by choosing a form for the order of the Legendre functions and hypergeometric parameter that is valid for all \(E\). This can be achieved by setting \(n\) and \(a\) to
\begin{equation}
    \label{eq:n_iep}
    n = \frac{1}{\gamma}\sqrt{\frac{l(l+1)}{R^2} + 4\gamma^2 - \left(E^2 + i\epsilon\right)}
\end{equation}
and
\begin{equation}
    \label{eq:a_iep}
    a = \frac{1}{2g\gamma}\sqrt{\frac{l(l+1)}{R^2} + g^2\gamma^2 - \left(E^2 + i\epsilon\right)}\:,
\end{equation}
where the limit \(\epsilon\rightarrow 0\) is implicit. Thus, without loss of generality, we can adopt the bound regime expressions.

\bibliographystyle{JHEP}
\bibliography{references.bib}

@article{CICOLI20241,
    author = "Cicoli, Michele and Conlon, Joseph P. and Maharana, Anshuman and Parameswaran, Susha and Quevedo, Fernando and Zavala, Ivonne",
    title = "{String cosmology: From the early universe to today}",
    eprint = "2303.04819",
    archivePrefix = "arXiv",
    primaryClass = "hep-th",
    doi = "10.1016/j.physrep.2024.01.002",
    journal = "Phys. Rept.",
    volume = "1059",
    pages = "1--155",
    year = "2024"
}

@article{PhysRevD.88.085038,
    author = "Ferrara, Sergio and Kallosh, Renata and Linde, Andrei and Porrati, Massimo",
    title = "{Minimal Supergravity Models of Inflation}",
    eprint = "1307.7696",
    archivePrefix = "arXiv",
    primaryClass = "hep-th",
    reportNumber = "CERN-PH-TH-2013-178, CERN-TH-13-178",
    doi = "10.1103/PhysRevD.88.085038",
    journal = "Phys. Rev. D",
    volume = "88",
    number = "8",
    pages = "085038",
    year = "2013"
}

@article{PhysRevD.96.123530,
    author = "Nakada, Hiroshi and Ketov, Sergei V.",
    title = "{Inflation from higher dimensions}",
    eprint = "1710.02259",
    archivePrefix = "arXiv",
    primaryClass = "hep-th",
    reportNumber = "IPMU17-0130",
    doi = "10.1103/PhysRevD.96.123530",
    journal = "Phys. Rev. D",
    volume = "96",
    number = "12",
    pages = "123530",
    year = "2017"
}

@article{dm1,
    author = "Brax, Philippe and Davis, Anne-Christine",
    title = "{Modified Gravity and the CMB}",
    eprint = "1109.5862",
    archivePrefix = "arXiv",
    primaryClass = "astro-ph.CO",
    doi = "10.1103/PhysRevD.85.023513",
    journal = "Phys. Rev. D",
    volume = "85",
    pages = "023513",
    year = "2012"
}

@article{dm2,
    author = "Perico, E. L. D. and Voivodic, R. and Lima, M. and Mota, D. F.",
    title = "{Cosmic voids in modified gravity scenarios}",
    eprint = "1905.12450",
    archivePrefix = "arXiv",
    primaryClass = "astro-ph.CO",
    doi = "10.1051/0004-6361/201935949",
    journal = "Astron. Astrophys.",
    volume = "632",
    pages = "A52",
    year = "2019"
}

@article{de1,
    author = "Joyce, Austin and Jain, Bhuvnesh and Khoury, Justin and Trodden, Mark",
    title = "{Beyond the Cosmological Standard Model}",
    eprint = "1407.0059",
    archivePrefix = "arXiv",
    primaryClass = "astro-ph.CO",
    doi = "10.1016/j.physrep.2014.12.002",
    journal = "Phys. Rept.",
    volume = "568",
    pages = "1--98",
    year = "2015"
}

@article{de2,
    author = "Klimchitskaya, Galina L. and Mostepanenko, Vladimir M.",
    title = "{The Nature of Dark Energy and Constraints on Its Hypothetical Constituents from Force Measurements}",
    eprint = "2403.05988",
    archivePrefix = "arXiv",
    primaryClass = "gr-qc",
    doi = "10.3390/universe10030119",
    journal = "Universe",
    volume = "10",
    number = "3",
    pages = "119",
    year = "2024"
}

@article{inf,
    author = "Bernardo, Heliudson and Costa, Renato and Nastase, Horatiu and Weltman, Amanda",
    title = "{Conformal inflation with chameleon coupling}",
    eprint = "1711.10408",
    archivePrefix = "arXiv",
    primaryClass = "hep-th",
    doi = "10.1088/1475-7516/2019/04/027",
    journal = "JCAP",
    volume = "04",
    pages = "027",
    year = "2019"
}

@article{scale1,
    author = "Shaposhnikov, Mikhail and Zenhausern, Daniel",
    title = "{Scale invariance, unimodular gravity and dark energy}",
    eprint = "0809.3395",
    archivePrefix = "arXiv",
    primaryClass = "hep-th",
    doi = "10.1016/j.physletb.2008.11.054",
    journal = "Phys. Lett. B",
    volume = "671",
    pages = "187--192",
    year = "2009"
}

@article{scale2,
    author = "Ferreira, Pedro G. and Hill, Christopher T. and Ross, Graham G.",
    title = "{No fifth force in a scale invariant universe}",
    eprint = "1612.03157",
    archivePrefix = "arXiv",
    primaryClass = "gr-qc",
    reportNumber = "FERMILAB-PUB-16-665-T",
    doi = "10.1103/PhysRevD.95.064038",
    journal = "Phys. Rev. D",
    volume = "95",
    number = "6",
    pages = "064038",
    year = "2017"
}

@article{conformal1,
    author = "Burrage, Clare and Copeland, Edmund J. and Millington, Peter and Spannowsky, Michael",
    title = "{Fifth forces, Higgs portals and broken scale invariance}",
    eprint = "1804.07180",
    archivePrefix = "arXiv",
    primaryClass = "hep-th",
    reportNumber = "IPPP/18/23, IPPP-18-23",
    doi = "10.1088/1475-7516/2018/11/036",
    journal = "JCAP",
    volume = "11",
    pages = "036",
    year = "2018"
}

@article{conformal2,
    author = "Brax, P. and Davis, A. C.",
    title = "{Conformal Inflation Coupled to Matter}",
    eprint = "1401.7281",
    archivePrefix = "arXiv",
    primaryClass = "astro-ph.CO",
    doi = "10.1088/1475-7516/2014/05/019",
    journal = "JCAP",
    volume = "05",
    pages = "019",
    year = "2014"
}

@article{brax_2021,
    author = "Brax, Philippe and Casas, Santiago and Desmond, Harry and Elder, Benjamin",
    title = "{Testing Screened Modified Gravity}",
    eprint = "2201.10817",
    archivePrefix = "arXiv",
    primaryClass = "gr-qc",
    doi = "10.3390/universe8010011",
    journal = "Universe",
    volume = "8",
    number = "1",
    pages = "11",
    year = "2021"
}

@article{khoury_2010,
    author = "Hinterbichler, Kurt and Khoury, Justin",
    title = "{Symmetron Fields: Screening Long-Range Forces Through Local Symmetry Restoration}",
    eprint = "1001.4525",
    archivePrefix = "arXiv",
    primaryClass = "hep-th",
    doi = "10.1103/PhysRevLett.104.231301",
    journal = "Phys. Rev. Lett.",
    volume = "104",
    pages = "231301",
    year = "2010"
}

@article{PhysRevD.99.043539,
    author = {Burrage, Clare and Copeland, Edmund J. and K{\"a}ding, Christian and Millington, Peter},
    title = "{Symmetron scalar fields: Modified gravity, dark matter, or both?}",
    eprint = "1811.12301",
    archivePrefix = "arXiv",
    primaryClass = "astro-ph.CO",
    doi = "10.1103/PhysRevD.99.043539",
    journal = "Phys. Rev. D",
    volume = "99",
    number = "4",
    pages = "043539",
    year = "2019"
}

@article{PhysRevD.99.104049,
    author = "Brax, Philippe and Fichet, Sylvain",
    title = "{Quantum Chameleons}",
    eprint = "1809.10166",
    archivePrefix = "arXiv",
    primaryClass = "hep-ph",
    doi = "10.1103/PhysRevD.99.104049",
    journal = "Phys. Rev. D",
    volume = "99",
    number = "10",
    pages = "104049",
    year = "2019"
}

@article{KADING2025101788,
    author = {K{\"a}ding, Christian},
    title = "{Frequency shifts induced by light scalar fields}",
    eprint = "2410.11567",
    archivePrefix = "arXiv",
    primaryClass = "hep-ph",
    doi = "10.1016/j.dark.2024.101788",
    journal = "Phys. Dark Univ.",
    volume = "47",
    pages = "101788",
    year = "2025"
}

@article{Hinterbichler:2011ca,
    author = "Hinterbichler, Kurt and Khoury, Justin and Levy, Aaron and Matas, Andrew",
    title = "{Symmetron Cosmology}",
    eprint = "1107.2112",
    archivePrefix = "arXiv",
    primaryClass = "astro-ph.CO",
    doi = "10.1103/PhysRevD.84.103521",
    journal = "Phys. Rev. D",
    volume = "84",
    pages = "103521",
    year = "2011"
}

@inbook{weinberg_2012,
	place={Cambridge},
	series={Cambridge Monographs on Mathematical Physics},
	title={One-dimensional solitons},
	booktitle={Classical Solutions in Quantum Field Theory: Solitons and Instantons in High Energy Physics},
	publisher={Cambridge University Press},
	author={Weinberg, Erick J.},
	year={2012},
	pages={6-37},
	collection={Cambridge Monographs on Mathematical Physics}
}

@article{garbrecht_2015_a,
    author = "Garbrecht, Bjorn and Millington, Peter",
    title = "{Green{\textquoteright}s function method for handling radiative effects on false vacuum decay}",
    eprint = "1501.07466",
    archivePrefix = "arXiv",
    primaryClass = "hep-th",
    reportNumber = "TUM-HEP-977-15",
    doi = "10.1103/PhysRevD.91.105021",
    journal = "Phys. Rev. D",
    volume = "91",
    pages = "105021",
    year = "2015"
}

@article{garbrecht_2015_b,
    author = "Garbrecht, Bjorn and Millington, Peter",
    title = "{Self-consistent solitons for vacuum decay in radiatively generated potentials}",
    eprint = "1509.08480",
    archivePrefix = "arXiv",
    primaryClass = "hep-ph",
    reportNumber = "TUM-HEP-1017-15",
    doi = "10.1103/PhysRevD.92.125022",
    journal = "Phys. Rev. D",
    volume = "92",
    pages = "125022",
    year = "2015"
}

@article{garbrecht_2018,
    author = "Garbrecht, Bjorn and Millington, Peter",
    title = "{Fluctuations about the Fubini-Lipatov instanton for false vacuum decay in classically scale invariant models}",
    eprint = "1804.04944",
    archivePrefix = "arXiv",
    primaryClass = "hep-th",
    reportNumber = "TUM-HEP-1136-18",
    doi = "10.1103/PhysRevD.98.016001",
    journal = "Phys. Rev. D",
    volume = "98",
    number = "1",
    pages = "016001",
    year = "2018"
}

@article{ai_2018,
    author = {Ai, Wen-Yuan and Garbrecht, Bj{\"o}rn and Millington, Peter},
    title = "{Radiative effects on false vacuum decay in Higgs-Yukawa theory}",
    eprint = "1807.03338",
    archivePrefix = "arXiv",
    primaryClass = "hep-th",
    reportNumber = "TUM-HEP-1151-18",
    doi = "10.1103/PhysRevD.98.076014",
    journal = "Phys. Rev. D",
    volume = "98",
    number = "7",
    pages = "076014",
    year = "2018"
}

@article{goldstone,
    author = "Goldstone, J. and Jackiw, R.",
    title = "{Quantization of Nonlinear Waves}",
    reportNumber = "MIT-CTP-443",
    doi = "10.1103/PhysRevD.11.1486",
    journal = "Phys. Rev. D",
    volume = "11",
    pages = "1486--1498",
    year = "1975"
}

@article{GARBRECHT2016105,
    author = "Garbrecht, Bjorn and Millington, Peter",
    title = "{Constraining the effective action by a method of external sources}",
    eprint = "1509.07847",
    archivePrefix = "arXiv",
    primaryClass = "hep-th",
    reportNumber = "TUM-HEP-1016-15",
    doi = "10.1016/j.nuclphysb.2016.02.022",
    journal = "Nucl. Phys. B",
    volume = "906",
    pages = "105--132",
    year = "2016"
}

@article{coleman_1977,
    author = "Coleman, Sidney R.",
    title = "{The Fate of the False Vacuum. 1. Semiclassical Theory}",
    reportNumber = "HUTP-77-A004",
    doi = "10.1103/PhysRevD.16.1248",
    journal = "Phys. Rev. D",
    volume = "15",
    pages = "2929--2936",
    year = "1977",
    note = "[Erratum: Phys.Rev.D 16, 1248 (1977)]"
}

@article{callan_1977,
    author = "Callan, Jr., Curtis G. and Coleman, Sidney R.",
    title = "{The Fate of the False Vacuum. 2. First Quantum Corrections}",
    reportNumber = "HUTP-77-A032",
    doi = "10.1103/PhysRevD.16.1762",
    journal = "Phys. Rev. D",
    volume = "16",
    pages = "1762--1768",
    year = "1977"
}

@article{burrage_2021,
    author = "Burrage, Clare and Elder, Benjamin and Millington, Peter and Saadeh, Daniela and Thrussell, Ben",
    title = "{Fifth-force screening around extremely compact sources}",
    eprint = "2104.14564",
    archivePrefix = "arXiv",
    primaryClass = "hep-ph",
    doi = "10.1088/1475-7516/2021/08/052",
    journal = "JCAP",
    volume = "08",
    pages = "052",
    year = "2021"
}

@article{bielski,
	author = {Sebastian Bielski and},
	doi = {10.1080/10652469.2012.690097},
	journal = {Integral Transforms and Special Functions},
	number = {4},
	pages = {331--337},
	publisher = {Taylor \& Francis},
	title = {Orthogonality relations for the associated Legendre functions of imaginary order},
	url = {https://doi.org/10.1080/10652469.2012.690097},
	volume = {24},
	year = {2013}
}

@article{universe10070297,
    author = {Fischer, Hauke and K{\"a}ding, Christian and Pitschmann, Mario},
    title = "{Screened Scalar Fields in the Laboratory and the Solar System}",
    eprint = "2405.14638",
    archivePrefix = "arXiv",
    primaryClass = "gr-qc",
    doi = "10.3390/universe10070297",
    journal = "Universe",
    volume = "10",
    number = "7",
    pages = "297",
    year = "2024"
}

@article{PhysRevD.107.044008,
    author = "Brax, Philippe and Davis, Anne-Christine and Elder, Benjamin",
    title = "{Screened scalar fields in hydrogen and muonium}",
    eprint = "2207.11633",
    archivePrefix = "arXiv",
    primaryClass = "hep-ph",
    doi = "10.1103/PhysRevD.107.044008",
    journal = "Phys. Rev. D",
    volume = "107",
    number = "4",
    pages = "044008",
    year = "2023"
}

@article{openquantum,
    author = {Burrage, Clare and K{\"a}ding, Christian and Millington, Peter and Min{\'a}{\v{r}}, Ji{\v{r}}{\'\i}},
    title = "{Open quantum dynamics induced by light scalar fields}",
    eprint = "1812.08760",
    archivePrefix = "arXiv",
    primaryClass = "hep-th",
    doi = "10.1103/PhysRevD.100.076003",
    journal = "Phys. Rev. D",
    volume = "100",
    number = "7",
    pages = "076003",
    year = "2019"
}

@book{olver,
	address = {Wellesley, MA},
	author = {Olver, F. W. J.},
	isbn = {1-56881-069-5},
	mrclass = {41-02 (33Cxx 41A60 65D20)},
	mrnumber = {MR1429619 (97i:41001)},
	note = {Reprint, with corrections, of original Academic Press edition, 1974},
	pages = {xviii+572},
	publisher = {A. K. Peters},
	title = "{Asymptotics and Special Functions}",
	year = {1997},
	zblno = {0982.41018}
}

@book{andrews,
	address = {Cambridge},
	author = {Andrews, George E. and Askey, Richard and Roy, Ranjan},
	isbn = {0-521-62321-9},
	mrclass = {33-01 (33-02)},
	mrnumber = {MR1688958 (2000g:33001)},
	mrreviewer = {Bruce C. Berndt},
	pages = {xvi+664},
	publisher = {Cambridge University Press},
	series = {Encyclopedia of Mathematics and its Applications},
	title = "{Special Functions}",
	volume = {71},
	year = {1999},
	zblno = {0920.33001}
}

@article{PhysRevD.7.1888,
    author = "Coleman, Sidney R. and Weinberg, Erick J.",
    title = "{Radiative Corrections as the Origin of Spontaneous Symmetry Breaking}",
    doi = "10.1103/PhysRevD.7.1888",
    journal = "Phys. Rev. D",
    volume = "7",
    pages = "1888--1910",
    year = "1973"
}

@article{yin_experimental_2025,
    author = "Yin, Peiran and others",
    title = "{Experimental constraints on the symmetron field with a magnetically levitated force sensor}",
    doi = "10.1038/s41550-024-02465-8",
    journal = "Nature Astron.",
    volume = "9",
    number = "4",
    pages = "598--607",
    year = "2025"
}

@article{derrick_comments_1964,
    author = "Derrick, G. H.",
    title = "{Comments on nonlinear wave equations as models for elementary particles}",
    doi = "10.1063/1.1704233",
    journal = "J. Math. Phys.",
    volume = "5",
    pages = "1252--1254",
    year = "1964"
}

@article{panda,
    author = {Panda, Cristian D. and Tao, Matthew J. and Ceja, Miguel and Khoury, Justin and Tino, Guglielmo M. and M{\"u}ller, Holger},
    title = "{Measuring gravitational attraction with a lattice atom interferometer}",
    eprint = "2310.01344",
    archivePrefix = "arXiv",
    primaryClass = "physics.atom-ph",
    doi = "10.1038/s41586-024-07561-3",
    journal = "Nature",
    volume = "631",
    number = "8021",
    pages = "515--520",
    year = "2024"
}

@article{PhysRevLett.93.171104,
    author = "Khoury, Justin and Weltman, Amanda",
    title = "{Chameleon fields: Awaiting surprises for tests of gravity in space}",
    eprint = "astro-ph/0309300",
    archivePrefix = "arXiv",
    doi = "10.1103/PhysRevLett.93.171104",
    journal = "Phys. Rev. Lett.",
    volume = "93",
    pages = "171104",
    year = "2004"
}

@article{PhysRevD.106.043528,
    author = "Perivolaropoulos, Leandros and Skara, Foteini",
    title = "{Gravitational transitions via the explicitly broken symmetron screening mechanism}",
    eprint = "2203.10374",
    archivePrefix = "arXiv",
    primaryClass = "astro-ph.CO",
    doi = "10.1103/PhysRevD.106.043528",
    journal = "Phys. Rev. D",
    volume = "106",
    number = "4",
    pages = "043528",
    year = "2022"
}

@article{PhysRevD.90.124041,
    author = "Llinares, Claudio and Pogosian, Levon",
    title = "{Domain walls coupled to matter: the symmetron example}",
    eprint = "1410.2857",
    archivePrefix = "arXiv",
    primaryClass = "astro-ph.CO",
    doi = "10.1103/PhysRevD.90.124041",
    journal = "Phys. Rev. D",
    volume = "90",
    number = "12",
    pages = "124041",
    year = "2014"
}

@article{PhysRevD.97.064015,
    author = "Brax, Philippe and Pitschmann, Mario",
    title = "{Exact solutions to nonlinear symmetron theory: One- and two-mirror systems}",
    eprint = "1712.09852",
    archivePrefix = "arXiv",
    primaryClass = "gr-qc",
    doi = "10.1103/PhysRevD.97.064015",
    journal = "Phys. Rev. D",
    volume = "97",
    number = "6",
    pages = "064015",
    year = "2018"
}

@article{PhysRevD.103.084013,
    author = "Pitschmann, Mario",
    title = "{Exact solutions to nonlinear symmetron theory: One- and two-mirror systems. II.}",
    eprint = "2012.12752",
    archivePrefix = "arXiv",
    primaryClass = "gr-qc",
    doi = "10.1103/PhysRevD.103.084013",
    journal = "Phys. Rev. D",
    volume = "103",
    number = "8",
    pages = "084013",
    year = "2021",
    note = "[Erratum: Phys.Rev.D 106, 109902 (2022)]"
}

@article{PhysRevD.99.024045,
    author = "Burrage, Clare and Elder, Benjamin and Millington, Peter",
    title = "{Particle level screening of scalar forces in 1+1 dimensions}",
    eprint = "1810.01890",
    archivePrefix = "arXiv",
    primaryClass = "hep-th",
    doi = "10.1103/PhysRevD.99.024045",
    journal = "Phys. Rev. D",
    volume = "99",
    number = "2",
    pages = "024045",
    year = "2019"
}

@article{PhysRevD.86.102003,
    author = "Upadhye, Amol",
    title = "{Dark energy fifth forces in torsion pendulum experiments}",
    eprint = "1209.0211",
    archivePrefix = "arXiv",
    primaryClass = "hep-ph",
    doi = "10.1103/PhysRevD.86.102003",
    journal = "Phys. Rev. D",
    volume = "86",
    pages = "102003",
    year = "2012"
}

@article{PhysRevLett.110.031301,
  title = "{Symmetron Dark Energy in Laboratory Experiments}",
  author = {Upadhye, Amol},
  journal = {Phys. Rev. Lett.},
  volume = {110},
  issue = {3},
  pages = {031301},
  numpages = {4},
  year = {2013},
  month = {Jan},
  publisher = {American Physical Society},
  doi = {10.1103/PhysRevLett.110.031301},
  url = {https://link.aps.org/doi/10.1103/PhysRevLett.110.031301}
}

@article{cronenberg_acoustic_2018,
    author = "Cronenberg, Gunther and Brax, Philippe and Filter, Hanno and Geltenbort, Peter and Jenke, Tobias and Pignol, Guillaume and Pitschmann, Mario and Thalhammer, Martin and Abele, Hartmut",
    title = "{Acoustic Rabi oscillations between gravitational quantum states and impact on symmetron dark energy}",
    eprint = "1902.08775",
    archivePrefix = "arXiv",
    primaryClass = "hep-ph",
    doi = "10.1038/s41567-018-0205-x",
    journal = "Nature Phys.",
    volume = "14",
    number = "10",
    pages = "1022--1026",
    year = "2018"
}

@article{PhysRevLett.107.111301,
    author = "Brax, Philippe and Pignol, Guillaume",
    title = "{Strongly Coupled Chameleons and the Neutronic Quantum Bouncer}",
    eprint = "1105.3420",
    archivePrefix = "arXiv",
    primaryClass = "hep-ph",
    doi = "10.1103/PhysRevLett.107.111301",
    journal = "Phys. Rev. Lett.",
    volume = "107",
    pages = "111301",
    year = "2011"
}

@article{PhysRevD.88.083004,
    author = "Brax, Philippe and Pignol, Guillaume and Roulier, Damien",
    title = "{Probing Strongly Coupled Chameleons with Slow Neutrons}",
    eprint = "1306.6536",
    archivePrefix = "arXiv",
    primaryClass = "quant-ph",
    doi = "10.1103/PhysRevD.88.083004",
    journal = "Phys. Rev. D",
    volume = "88",
    pages = "083004",
    year = "2013"
}

@article{cannex,
    author = "Sedmik, Ren{\'e} I. P.",
    editor = "Mostepanenko, V. M. and Velichko, E. N.",
    title = "{Casimir and non-Newtonian force experiment (CANNEX): Review, status, and outlook}",
    doi = "10.1142/S0217751X20400084",
    journal = "Int. J. Mod. Phys. A",
    volume = "35",
    number = "02n03",
    pages = "2040008",
    year = "2020"
}

@article{Devoto_2022,
    author = "Devoto, Federica and Devoto, Simone and Di Luzio, Luca and Ridolfi, Giovanni",
    title = "{False vacuum decay: an introductory review}",
    eprint = "2205.03140",
    archivePrefix = "arXiv",
    primaryClass = "hep-ph",
    doi = "10.1088/1361-6471/ac7f24",
    journal = "J. Phys. G",
    volume = "49",
    number = "10",
    pages = "103001",
    year = "2022"
}

@article{PhysRevLett.109.041301,
    author = "Upadhye, Amol and Hu, Wayne and Khoury, Justin",
    title = "{Quantum Stability of Chameleon Field Theories}",
    eprint = "1204.3906",
    archivePrefix = "arXiv",
    primaryClass = "hep-ph",
    doi = "10.1103/PhysRevLett.109.041301",
    journal = "Phys. Rev. Lett.",
    volume = "109",
    pages = "041301",
    year = "2012"
}

@article{universe7070234,
    author = "Sedmik, Ren{\'e} I. P. and Pitschmann, Mario",
    title = "{Next Generation Design and Prospects for Cannex}",
    eprint = "2107.07645",
    archivePrefix = "arXiv",
    primaryClass = "physics.ins-det",
    doi = "10.3390/universe7070234",
    journal = "Universe",
    volume = "7",
    number = "7",
    pages = "234",
    year = "2021"
}

@article{Burrage_2015,
    author = "Burrage, Clare and Copeland, Edmund J. and Hinds, E. A.",
    title = "{Probing Dark Energy with Atom Interferometry}",
    eprint = "1408.1409",
    archivePrefix = "arXiv",
    primaryClass = "astro-ph.CO",
    doi = "10.1088/1475-7516/2015/03/042",
    journal = "JCAP",
    volume = "03",
    pages = "042",
    year = "2015"
}

@article{clements2023detectingdarkdomainwalls,
    author = "Clements, Kate and Elder, Benjamin and Hackermueller, Lucia and Fromhold, Mark and Burrage, Clare",
    title = "{Detecting Dark Domain Walls}",
    eprint = "2308.01179",
    archivePrefix = "arXiv",
    primaryClass = "gr-qc",
    doi = "10.1103/PhysRevD.109.123023",
    journal = "Phys. Rev. D",
    volume = "109",
    number = "12",
    pages = "123023",
    year = "2024"
}

@article{astronomy2020009,
    author = {K{\"a}ding, Christian},
    title = "{Lensing with Generalized Symmetrons}",
    eprint = "2304.05875",
    archivePrefix = "arXiv",
    primaryClass = "astro-ph.CO",
    doi = "10.3390/astronomy2020009",
    journal = "Astronomy",
    volume = "2",
    number = "2",
    pages = "128--140",
    year = "2023"
}

@article{PhysRevLett.117.211102,
    author = "Burrage, Clare and Copeland, Edmund J. and Millington, Peter",
    title = "{Radiative Screening of Fifth Forces}",
    eprint = "1604.06051",
    archivePrefix = "arXiv",
    primaryClass = "gr-qc",
    doi = "10.1103/PhysRevLett.117.211102",
    journal = "Phys. Rev. Lett.",
    volume = "117",
    number = "21",
    pages = "211102",
    year = "2016"
}

@article{universe11050158,
    author = "Bachs-Esteban, Joan and Lopes, Il{\'\i}dio and Rubio, Javier",
    title = "{Screening Mechanisms on White Dwarfs: Symmetron and Dilaton}",
    eprint = "2505.05871",
    archivePrefix = "arXiv",
    primaryClass = "gr-qc",
    doi = "10.3390/universe11050158",
    journal = "Universe",
    volume = "11",
    number = "5",
    pages = "158",
    year = "2025"
}

@article{PhysRevD.101.083501,
    author = "Chiow, Sheng-wey and Yu, Nan",
    title = "{Constraining symmetron dark energy using atom interferometry}",
    eprint = "1911.00441",
    archivePrefix = "arXiv",
    primaryClass = "astro-ph.CO",
    doi = "10.1103/PhysRevD.101.083501",
    journal = "Phys. Rev. D",
    volume = "101",
    number = "8",
    pages = "083501",
    year = "2020"
}

@article{PhysRevLett.123.061102,
    author = "Sabulsky, Dylan O. and Dutta, Indranil and Hinds, E. A. and Elder, Benjamin and Burrage, Clare and Copeland, Edmund J.",
    title = "{Experiment to detect dark energy forces using atom interferometry}",
    eprint = "1812.08244",
    archivePrefix = "arXiv",
    primaryClass = "physics.atom-ph",
    doi = "10.1103/PhysRevLett.123.061102",
    journal = "Phys. Rev. Lett.",
    volume = "123",
    number = "6",
    pages = "061102",
    year = "2019"
}

@article{PhysRevD.97.084050,
    author = "Brax, Philippe and Davis, Anne-Christine and Elder, Benjamin and Wong, Leong Khim",
    title = "{Constraining screened fifth forces with the electron magnetic moment}",
    eprint = "1802.05545",
    archivePrefix = "arXiv",
    primaryClass = "hep-ph",
    doi = "10.1103/PhysRevD.97.084050",
    journal = "Phys. Rev. D",
    volume = "97",
    number = "8",
    pages = "084050",
    year = "2018"
}

@article{BRAX2023101294,
    author = "Brax, Philippe and Fichet, Sylvain",
    title = "{Scalar-mediated quantum forces between macroscopic bodies and interferometry}",
    eprint = "2203.01342",
    archivePrefix = "arXiv",
    primaryClass = "hep-th",
    doi = "10.1016/j.dark.2023.101294",
    journal = "Phys. Dark Univ.",
    volume = "42",
    pages = "101294",
    year = "2023"
}

@article{dong_symmetron_2014,
    author = "Dong, Ruifeng and Kinney, William H. and Stojkovic, Dejan",
    title = "{Symmetron Inflation}",
    eprint = "1307.4451",
    archivePrefix = "arXiv",
    primaryClass = "astro-ph.CO",
    doi = "10.1088/1475-7516/2014/01/021",
    journal = "JCAP",
    volume = "01",
    pages = "021",
    year = "2014"
}

@article{PhysRevD.108.024007,
    author = {H{\"o}g{\r{a}}s, Marcus and M{\"o}rtsell, Edvard},
    title = "{Impact of symmetron screening on the Hubble tension: New constraints using cosmic distance ladder data}",
    eprint = "2303.12827",
    archivePrefix = "arXiv",
    primaryClass = "astro-ph.CO",
    doi = "10.1103/PhysRevD.108.024007",
    journal = "Phys. Rev. D",
    volume = "108",
    number = "2",
    pages = "024007",
    year = "2023"
}

@article{PhysRevD.105.103536,
    author = "Solomon, Rance and Agarwal, Garvita and Stojkovic, Dejan",
    title = "{Environment dependent electron mass and the Hubble constant tension}",
    eprint = "2201.03127",
    archivePrefix = "arXiv",
    primaryClass = "hep-ph",
    doi = "10.1103/PhysRevD.105.103536",
    journal = "Phys. Rev. D",
    volume = "105",
    number = "10",
    pages = "103536",
    year = "2022"
}

@article{schwinger_brownian_1961,
    author = "Schwinger, Julian S.",
    title = "{Brownian motion of a quantum oscillator}",
    doi = "10.1063/1.1703727",
    journal = "J. Math. Phys.",
    volume = "2",
    pages = "407--432",
    year = "1961"
}

@article{Keldysh:1964ud,
    author = "Keldysh, L. V.",
    title = "{Diagram Technique for Nonequilibrium Processes}",
    doi = "10.1142/9789811279461_0007",
    journal = "Sov. Phys. JETP",
    volume = "20",
    pages = "1018--1026",
    year = "1965"
}

\end{document}